\documentclass{mn2e}
\usepackage{epsf,times}
\usepackage{longtable}
\newcommand{\etal}{{et al}\/.}
\begin{document}
\title[Nuclei of 3CRR radio sources]{The active nuclei of $z<1.0$
  3CRR radio sources}
\author[M.J.~Hardcastle \etal]{M.J.\ Hardcastle$^1$, D.A. Evans$^{2,3}$
  and J.H. Croston$^1$\\
$^1$ School of Physics,
  Astronomy and Mathematics, University of
Hertfordshire, College Lane, Hatfield, Hertfordshire AL10 9AB\\
$^2$ Massachusetts Institute of Technology, Kavli Institute for
Astrophysics and Space Research, 77 Massachusetts Avenue, Cambridge, MA
02139, USA\\
$^3$ Harvard-Smithsonian Center for Astrophysics, 60 Garden Street,
  Cambridge, MA~02138, USA}
\maketitle
\begin{abstract}
We combine {\it Chandra} and {\it XMM-Newton} X-ray data from our
previous papers with new X-ray observations and with {\it Spitzer}
mid-infrared data in order to study the nature of the nuclei of radio
galaxies and radio-loud quasars with $z<1.0$ from the 3CRR sample. The
significant increase in sample size over our previous work, the
reduction of bias in the sample as a result of new observations, and
the availability of more mid-infrared data allow us to show
conclusively that almost all objects classed as low-excitation radio
galaxies in optical spectroscopic studies lack a radiatively efficient
active nucleus. We show that the distribution of absorbing columns in
the narrow-line radio galaxies differs from the population of
X-ray-selected radio-quiet type-2 quasars and from that in local
Seyfert 2s. We comment on the current evidence for the nature of the
soft X-ray component in radio-galaxy nuclear spectra, concluding that
a jet origin for this component is very hard to evade. Finally, we
discuss the recently discovered `fundamental plane' of black hole
activity, showing that care must be taken when placing radio-loud AGN
on such diagnostic diagrams.
\end{abstract}

\section{Introduction}

The nature of the active nuclei of radio galaxies and radio-loud
quasars has been a puzzle ever since large samples with optical
identification and spectroscopy began to be constructed in the 1970s.
Early work connected extragalactic radio sources with a bewildering
variety of host objects, ranging in the local universe from recent
merger remnants like the host galaxy of Centaurus A, NGC 5128, to
quiescent ellipticals like M87, and in the more distant universe from
undistinguished ellipticals with only stellar features in their
optical spectra to the most powerful quasars known. Key steps in the
understanding of this diversity included the discovery by Fanaroff \&
Riley (1974) that large-scale radio structure has a strong
relationship to radio luminosity: their division of radio sources into
centre-brightened (FRI) and edge-brightened (FRII) classes clearly
encodes some important jet physics. Equally important, though far less
widely cited, was the discovery by Hine \& Longair (1979) that the
optical emission lines from the nuclei of radio galaxy hosts could
also be classified as `weak' or `strong', with a relationship to radio
luminosity (weak-lined objects, hereafter low-excitation radio
galaxies or LERGs tend to have low radio luminosity; strong-lined
objects, hereafter high-excitation radio galaxies or HERGs, tend to
have high radio luminosities) but, crucially, no one-to-one
correspondence with the Fanaroff-Riley morphological classes.

The development of unified models, beginning with the realization that
both relativistic jets (e.g. Scheuer \& Readhead 1979; Orr \& Browne
1982) and active nuclei (e.g. Antonucci 1982) would have different
appearances depending on the orientation of the observer, and
culminating in quantitative constraints on source properties through
population statistics (e.g. Barthel 1989; Urry, Padovani \& Stickel
1991; Padovani \& Urry 1992; Hardcastle \etal\ 2003) greatly
simplified the picture (see Antonucci 1993 and Urry \& Padovani 1995
for contemporary reviews). It became clear that quasars and high-power
narrow-line radio galaxies (NLRGs) were likely to be the same
population, seen at different orientations, and that low-power radio
galaxies, which are generally LERGs (Hine \& Longair 1979) could form
the parent population of the mostly lineless BL Lac objects. A
tendency to think of these models respectively as `FRII' and `FRI'
unification, after the objects which dominate the HERG and LERG
populations, has confused the literature ever since, and led to much
work on physical differences in the nuclei of FRI and FRII radio
galaxies. In fact as Hine \& Longair showed, and as has been
repeatedly pointed out over the ensuing decades (e.g. Barthel 1994;
Laing \etal\ 1994; Jackson \& Rawlings 1997; Chiaberge, Capetti \&
Celotti 2002; Hardcastle 2004; Whysong \& Antonucci 2004) there is a
population of FRII radio galaxies with low-excitation optical spectra,
which modellers should not ignore; in fact, such low-excitation FRIIs
are required to participate in low-luminosity unified models in order
to explain the numerous BL Lac objects with FRII radio structure (e.g.
Rector \& Stocke 2001). There are clearly also radio sources with FRI
structure but high-excitation optical characteristics (e.g. the
broad-line FRI radio galaxy 3C\,120, or the FRI quasar of Blundell \&
Rawlings 2001), although such objects are rarer in radio-selected
samples.

The key unanswered question is therefore not `what causes the
differences between FRI and FRII radio structures?' -- that can be
adequately explained by the interactions between jets of different
powers and their environments, see e.g. Bicknell (1995) -- but `what
is the nature and the cause of the differences between the
low-excitation and high-excitation active nuclei?'. This question has
taken on a new urgency with the advent of large surveys which provide
populations of homogeneously selected AGN, both radio-loud and
radio-quiet, and show significant differences between their host
galaxies and environments as a function of radio power (e.g. Best
\etal\ 2006). Recently a consensus has started to emerge (Chiaberge
\etal\ 2002; Whysong \& Antonucci 2004; Hardcastle, Evans \& Croston
2006, hereafter H06) that the LERGs in fact lack any of the
conventional apparatus of an AGN -- radiatively efficient accretion
disc, X-ray emitting corona, and obscuring, mid-infra-red luminous
torus -- and that their radio through X-ray nuclear emission, and even
such nuclear emission lines as are seen, can be explained purely as a
result of the properties of the small-scale jet. In this picture some
feature of the AGN and/or its fuel supply must account for the
difference between the LERGs and the HERGs (i.e. NLRGs, broad-line
radio galaxies (BLRGs) and quasars). We recently proposed (Hardcastle,
Evans \& Croston 2007, hereafter H07) that the low-excitation objects
are those fuelled by accretion directly from the hot intergalactic
medium, as is required in models in which cooling from the hot phase
triggers AGN activity and re-heats the cold gas, while the
high-excitation objects (where large amounts of cold gas close to the
AGN are required in order to produce the observed torus and the
inferred thin, radiatively efficient accretion disc) are produced by
accretion of cold gas, presumably driven by interaction or mergers. We
showed quantitatively that `hot-mode' accretion could power nearby FRI
LERGs (an aspect of the model subsequently tested in more detail by
Balmaverde \etal\ 2008) but that it was not sufficient to power
the most powerful NLRGs. The H07 model appears to be supported by studies
relating the AGN properties and environments of radio sources in large
surveys (e.g. Tasse \etal\ 2008).

However, more work remains to be done on the nature of AGN in both
low- and high-excitation radio sources. To date studies that have
shown evidence for radiative inefficiency in LERGs have concentrated
on a single wavelength, either optical (Chiaberge \etal\ 2002),
infra-red (Whysong \& Antonucci 2004; Ogle \etal\ 2006) or X-ray
(H06). At the same time, they have necessarily relied on small
samples, since the relevant data are slow to accumulate for all-sky
samples of radio sources, such as the 3CRR sample (Laing, Riley \&
Longair 1983), the sample that formed the basis of our earlier work
(Evans \etal\ 2006; H06; Belsole \etal\ 2006). Over the past few years
we have been carrying out a programme of X-ray observations of 3CRR
objects, primarily with {\it XMM-Newton} (see Evans \etal\ 2008 for a
discussion of some of these). At the same time, observations by others
in both X-ray and mid-infra-red (using {\it Spitzer}) have
substantially increased the coverage of the 3CRR sample from the
situation in 2006, which allows us for the first time to take a truly
multi-wavelength view of a large sample. In the present paper we
therefore present an update on the $z<1.0$ 3CRR sample discussed in our
previous papers. Using newly analysed and existing X-ray and infra-red
data, we revisit the nature of low-excitation objects and the
relationship between the high-excitation population and conventional
radio-quiet AGN. We discuss the distribution of absorbing column in
the narrow-line radio galaxies and their relationship to the
population of radio-quiet type-2 quasars, the current state of the
evidence for a jet origin of the soft X-ray component in radio-galaxy
nuclear spectra, and the correct approach to positioning radio
galaxies on the `fundamental plane' of black hole activity (e.g.
Merloni, Heinz \& di Matteo 2003).

Throughout the paper, we use a
concordance cosmology with $H_0 = 70$ km s$^{-1}$ Mpc$^{-1}$,
$\Omega_{\rm m} = 0.3$ and $\Omega_\Lambda = 0.7$. Where luminosities
tabulated by others are used, they may be assumed to have been
corrected to this cosmology.

\section{Data and analysis}

\subsection{Sample}

Our parent sample is the 3CRR catalogue (Laing, Riley \& Longair 1983;
hereafter LRL). In this paper, to ensure good X-ray coverage, we
consider all the 3CRR objects with $z<1.0$. This redshift range allows
us to include data from the previous X-ray work by Evans \etal\
(2006), H06 and Belsole \etal\ (2006). In the infrared (IR), the 3CRR
objects in this redshift range have been studied by Ogle \etal\ (2006)
and Cleary \etal\ (2007). To obtain the largest possible sample, we
supplement the published X-ray and radio data with new data from the
{\it Chandra}, {\it XMM-Newton} and {\it Spitzer} archives, as
described in the following subsections. By definition, 3CRR objects
have a known low-frequency radio luminosity, and all now have a known
redshift and a measurement of core flux density at or near 5
GHz\footnote{See http://3crr.extragalactic.info/ for a compilation of
these values.}: in addition, emission-line fluxes and galaxy
magnitudes are available for many of them. LRL qualitatively
classified the emission-line types for all the objects in the sample,
and for consistency with our earlier work we adopt their
classifications here. Emission-line luminosities for the sample were
compiled by Willott \etal\ (1999), although emission-line studies of
3CRR objects are still hampered by the lack of homogeneous
optical/near-IR spectroscopic information.

There are 135 sources in the 3CRR sample with $z<1.0$, excluding the
starburst galaxy M82 (3C\,231). Of those, at the time of writing, 73
have been observed in imaging spectroscopy modes with {\it Chandra}
and 39 with {\it XMM}, giving a total of 89 (66 per cent) with
current-generation X-ray observations. We choose to use only {\it
Chandra} and {\it XMM} data to avoid the problems of contamination
imposed by the very low resolution of instruments such as {\it ASCA}.
100 $z<1.0$ 3CRR sources have been observed with {\it Spitzer} in the
IRS `Staring' mode, which gives us access to mid-IR photometry,
and there are 69 objects with both X-ray and IRS-S observations.
Nearly all the archival data are publicly available. In terms of
emission-line classifications, 27/34 LERGs, 40/66 NLRGs, 7/10 BLRGs and
15/24 quasars have been observed in the X-ray, while 25/34 LERGs, 45/66
NLRGs, 8/10 BLRGs and 22/24 quasars have IRS-S data. From this it can be
seen that there is no longer significant bias in terms of
emission-line type in the X-ray sample, partly as a result of our
observing programmes specifically aimed at removing that bias (see the
following subsection). In the mid-IR, there is a bias towards
broad-line objects that can be understood when one considers that
several large-sample {\it Spitzer} studies have been carried out with
the aim of testing unified models.

\begin{figure}
\epsfxsize 8.5cm
\epsfbox{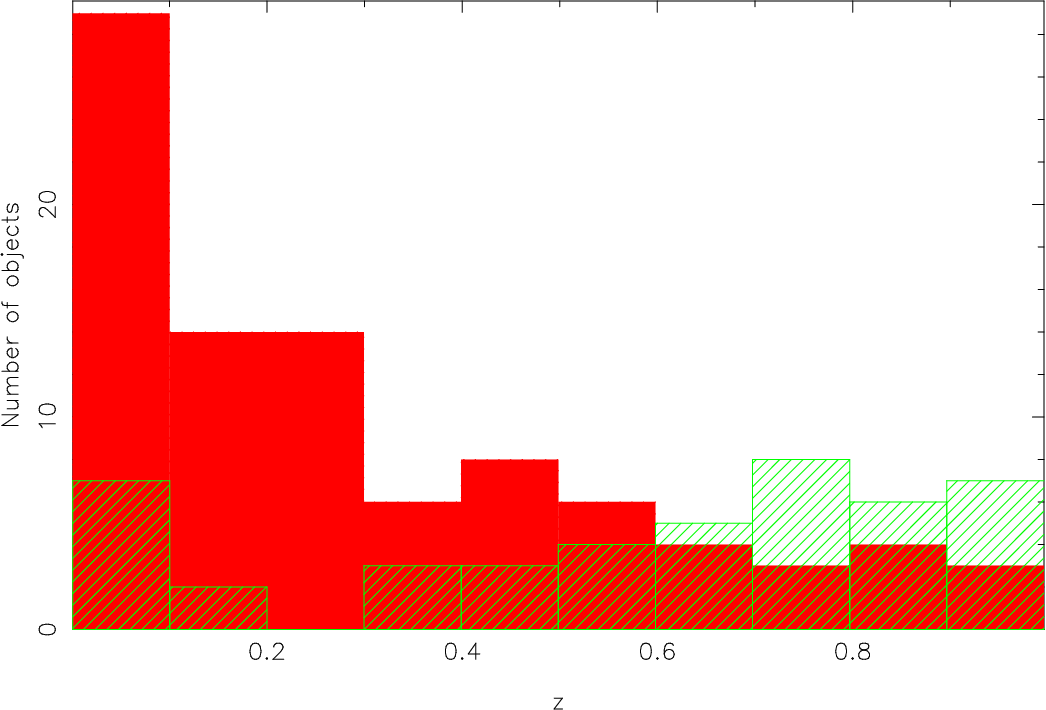}
\vskip 6pt
\epsfxsize 8.5cm
\epsfbox{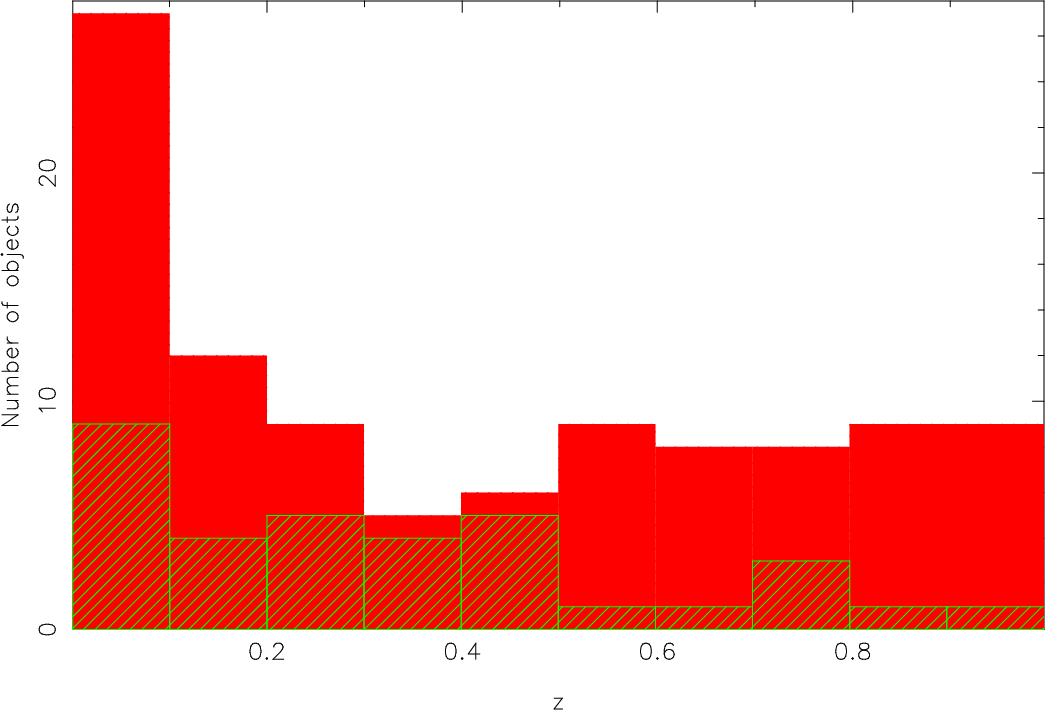}
\caption{Redshift distributions for observed (red) and unobserved
  (green) sources with {\it Chandra} and {\it XMM-Newston} (top) and with the {\it Spitzer} IRS (bottom).}
\label{zdist}
\end{figure}

The samples observed are, however, biased in terms of their redshift
distribution. Fig.\ \ref{zdist} shows that, while X-ray observations
approach completeness for the $z<0.5$ subsample, they are very much
less complete for $z>0.5$. By contrast, the {\it Spitzer} data are
more complete at high redshifts than at intermediate ones. Again, this
seems likely to be a consequence of the different purposes for which
the X-ray and IR data have been taken.

\subsection{X-ray data}

The majority of the X-ray data comes from earlier work by Evans \etal\
(2006), H06, Belsole \etal\ (2006) and Evans
\etal\ (2008). Belsole \etal\ (2006) did not determine upper limits on
heavily-absorbed nuclear components for those sources where they did
not fit a two-component model, and so we re-analysed their data for
such sources. We also scaled their $1\sigma$ error estimates to 90 per
cent confidence errors for consistency with the approach taken by
the other papers.

However, a significant fraction of our sample (31 objects) were too
recently observed to be covered by these papers. Observational details
for these objects are listed in Table \ref{sources}. A significant
fraction of the data comes from the {\it Chandra} 3CR snapshot survey
(Massaro \etal\ in prep.; data for 3C\,33.1, 132, 153, 171, 184.1,
293, 300, 315 and 381) while some is from our dedicated {\it XMM}
programme targeting LERGs and NLRGs (4C\,12.03, 3C\,16, 3C\,20,
4C\,14.11, 3C\,244.1, 3C\,349). For these objects we analysed the data
(either from the archives or from our own observations) in the manner
we described in H06, using {\sc ciao} 3.4 and CALDB 3.4 for the {\it
Chandra} data, and {\sc sas} 7.1.0 for the {\it XMM} data. Data preparation
and filtering and spectral extraction all exactly
duplicate the approach of H06, and are described there.

Spectral fitting, in the energy range 0.4-7.0 keV (for {\it Chandra}
data) or 0.3-8.0 keV ({\it XMM-Newton}), was also exactly the same as
for H06. That is, we first fitted a single power law with fixed
Galactic absorption to the data (see Table \ref{sources} for the
Galactic column densities used). For those sources where a second
component of X-ray emission was clearly required (seen in large
residuals and poor $\chi^2$ values) we then added a second power law
with a free, but initially large, absorbing column at the redshift of
the source. This gave rise to good fits in almost all the sources
where the $\chi^2$ was initially poor, though occasionally it was
necessary to fix the index of the unabsorbed power-law to $\Gamma =
2.0$ to get good constraints on its normalization. If a single power
law provided a good fit to the data, we added a heavily absorbed power
law with fixed $\Gamma = 1.7$ and an absorbing column $N_{\rm H} =
10^{23}$ cm$^{-2}$ at the redshift of the source and re-fitted for the
normalization of this new component. If the 90 per cent uncertainty on
the normalization was consistent with zero, we treated the upper bound
as an upper limit on a heavily absorbed component. If, on the other
hand, the fit was improved with a non-zero normalization for this
component, we allowed the absorbing column and, if well-constrained,
the photon index of the second power law to vary, and treated the
resulting model as a detection of a second component. Throughout the
rest of the paper we refer, following H06, to the first component as
the `unabsorbed' component -- since it has little or no intrinsic
absorption -- and the second, heavily absorbed component as the
`accretion-related' component. We refer to the absorption-corrected
luminosities of these components as $L_{\rm Xu}$ and $L_{\rm Xa}$
respectively. In quasars and BLRGs where a single power-law model is
fitted, and where we have reason to believe that we may be seeing the
accretion disc directly, we use $L_{\rm Xu}$ as our best estimate of
$L_{\rm Xa}$: thus for a few sources $L_{\rm Xu} = L_{\rm Xa}$ by
definition, but in most cases the two luminosities are largely
independent.

Results of the spectral fits for each of the newly analysed sources
are given in Table \ref{results}; as in H06, errors on the fitted
parameters are 90 per cent confidence for one interesting parameter
($\Delta\chi^2 = 2.7$), though in all other contexts that errors are
quoted in this paper they are the conventional $1\sigma$ errors
(corresponding to $\Delta\chi^2 = 1$). Individual sources are
discussed, and references to previous work given, in Appendix A.

\begin{table*}
\caption{Basic parameters and X-ray observational information for
  sources in the sample with new X-ray data. For {\it XMM-Newton} data
  the livetimes are for the MOS1, MOS2 and pn instruments, in
  that order. Where multiple observation IDs are quoted the livetime
  is the combined total. Galactic $N_{\rm H}$ values are derived from
  the {\sc colden} software.}

\label{sources}
\begin{tabular}{lrrrrll}
\hline
Source&$z$&Emission-&Galactic $N_{\rm H}$&Telescope&Observation&Livetime\\
&&line type&($\times 10^{20}$ cm$^{-2}$)&&ID&(s)\\
\hline
4C\,12.03 & 0.156   & LERG & 5.26 &{\it XMM-Newton} &0551760101 &10324, 10974, 5307\\
3C\,6.1   & 0.8404  & NLRG &17.49 &{\it Chandra} &3009, 4363 &56390\\
3C\,16    & 0.405   & LERG & 4.89 &{\it XMM-Newton} &0551760201 &12179, 13197, 5049\\
3C\,20    & 0.174   & NLRG &17.93 &{\it XMM-Newton} &0551760501 &25303, 26532, 14182\\
3C\,33.1  & 0.181   & BLRG &22.50 &{\it Chandra} &9295 &8068\\
3C\,48    & 0.367   & Q    & 4.80 &{\it Chandra} &3097 &9225\\
3C\,61.1  & 0.186   & NLRG & 7.57 &{\it XMM-Newton} &0500910101 &25388, 25239, 13143\\
3C\,76.1  & 0.0324  & LERG &10.10 &{\it XMM-Newton} &0201860201 &17762, 18165, 10961\\
4C\,14.11 & 0.206   & LERG &15.51 &{\it XMM-Newton} &0501620101 &15791, 15788, 9533\\
3C\,132   & 0.214   & NLRG &22.14 &{\it Chandra} &9329 &7692\\
3C\,153   & 0.2769  & NLRG &17.40 &{\it Chandra} &9302 &8065\\
3C\,171   & 0.2384  & NLRG & 6.60 &{\it Chandra} &10303 &59461\\
3C\,184.1 & 0.1187  & NLRG & 3.35 &{\it Chandra} &9305 &8022\\
3C\,220.1 & 0.61    & NLRG & 1.93 &{\it Chandra} &839 &18922\\
3C\,228   & 0.5524  & NLRG & 3.28 &{\it Chandra} &2095, 2453 &24391\\
3C\,234   & 0.1848  & NLRG & 1.90 &{\it XMM-Newton} &0405340101 &35698, 36050, 28758\\
3C\,244.1 & 0.428   & NLRG & 0.67 &{\it XMM-Newton} &0501621501 &8646, 8867, 2417\\
3C\,274.1 & 0.422   & NLRG & 2.06 &{\it Chandra} &0551760601 &15246\\
3C\,277.2 & 0.766   & NLRG & 1.93 &{\it XMM-Newton} &0082990101 &41089, 41320, 29950\\
3C\,288   & 0.246   & LERG & 0.81 &{\it Chandra} &9275 &39642\\
3C\,293   & 0.0452  & LERG & 1.29 &{\it Chandra} &9310 &7814\\
3C\,300   & 0.272   & NLRG & 2.44 &{\it XMM-Newton} &0500910601, 0500910901, 500911101 &19699, 42415, 42344\\
3C\,315   & 0.1083  & NLRG & 4.62 &{\it Chandra} &9313 &7670\\
3C\,325   & 0.86    & Q    & 1.74 &{\it Chandra} &4818, 6267 &58308\\
3C\,349   & 0.205   & NLRG & 1.99 &{\it XMM-Newton} &0501620301 &22533, 22089, 15170\\
3C\,381   & 0.1605  & BLRG & 5.87 &{\it Chandra} &9317 &8065\\
3C\,427.1 & 0.572   & LERG &11.60 &{\it Chandra} &2194 &39452\\
3C\,433   & 0.1016  & NLRG & 9.15 &{\it Chandra} &7881 &37172\\
3C\,442A  & 0.027   & LERG & 5.08 &{\it Chandra} &5635 &27006\\
3C\,457   & 0.428   & NLRG & 5.06 &{\it XMM-Newton} &0502500101 &52172, 52185, 29735\\

\hline
\end{tabular}
\end{table*}

\begin{table*}
\caption{Results of spectral fitting. For each source the fits, or
  upper limits, for an absorbed and unabsorbed power law are shown.
  The model described is the best-fitting model (PL = power law,
  ABS(PL) = absorbed power law, TH = thermal component, GAU =
  Gaussian). Components of the fit other than power-law components are
  discussed in Appendix A. Numerical values marked with a dagger were
  frozen in the fit (either to derive upper limits or because the data
  were not good enough to constrain them). 1-keV flux densities and
  luminosities are the unabsorbed values in all cases. Only pn count
  rates are shown for {\it XMM} upper limits.}
\label{results}
\begin{tabular}{lllllllll}
\hline
Source&Net counts&Model&$\chi^2/$d.o.f.&Component&1-keV flux&Photon
&log$_{10}$&$N_{\rm H}$\\
&&&&&(nJy)&index&luminosity&($\times 10^{22}$\\
&&&&&&&(ergs s$^{-1}$)&cm$^{-2}$)\\
\hline
4C\,12.03 & $<$45 & NONE &-- &PL&$<$3.2&--&$<$41.91 &--\\
&&&&ABS(PL)&$<$26.5&--&$<$43.02 &10.0\dag\\
3C\,6.1   & 2497 $\pm$ 51& PL &119.4/101 &PL&36.7$_{-1.8}^{+1.7}$ & 1.44$_{-0.06}^{+0.06}$ & 44.92 &--\\
&&&&ABS(PL)&$<$8.5&--&$<$44.17 &10.0\dag\\
3C\,16    & $<$30 &  &-- &PL&$<$2.4&--&$<$42.74 &--\\
&&&&ABS(PL)&$<$15.1&--&$<$43.69 &10.0\dag\\
3C\,20    & 490 $\pm$ 26, 530 $\pm$ 26, & PL+ABS(PL) &87.0/80 &PL&5.7$_{-1.3}^{+1.2}$ & 1.53$_{-0.61}^{+0.53}$ & 42.56 &--\\
&819 $\pm$ 34&&&ABS(PL)&211$_{-48}^{+309}$ & 1.65$_{-0.19}^{+0.21}$ & 44.05 &18.2$_{-3.9}^{+5.4}$\\
3C\,33.1  & 761 $\pm$ 28& PL+ABS(PL) &34.6/31 &PL&7.5$_{-3.6}^{+3.4}$ & 2.00\dag & 42.43 &--\\
&&&&ABS(PL)&135$_{-34}^{+122}$ & 0.89$_{-0.37}^{+0.33}$ & 44.38 &4.1$_{-1.4}^{+1.6}$\\
3C\,48    & 6471 $\pm$ 83& PL &227.3/145 &PL&518$_{-11}^{+11}$ & 1.93$_{-0.04}^{+0.04}$ & 45.00 &--\\
3C\,61.1  & 76 $\pm$ 14, 96 $\pm$ 14, & PL+ABS(PL) &8.7/8 &PL&2.2$_{-1.2}^{+1.2}$ & 2.00\dag & 41.92 &--\\
&139 $\pm$ 19&&&ABS(PL)&148$_{-52}^{+70}$ & 1.70\dag & 43.93 &56.0$_{-20.2}^{+26.7}$\\
3C\,76.1  & 93 $\pm$ 14, 103 $\pm$ 14, & TH+PL &16.9/15 &PL&3.9$_{-1.6}^{+2.1}$ & 0.95$_{-0.24}^{+0.27}$ & 41.28 &--\\
&223 $\pm$ 22&&&ABS(PL)&$<$21.7&--&$<$41.26 &10.0\dag\\
4C\,14.11 & 99 $\pm$ 14, 94 $\pm$ 14, & PL &12.9/14 &PL&8.0$_{-1.1}^{+1.1}$ & 1.30$_{-0.20}^{+0.20}$ & 43.01 &--\\
&174 $\pm$ 16&&&ABS(PL)&$<$8.4&--&$<$42.78 &10.0\dag\\
3C\,132   & 43 $\pm$ 7& PL+ABS(PL) &-- &PL&$<$1.9&--&$<$41.99 &--\\
&&&&ABS(PL)&22.6$_{-8.4}^{+9.5}$ & 1.70\dag & 43.25 &4.7$_{-1.8}^{+2.8}$\\
3C\,153   & $<$7 & NONE &-- &PL&$<$1.1&--&$<$41.99 &--\\
&&&&ABS(PL)&$<$5.6&--&$<$42.89 &10.0\dag\\
3C\,171   & 1348 $\pm$ 38& PL+ABS(PL) &50.6/57 &PL&1.1$_{-0.3}^{+0.4}$ & 2.00\dag & 41.86 &--\\
&&&&ABS(PL)&118$_{-28}^{+31}$ & 1.67$_{-0.23}^{+0.20}$ & 44.08 &8.5$_{-1.3}^{+1.4}$\\
3C\,184.1 & 544 $\pm$ 24& PL+ABS(PL) &18.7/23 &PL&3.7$_{-1.8}^{+1.7}$ & 2.00\dag & 41.73 &--\\
&&&&ABS(PL)&64$_{-24}^{+65}$ & 0.57$_{-0.34}^{+0.30}$ & 43.91 &3.7$_{-1.3}^{+1.6}$\\
3C\,220.1 & 1034 $\pm$ 33& PL &32.4/42 &PL&31.2$_{-1.8}^{+1.7}$ & 1.52$_{-0.08}^{+0.08}$ & 44.50 &--\\
&&&&ABS(PL)&$<$13.1&--&$<$44.04 &10.0\dag\\
3C\,228   & 594 $\pm$ 25& PL+ABS(PL) &21.0/24 &PL&16.4$_{-14.9}^{+2.1}$ & 2.07$_{-0.33}^{+7.93}$ & 43.86 &--\\
&&&&ABS(PL)&10.0$_{-8.6}^{+9.6}$ & 1.70\dag & 43.65 &5.9\dag\\
3C\,234   & 1553 $\pm$ 41, 1581 $\pm$ 41, & PL+TH+ABS(PL+GAU) &362.5/320 &PL&22.5$_{-1.1}^{+1.2}$ & 2.06$_{-0.16}^{+0.13}$ & 42.89 &--\\
&4636 $\pm$ 70&&&ABS(PL)&263$_{-54}^{+190}$ & 1.39$_{-0.16}^{+0.10}$ & 44.36 &28.1$_{-2.0}^{+2.6}$\\
3C\,244.1 & 46 $\pm$ 10, 27 $\pm$ 9, & PL &4.6/5 &PL&4.9$_{-1.4}^{+1.4}$ & 1.71$_{-0.54}^{+0.55}$ & 43.25 &--\\
&36 $\pm$ 7&&&ABS(PL)&$<$13.2&--&$<$42.92 &10.0\dag\\
3C\,274.1 & 27 $\pm$ 8& PL &7.4/9 &PL&4.7$_{-0.7}^{+0.7}$ & 1.61$_{-0.26}^{+0.26}$ & 43.27 &--\\
&&&&ABS(PL)&$<$10.2&--&$<$43.56 &10.0\dag\\
3C\,277.2 & 106 $\pm$ 14, 91 $\pm$ 13, & PL &13.0/18 &PL&3.0$_{-0.4}^{+0.4}$ & 1.59$_{-0.23}^{+0.24}$ & 43.67 &--\\
&223 $\pm$ 21&&&ABS(PL)&$<$4.6&--&$<$43.81 &10.0\dag\\
3C\,288   & $<$17 & NONE &-- &PL&$<$0.4&--&$<$41.41 &--\\
&&&&ABS(PL)&$<$2.9&--&$<$42.48 &10.0\dag\\
3C\,293   & 173 $\pm$ 14& PL+ABS(PL) &2.7/5 &PL&$<$2.2&--&$<$40.62 &--\\
&&&&ABS(PL)&254$_{-67}^{+85}$ & 1.70\dag & 42.87 &13.1$_{-3.8}^{+5.1}$\\
3C\,300   & 1083 $\pm$ 36, 648 $\pm$ 28, & PL+ABS(PL) &97.7/112 &PL&21.1$_{-0.8}^{+0.8}$ & 1.78$_{-0.06}^{+0.06}$ & 43.40 &--\\
&641 $\pm$ 28&&&ABS(PL)&$<$2.3&--&$<$42.49 &10.0\dag\\
3C\,315   & $<$9 & NONE &-- &PL&$<$1.4&--&$<$41.20 &--\\
&&&&ABS(PL)&$<$12.6&--&$<$42.36 &10.0\dag\\
3C\,325   & 670 $\pm$ 27& PL+ABS(PL) &27.4/27 &PL&$<$1.0&--&$<$43.16 &--\\
&&&&ABS(PL)&15.5$_{-4.1}^{+6.1}$ & 1.45$_{-0.18}^{+0.22}$ & 44.56 &2.9$_{-0.8}^{+0.9}$\\
3C\,349   & 706 $\pm$ 31, 726 $\pm$ 31, & PL+ABS(PL) &126.6/112 &PL&3.1$_{-1.5}^{+0.8}$ & 2.00\dag & 41.82 &--\\
&1466 $\pm$ 41&&&ABS(PL)&56.5$_{-5.9}^{+6.3}$ & 1.39$_{-0.15}^{+0.16}$ & 43.87 &1.2$_{-0.2}^{+0.2}$\\
3C\,381   & 257 $\pm$ 16& PL+ABS(PL) &11.5/8 &PL&7.8$_{-1.8}^{+1.7}$ & 2.38$_{-0.85}^{+0.76}$ & 42.11 &--\\
&&&&ABS(PL)&489$_{-128}^{+170}$ & 1.70\dag & 44.31 &30.5$_{-6.9}^{+8.7}$\\
3C\,427.1 & $<$27 & NONE &-- &PL&$<$0.5&--&$<$42.45 &--\\
&&&&ABS(PL)&$<$2.4&--&$<$43.24 &10.0\dag\\
3C\,433   & 2550 $\pm$ 52& PL+ABS(PL) &-- &PL&1.1$_{-0.5}^{+0.5}$ & 2.00\dag & 41.06 &--\\
&&&&ABS(PL)&477$_{-117}^{+118}$ & 1.63$_{-0.26}^{+0.21}$ & 43.92 &9.3$_{-1.2}^{+1.2}$\\
3C\,442A  & 112 $\pm$ 12& PL &11.7/12 &PL&3.4$_{-0.8}^{+0.9}$ & 0.87$_{-0.22}^{+0.21}$ & 41.10 &--\\
&&&&ABS(PL)&$<$6.2&--&$<$40.78 &10.0\dag\\
3C\,457   & 546 $\pm$ 32, 468 $\pm$ 29, & PL+ABS(PL+GAU) &49.4/46 &PL&3.3$_{-0.3}^{+0.4}$ & 1.22$_{-0.29}^{+0.26}$ & 43.35 &--\\
&881 $\pm$ 33&&&ABS(PL)&95$_{-8}^{+104}$ & 1.67$_{-0.09}^{+0.09}$ & 44.56 &34.2$_{-4.3}^{+4.7}$\\

\hline
\end{tabular}
\end{table*}

\subsection{Infrared data}

Our approach in determining mid-IR luminosities follows that described
by Ogle \etal\ (2006). They measured the rest-frame 15-$\mu$m
luminosity, on the basis that this samples the continuum without being
contaminated by any of the known spectral features in the mid-IR band.
Cleary \etal\ (2007) also tabulate 15-$\mu$m luminosities for the
sources in their sample, unfortunately without quoting errors (for the
purposes of regression we assign IR luminosities from this source an
arbitrary error of 0.01 dex, or $\sim 2$ per cent). We have measured
new values of the 15-$\mu$m flux density only for the 32 objects that have
available {\it Spitzer} IRS data and that do not have a quoted value
either in Ogle \etal\ or Cleary \etal; otherwise we use published
data.

To measure the flux densities we used the `post-BCD' spectra available
from the {\it Spitzer} archive. Although in general these are not
recommended for science analysis, the only significant difference
between the BCD and post-BCD IRS data is that the post-BCD data are
background-subtracted: two positions are observed for each spectrum
and the background subtraction for each channel and spectral order
involves subtracting the image from one position from the image at the
other position. Since the automated procedure used to produce the
post-BCD data is exactly what we would have done ourselves in any
case, we feel justified in using post-BCD data for the analysis. We
extracted spectra from all available orders using {\sc spice} and then
merged them (taking the weighted mean of duplicated data points where
the wavelength ranges of different modules overlapped) to give a
single table of flux density and error as a function of frequency: we
then used the weighted mean of the flux density around rest-frame 15
$\mu$m (exactly as described by Ogle \etal\ 2006) to determine the
15-$\mu$m flux density and its error. We verified for a number of test
cases that this gave flux densities in agreement with those tabulated
by Ogle \etal\ (2006). The flux densities we measure and their errors
are shown in Table \ref{irfluxes}.

\begin{table}
\caption{Newly measured {\it Spitzer} 15-$\mu$m flux densities for
3CRR sources. Errors are $1\sigma$ errors derived from the weighted
mean used to determine the flux densities.}
\label{irfluxes}
\begin{center}
\begin{tabular}{lrr}
\hline
Source&Flux density&Error\\
&(mJy)&(mJy)\\
\hline
3C\,20  &  9.92  &  0.08 \\
3C\,33.1  &  34.70  &  0.16 \\
3C\,47  &  34.39  &  0.35 \\
3C\,48  &  110.91  &  0.46 \\
3C\,79  &  42.08  &  0.34 \\
3C\,109  &  120.02  &  0.35 \\
3C\,228  &  0.99  &  0.18 \\
3C\,249.1  &  37.86  &  0.12 \\
3C\,295  &  4.36  &  0.05 \\
3C\,346  &  5.45  &  0.05 \\
3C\,351  &  78.06  &  0.20 \\
3C\,321  &  161.91  &  0.26 \\
3C\,293  &  19.87  &  0.05 \\
3C\,225B  &  0.70  &  0.18 \\
3C\,226  &  15.65  &  0.22 \\
3C\,314.1  &  0.15  &  0.03 \\
3C\,341  &  16.88  &  0.07 \\
3C\,386  &  2.47  &  0.04 \\
3C\,31  &  17.19  &  0.06 \\
3C\,66B  &  4.76  &  0.04 \\
3C\,76.1  &  1.85  &  0.07 \\
3C\,83.1B  &  5.27  &  0.05 \\
3C\,84  &  1146.04  &  0.42 \\
3C\,264  &  10.32  &  0.07 \\
3C\,272.1  &  24.74  &  0.05 \\
3C\,274  &  42.96  &  0.25 \\
3C\,296  &  0.25  &  0.05 \\
3C\,310  &  0.84  &  0.06 \\
3C\,338  &  2.40  &  0.04 \\
NGC 6251  &  27.80  &  0.03 \\
3C\,449  &  0.17  &  0.03 \\
3C\,465  &  3.17  &  0.05 \\

\hline
\end{tabular}
\end{center}
\end{table}

\section{Results}
\label{sec:results}

The results of our analysis are summarized as a table of luminosities
for different aspects of the AGN activity (Table \ref{lumtab}). Our
X-ray luminosities or upper limits are measured in the 2--10 keV band,
while we have luminosity densities for the whole source at 178 MHz,
the radio core at 5 GHz, and the AGN at 15 $\mu$m. We also have
emission-line powers in the [OIII] and [OII] lines. At redshifts
$z<0.3$ we use the virtually complete, homogeneous database of
Buttiglione \etal\ (2009) for [OIII], while emission-line powers for
higher-redshift objects and for [OII] are taken from the online
table\footnote{Now at
http://www.science.uottawa.ca/$\sim$cwillott/3crr/3crr.html .} of
Willott \etal\ (1999), converted to the cosmology used here. For
convenience we convert all luminosity densities into $\nu L_\nu$ form
so that all numbers are roughly commensurate. Missing entries in the
table indicate a lack of observations in the corresponding band: only
the radio luminosities are complete for all objects.

To derive results about the physics of the radio-loud AGN in our
sample we must consider the relationships between the various
luminosities we have available. We emphasise that we already know that
{\it none} of these luminosities is expected to provide a guaranteed
insight into the physical processes in the central engine. The
low-frequency radio luminosity is related to time-averaged jet kinetic
power, but also to the age of the source and to properties of the
external environment. The core radio luminosity is related to the
instantaneous jet power, but must necessarily be strongly affected by
beaming for all sources if unified models are correct. The mid-IR
luminosities tell us something about the emission from a torus, if
present, but it is widely thought (e.g.\ Heckman, Chambers \& Postman
1992; Hes, Barthel \& Hoekstra 1995; Cleary \etal\ 2007) that there is
an orientation-dependent component to the mid-IR luminosity of
3CRR sources, a point we will return to later in the paper, while at
low luminosities there is the potential for contamination with other
sources of mid-IR emission. The optical emission lines provide some
measure of the accretion disc photoionizing luminosity, if a disc is
present, but there is disagreement in the literature (e.g.\ Jackson \&
Browne 1990; Hes, Barthel \& Fosbury 1993; Simpson 1998) about whether
these are radiated isotropically and which is the best proxy for
non-stellar continuum strength, while again there is the possibility
of contamination by emission-line material not photoionized by the
accretion disc (e.g. either shock ionization or photoionization by the
jet); moreover, the dependence of line luminosity on photoionizing
luminosity is non-linear in general and depends on the detailed
properties of the photoionizing spectrum and the material being
ionized (e.g. Tadhunter \etal\ 1998). Finally, our X-ray luminosities
for the absorbed and unabsorbed components suffer from contamination
by each other (especially if there is little absorption) and from
uncertainty about the level of upper limits on absorbed emission; and,
while an absorbed nuclear component seems likely to give us
information about the luminosity of the accretion disc, there is still
disagreement about the nature of the unabsorbed components, although
it has been clear for some time that the jet is likely to dominate the
X-rays in core-dominated quasars (e.g. Worrall \etal\ 1987; for more
discussion of the possible origins of this component see Section
\ref{jetrelated}).

In this section of the paper we therefore consider the relationships
between the different luminosities, and see what conclusions can be
drawn from them, without attempting to test particular models in
detail. At the end of the section we summarize our view of the
implications of these relationships for the physics of radio-loud AGN.
We address several more detailed aspects of the results in Section
\ref{discussion}.

Throughout the section we use consistent methods for correlation and
regression analysis. Since 3CRR is a flux-limited sample at 178 MHz,
and since there are also effectively flux or flux-density limits at all other
wavebands, we expect to see correlations at some level in any
luminosity-luminosity plot we can produce. We account for this by
testing for partial correlation in the presence of redshift in all our
correlation analysis: we do this using the code of Akritas \& Seibert
(1996), which can take into account the presence of upper limits in
the data. Where we carry out linear regression we use a
Bayesian code, based on the Markov Chain Monte Carlo
algorithm, which takes account of both errors and upper limits and
which determines, as well as slope and intercept, the width of a
log-normal dispersion about the regression line: it is the width of
this dispersion (rather than the formal uncertainty) that is indicated
on the plots. We adopt uniform
(uninformative) priors for all three fitted quantities. Correlation
coefficients and parameters for the regression lines are tabulated in
Tables \ref{corsum} and \ref{regsum}.

\subsection{X-ray data and radio/X-ray correlations}
\label{rxc}

As discussed above, for each X-ray observed source we have a
measurement of, or limit on, the luminosity of an unabsorbed or weakly
absorbed component ($L_{\rm Xu}$) and an `accretion-related' component
($L_{\rm Xa}$), defined in the same way as in H06.

\begin{figure}
\epsfxsize 8.5cm
\epsfbox{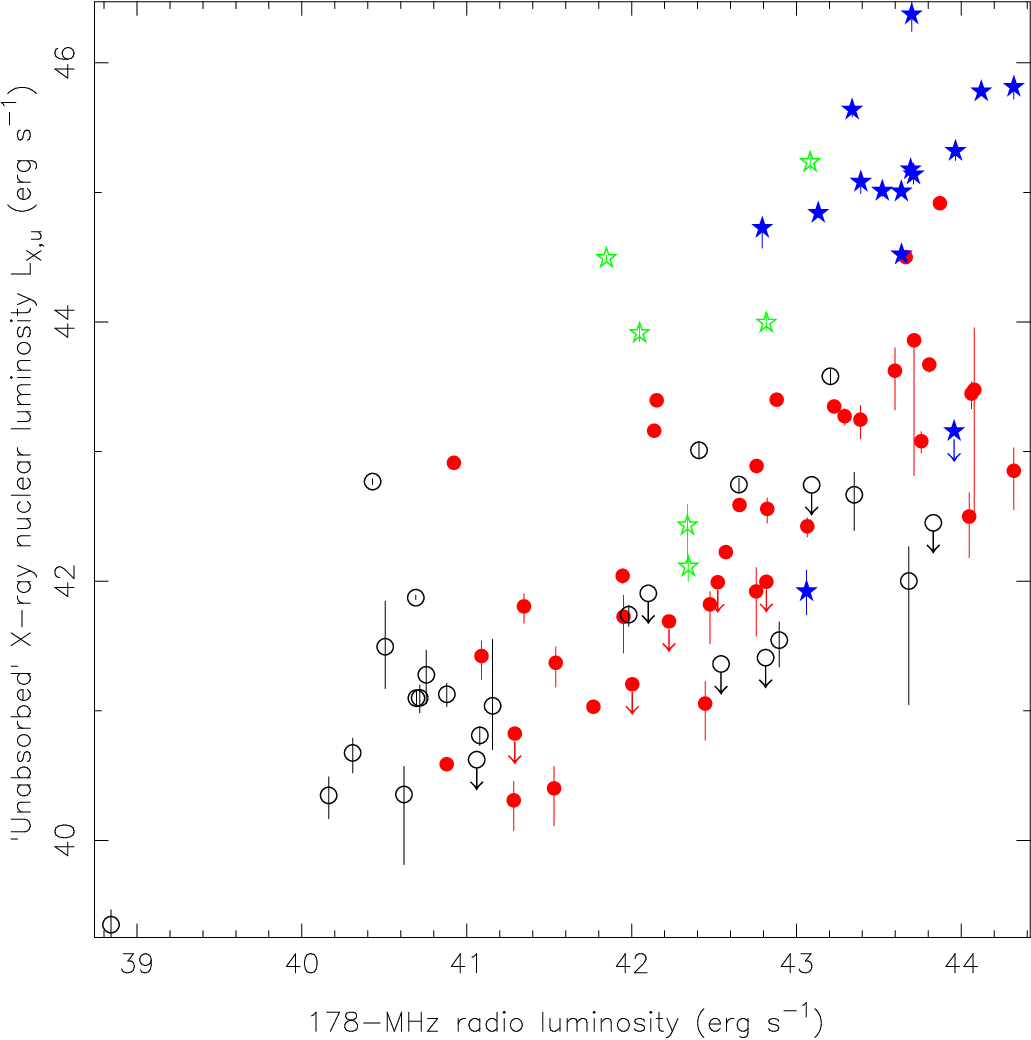}
\caption{X-ray luminosity for the unabsorbed component, $L_{\rm
  Xu}$, as a function of 178-MHz total radio luminosity for the
  $z<1.0$ 3CRR sample. Black open circles indicate LERGs, red filled
  circles NLRGs, green open stars
  BLRGs and blue filled stars quasars.}
\label{lxu-lr}
\end{figure}

\begin{figure}
\epsfxsize 8.5cm
\epsfbox{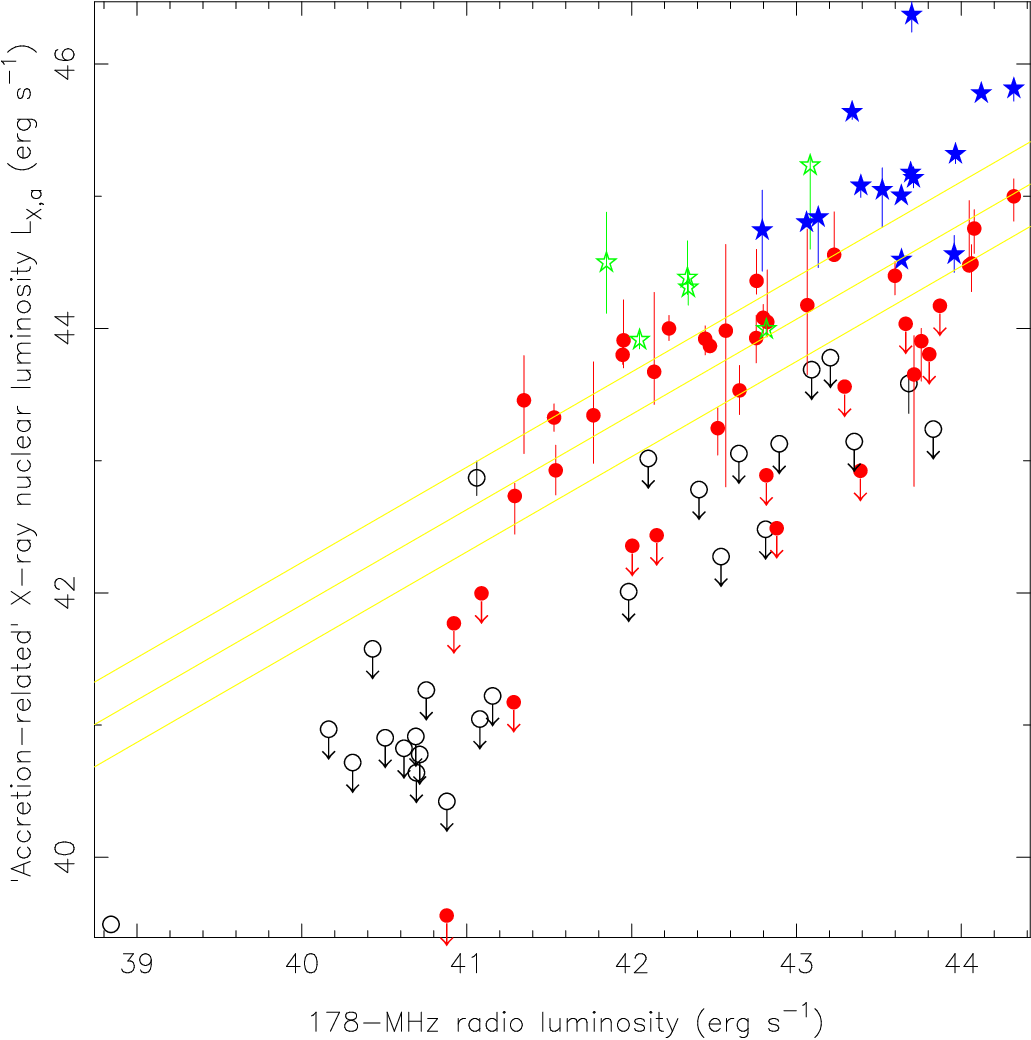}
\caption{X-ray luminosity for the `accretion-related' component,
  $L_{\rm Xa}$, as a function of 178-MHz total radio luminosity for
  the $z<1.0$ 3CRR sample. Symbols as in Fig.\ \ref{lxu-lr}.
  Regression is for detected NLRGs only. As elsewhere in the paper,
  upper limits assume $N_{\rm H} = 10^{23}$ cm$^{-2}$. It can be seen
  that the upper limits for the LERGs lie systematically below the
  regression line.}
\label{lxa-lr}
\end{figure}

In Figs \ref{lxu-lr} and \ref{lxa-lr} (corresponding to figs 2 and 3
of H06) we plot these two quantities as a function of the total radio
luminosity at 178 MHz, $L_{\rm 178}$. It will be seen that there is a
good deal more scatter in the $L_{\rm Xu}-L_{\rm 178}$ relation at
high radio luminosities than in the corresponding relation for $L_{\rm
  Xa}$. Indeed, the null hypothesis of no intrinsic correlation cannot
be rejected at the $3\sigma$ level for the $L_{\rm Xu}-L_{\rm 178}$
relation, whereas $L_{\rm Xa}$ and $L_{\rm 178}$ are significantly
correlated even in the presence of the common correlation with
redshift (as found by H06 for their smaller sample). This suggests no
very strong physical relationship between the unabsorbed X-ray power,
before correcting for beaming, and the power of the AGN as measured by
the total radio luminosity. Individual emission-line classes of
source, such as the NLRGs, appear better correlated on the $L_{\rm
  Xu}-L_{\rm 178}$ plot, but partial correlation analysis shows that
we cannot rule out the null hypothesis of no intrinsic correlation for
any of the emission-line classes on Fig.\ \ref{lxu-lr}. By contrast,
the NLRGs (though no other individual class) are significantly
correlated on the partial correlation test in Fig.\ \ref{lxa-lr}.
Separation of the sources by emission-line type can also be seen in
both plots: quasars and BLRGs tend to have much higher $L_{\rm Xu}$,
and somewhat higher $L_{\rm Xa}$, for a given $L_{\rm 178}$, while the
upper limits on $L_{\rm Xa}$ for the LERGs lie systematically below
the luminosities for detected NLRGs. We illustrate this by
  plotting the regression line for the NLRGs only on
  Fig.\ \ref{lxa-lr}.

\begin{figure}
\epsfxsize 8.5cm
\epsfbox{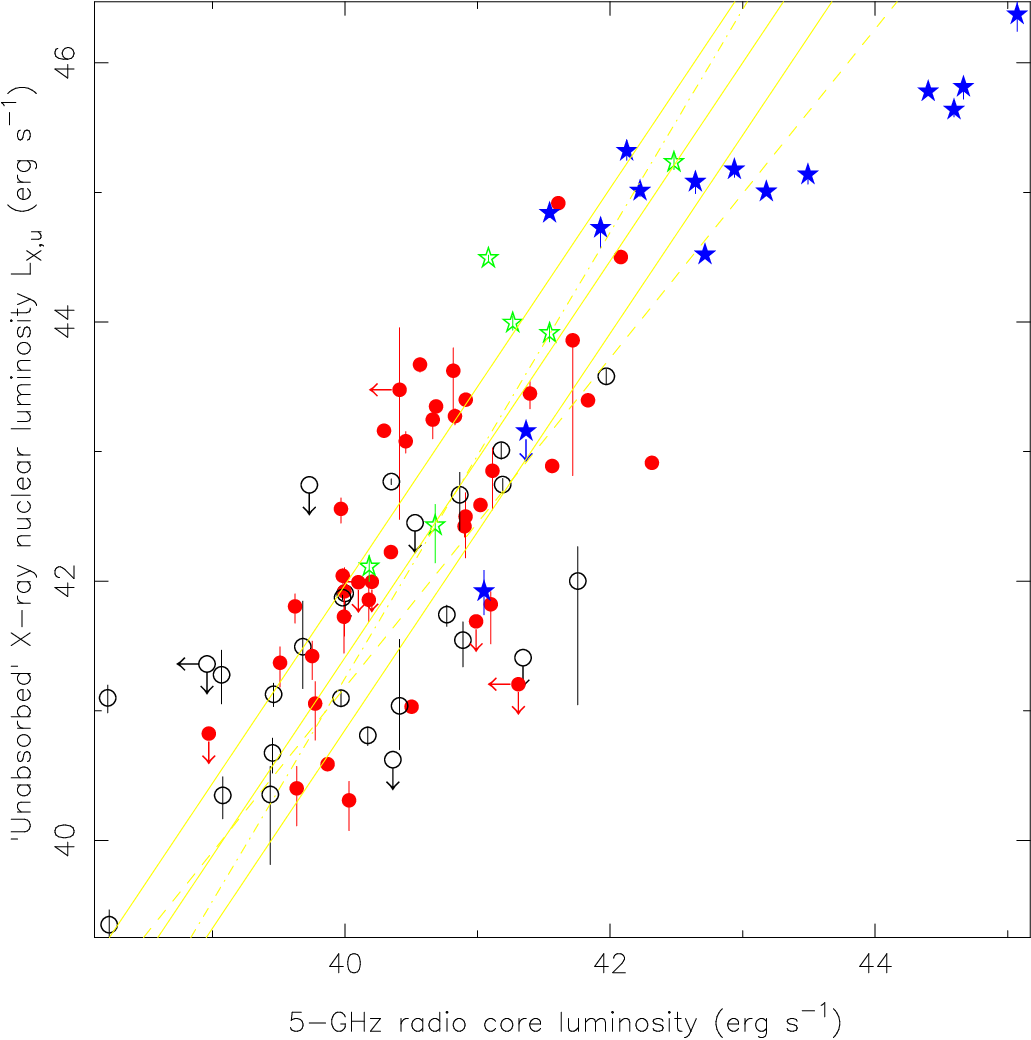}
\caption{X-ray luminosity for the unabsorbed component, $L_{\rm
  Xu}$, as a function of 5-GHz core radio luminosity for the
  $z<1.0$ 3CRR sample. Regression is for NLRGs and LERGs; dashed and
  dotted lines represent the results of individual regressions for the
  LERGs and NLRGs respectively (scatter not shown for clarity). Symbols as in Fig.\ \ref{lxu-lr}.}
\label{lxu-lc}
\end{figure}

\begin{figure}
\epsfxsize 8.5cm
\epsfbox{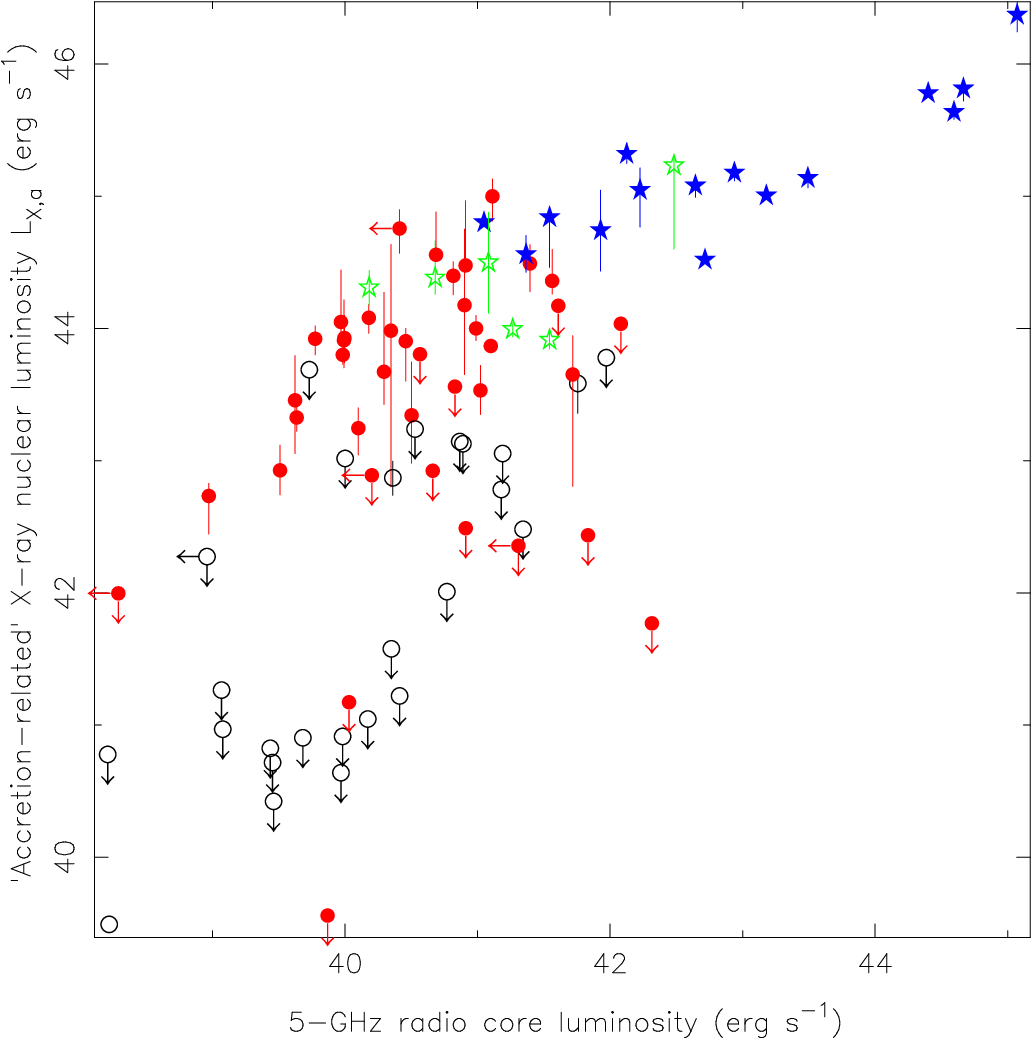}
\caption{X-ray luminosity for the `accretion-related' component, $L_{\rm
  Xa}$, as a function of 5-GHz core radio luminosity for the
  $z<1.0$ 3CRR sample. Symbols as in Fig.\ \ref{lxu-lr}.}
\label{lxa-lc}
\end{figure}

Figs \ref{lxu-lc} and \ref{lxa-lc} show the corresponding plots of the
X-ray quantities as a function of the 5-GHz nuclear luminosity,
$L_{\rm 5}$. Both of these correlations are formally significant in
the presence of common correlation with redshift, but the correlation
between $L_{\rm Xu}$ and $L_5$ is by far the stronger of the two
(Table \ref{corsum}),
though there is still a good deal of scatter in the relationship, and
the correlation between $L_{\rm Xa}$ and $L_5$ vanishes if the
quasars are excluded -- the quasars in the top right of both figures
are objects for which we cannot perform a good separation of $L_{\rm
Xu}$ and $L_{\rm Xa}$ -- while the correlation of Fig.\ \ref{lxu-lc}
remains significant. The LERGs and NLRGs have individually
  significant correlations between $L_{\rm Xu}$ and $L_5$, while no
  individual emission-line class has a correlation between $L_{\rm
    Xa}$ and $L_5$. Regression analysis shows that the regression
  lines for  LERGs and NLRGs are consistent (Fig.\ \ref{lxu-lc}; Table
  \ref{regsum}), whereas by eye they are
widely separated on the plot involving $L_{\rm Xa}$.

As we have argued previously (e.g. Hardcastle \& Worrall 1999) much of the
dispersion in $L_5$ must originate in beaming if unified models are correct.
We would expect beaming to introduce two to three orders of magnitude
scatter in any correlation between beamed and unbeamed quantitues for the
LERG population (for plausible Lorentz factors), and so any strong
correlation between $L_5$ and some other quantity without this scatter
forces us to the conclusion that much of that other quantity is also
beamed, and so must originate in the jet. Beaming will introduce
considerably less scatter for the NLRGs, which occupy a restricted range
of angles to the line of sight, but where they lie on the same correlation
as the LERGs, the argument for a jet origin can be applied to the entire
population. The data are thus consistent with the model
in which $L_{\rm Xu}$ originates in the jet while $L_{\rm Xa}$
originates in the accretion disc. Our inability to separate the two
sources of X-ray luminosity in the BLRGs and quasars accounts for
their tendency to lie above the regression line for the NLRGs in
Fig.\ \ref{lxa-lr}. The fact that the quasars with the largest radio
core luminosities seem to have lower X-ray to radio ratios in Fig.
\ref{lxu-lc} (illustrated by their deviation from the regression
line for NLRGs and LERGs) is also seen in the {\it ROSAT} analysis of
Hardcastle \& Worrall (1999); it is possible that it reflects
different effective speeds for the beaming of the radio and X-ray
emission components (e.g. Ghisellini \& Maraschi 1989). We return to the question of the origin of the unabsorbed component in
Section \ref{jetrelated}.

\begin{figure}
\epsfxsize 8.5cm
\epsfbox{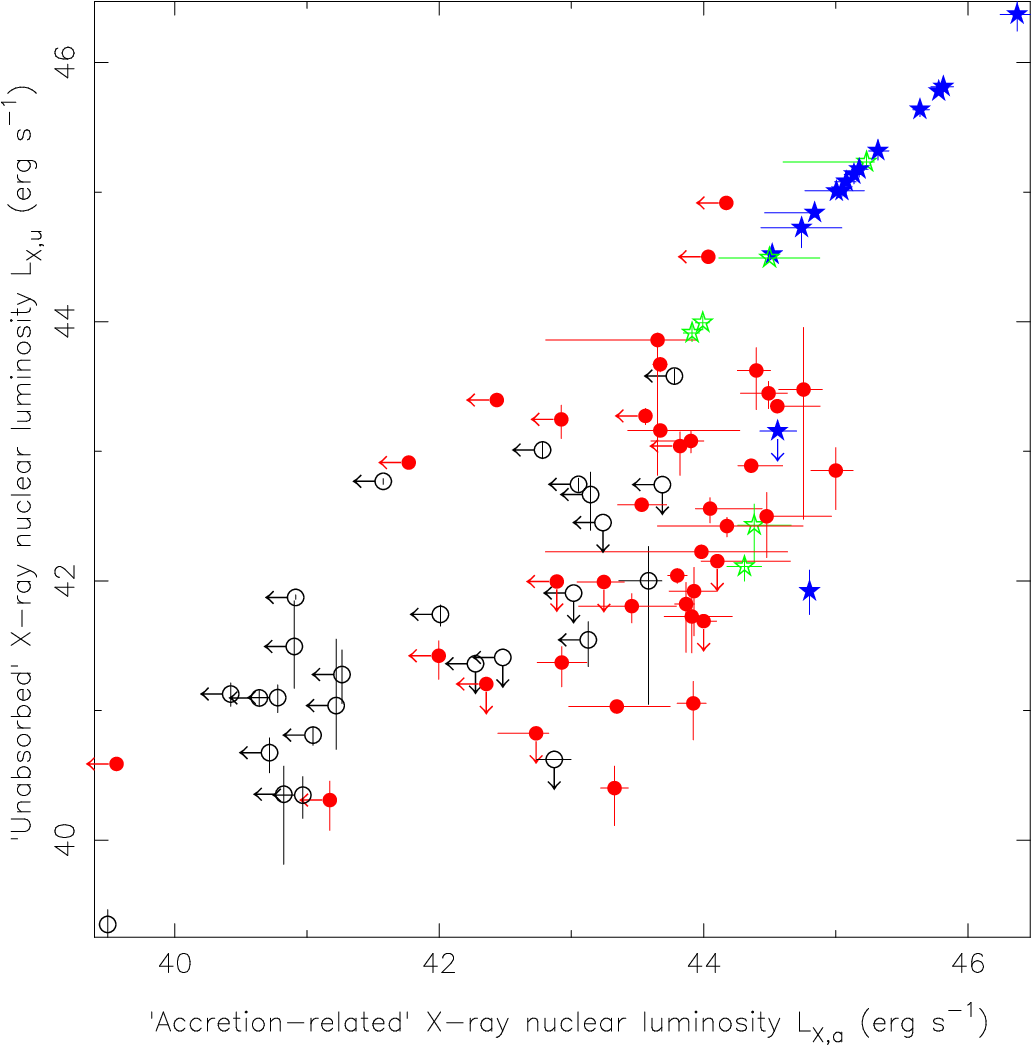}
\caption{X-ray luminosity for the unabsorbed component, $L_{\rm Xu}$, plotted against
  the `accretion-related' component, $L_{\rm
  Xa}$, for the
  $z<1.0$ 3CRR sample. Symbols as in Fig.\ \ref{lxu-lr}.}
\label{lxu-lxa}
\end{figure}

Finally, we plot the two X-ray luminosities against each other (Fig.
\ref{lxu-lxa}). No significant correlation between the two luminosities
exists for NLRGs when the common correlation with redshift is taken
into account. For some BLRGs and quasars our best estimates of $L_{\rm
  Xu}$ and $L_{\rm Xa}$ are equal and so these appear well correlated.

\subsection{Radio/IR correlations}
\label{radio-ir}

\begin{figure}
\epsfxsize 8.5cm
\epsfbox{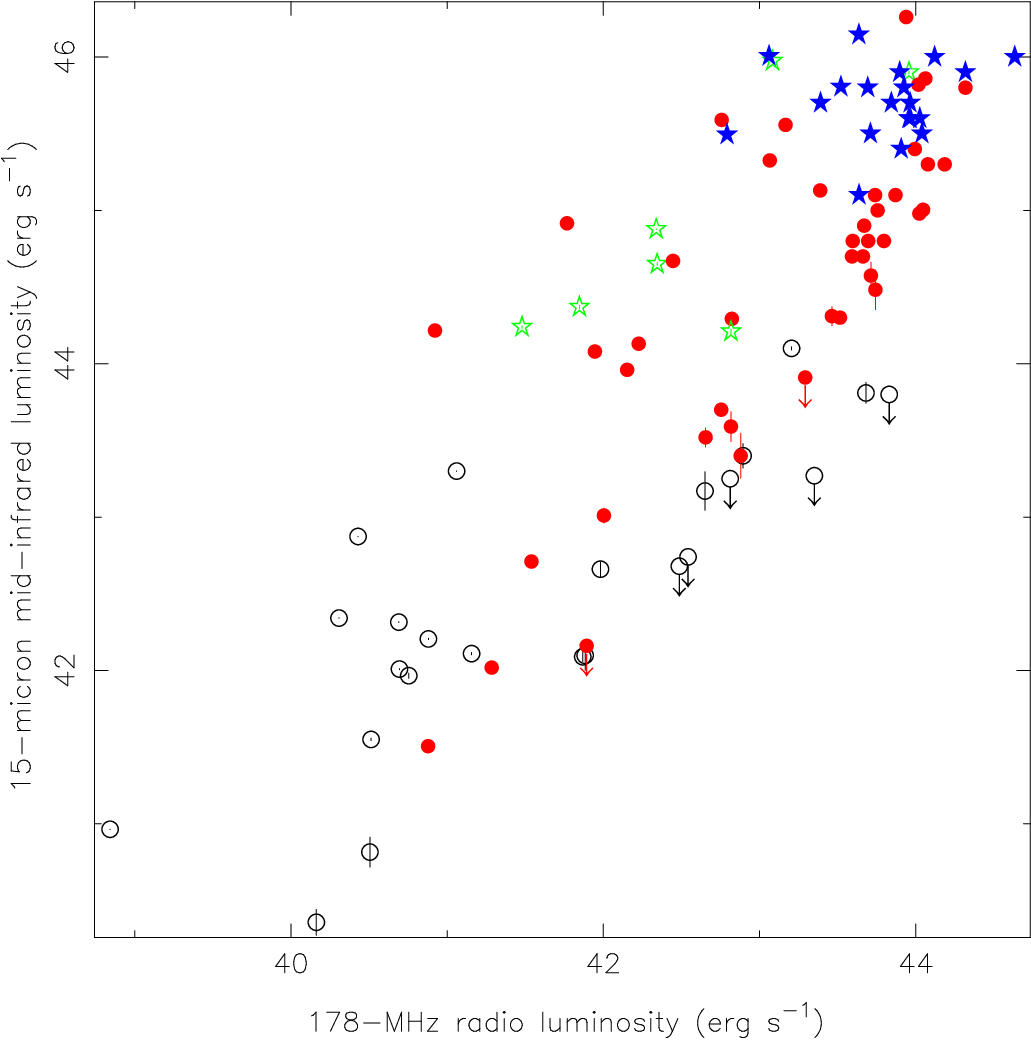}
\caption{15-$\mu$m infrared luminosity as a function of 178-MHz total radio luminosity for the
  $z<1.0$ 3CRR sample. Symbols as in Fig.\ \ref{lxu-lr}.}
\label{li-lr}
\end{figure}

\begin{figure}
\epsfxsize 8.5cm
\epsfbox{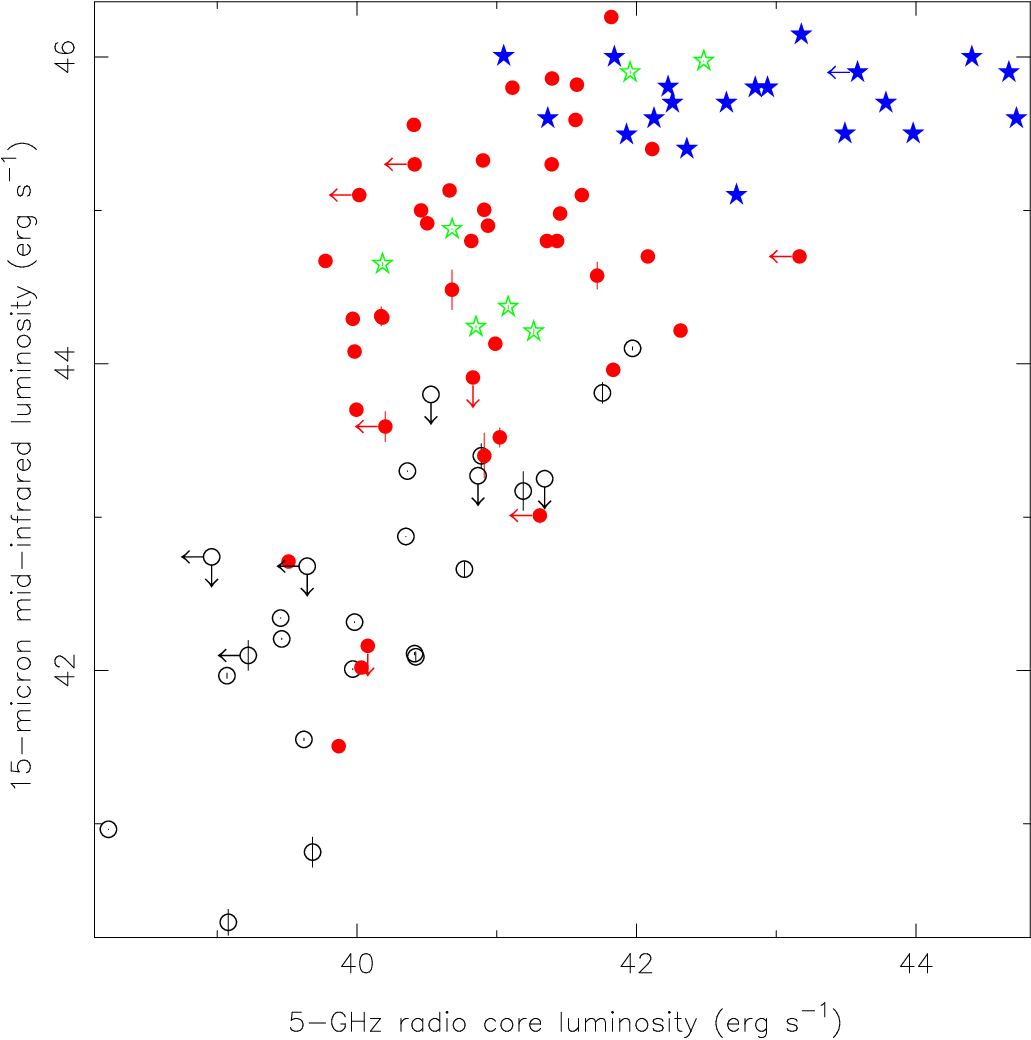}
\caption{15-$\mu$m infrared luminosity as a function of 5-GHz core radio luminosity for the
  $z<1.0$ 3CRR sample. Symbols as in Fig.\ \ref{lxu-lr}.}
\label{li-lc}
\end{figure}

In Figs \ref{li-lr} and \ref{li-lc} we plot the relationships between
total 15-$\mu$m luminosity ($L_{\rm IR}$) and the radio total and core
luminosity for the objects in our sample with IR flux density
measurements. Both these plots show correlations that are formally
highly significant even in the presence of the common correlation with
redshift. Fig.\ \ref{li-lr} may be compared with figs 8 and 9 of
Cleary \etal\ (2007). The clearest trend in this figure is for the
broad-line objects to lie systematically above the NLRGs at a given
radio luminosity/redshift. This trend was also noted by Cleary
\etal\ (2007), and, in a sample of high-$z$ 3CRR objects, by Haas
\etal\ (2008). We also see that the LERGs tend to lie below the NLRGs.
Cleary \etal\ explain the tendency for the BLRGs and quasars to have
higher IR luminosities than the NLRGs in terms of two effects: a
contamination of the IR by non-thermal (i.e. synchrotron) emission,
and excess line-of-sight absorption in the NLRGs that is not present
in the quasars. If these effects are accounted for, Cleary \etal\ show
that the correlation is improved: since it is already strong, these
observations imply that there is a physical relationship between
$L_{\rm IR}$ and $L_{\rm 178}$ for the high-excitation objects, but
that the same relationship does not hold for the LERGs. Dicken
\etal\ (2008) see no evidence for a tendency for quasars to lie above
radio galaxies in the IR in their analysis of an independent, complete
sample, but we note that they use total radio luminosities at 5 GHz to
estimate radio power; although their selection criteria at 2.7 GHz are
intended to exclude objects dominated by beaming, any tendency for
quasars to have higher 5-GHz luminosities than radio galaxies of
similar low-frequency luminosity would tend to suppress a difference
between the classes on a plot analogous to Fig. \ref{li-lr}.

In Fig.\ \ref{li-lc} the distinction between broad-line objects and
NLRGs is not so apparent, since the broad-line objects have
systematically higher $L_5$ and $L_{\rm IR}$. This is certainly
consistent with the idea that non-thermal emission in the IR plays a
significant role in the BLRGs and quasars. As with Fig.\ \ref{lxu-lc},
we see a flattening of the correlation at the highest radio
luminosities. In this figure the partial correlation for the LERGs
alone is significant; this is not true for any other individual class
of source.

\subsection{IR/X-ray correlations}
\label{xray-ir}

\begin{figure}
\epsfxsize 8.5cm
\epsfbox{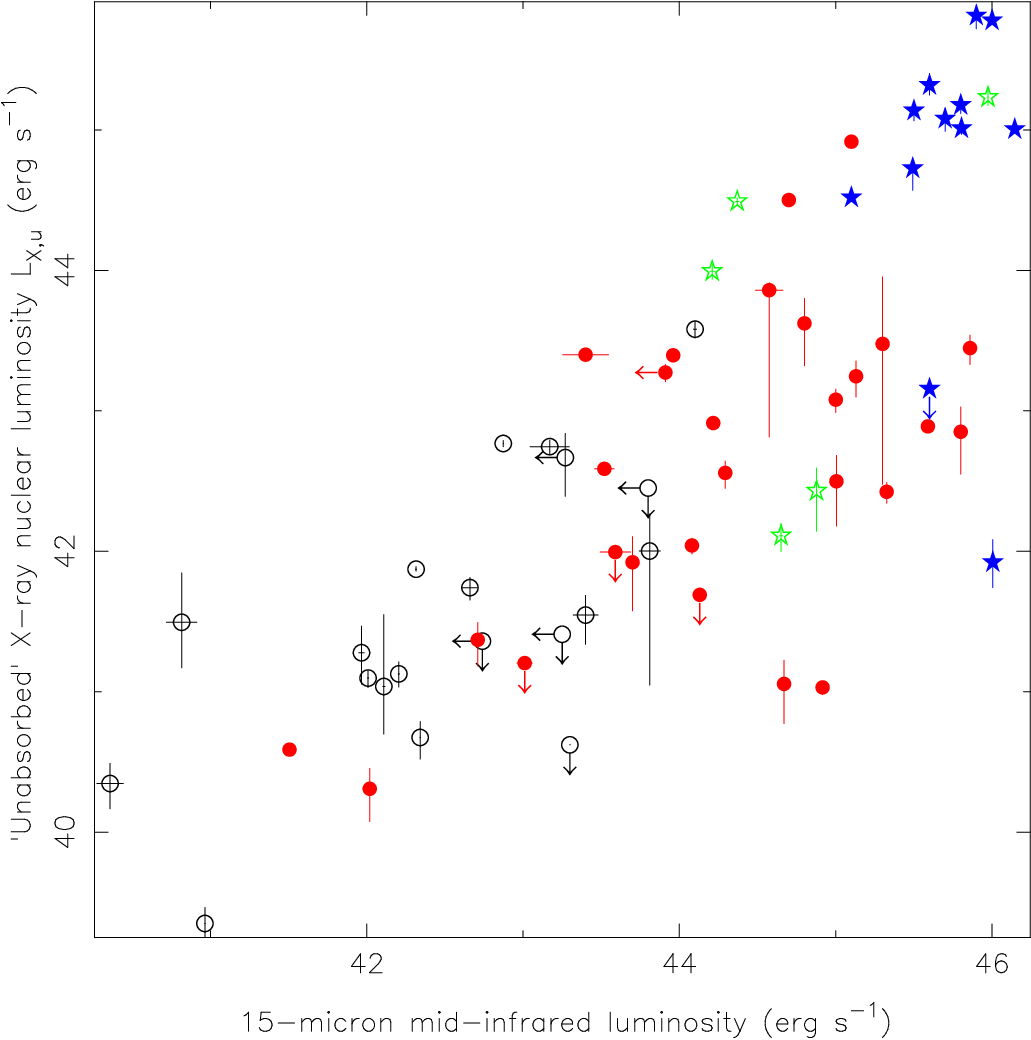}
\caption{Unabsorbed X-ray luminosity, $L_{\rm
  Xu}$, as a function of total 15-$\mu$m infrared luminosity for the
  $z<1.0$ 3CRR sample. Symbols as in Fig.\ \ref{lxu-lr}.}
\label{li-lxu}
\end{figure}

\begin{figure}
\epsfxsize 8.5cm
\epsfbox{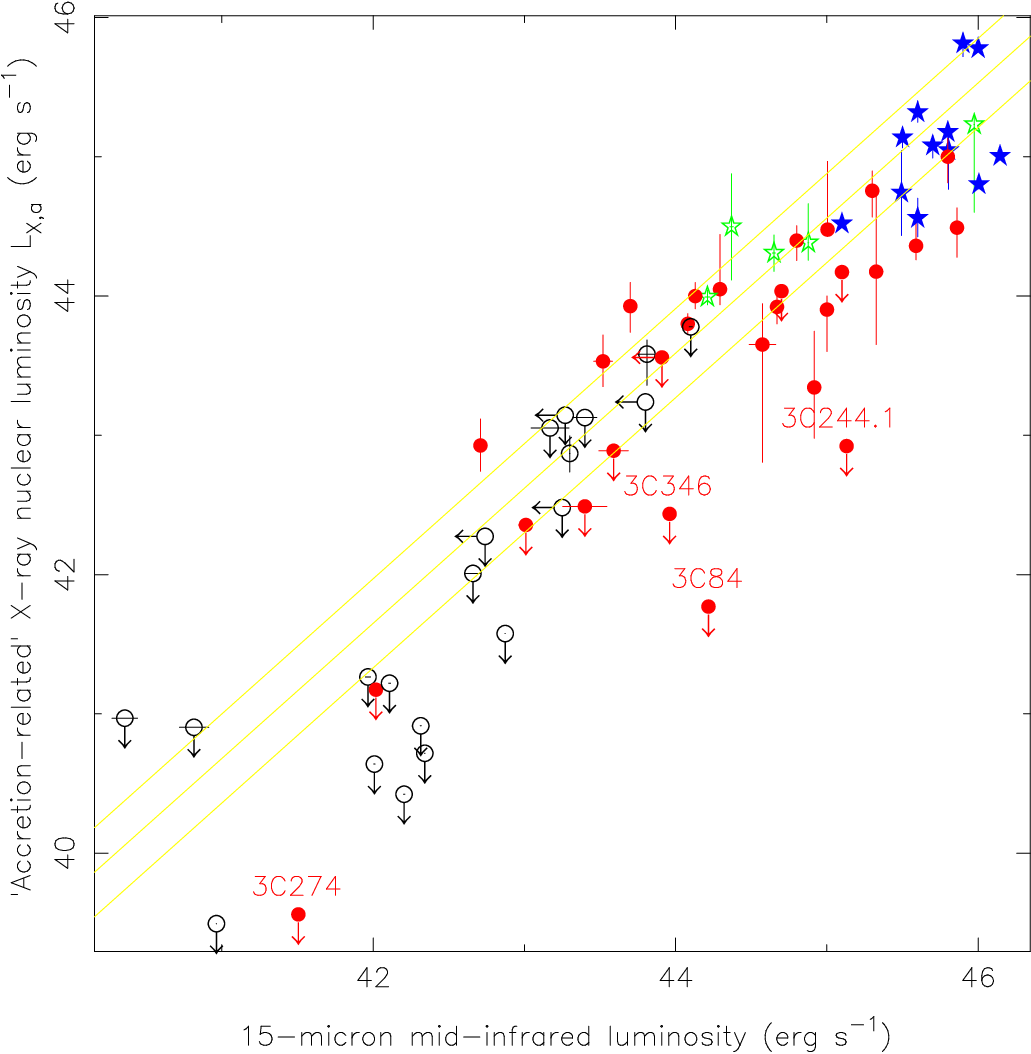}
\caption{`Accretion-related' X-ray luminosity, $L_{\rm
  Xa}$ as a function of total 15-$\mu$m infrared luminosity for the
  $z<1.0$ 3CRR sample. Regression is for all X-ray-detected objects.
  Symbols as in Fig.\ \ref{lxu-lr}. Several NLRG outliers from the
  regression are labelled on this figure and discussed in the text.}
\label{li-lxa}
\end{figure}

Figs \ref{li-lxu} and \ref{li-lxa} show the relationships between
15-$\mu$m IR luminosity and the unabsorbed and `accretion-related'
X-ray emission components, respectively. These are both formally
significant correlations, but the correlation between $L_{\rm IR}$ and
$L_{\rm Xa}$ is much the stronger of the two, and has a slope
consistent with unity (Tables \ref{corsum} and \ref{regsum}). (For the regression
we only consider objects with measurements of $L_{\rm Xa}$: it is not
appropriate to incorporate the upper limits on $L_{\rm Xa}$ in the
regression because these are dependent on the assumed absorbing
column, a point we return to in Section \ref{comptonthick}.) This
strong, approximately linear correlation (whose significance is
  improved still further if the
LERGs are excluded) is very strong evidence that
$L_{\rm IR}$ and $L_{\rm Xa}$ are measuring closely related
quantities, linked by the overall power of the accretion disc. Quasars
and BLRGs lie above NLRGs of similar $L_{\rm IR}$ in Fig.\ \ref{li-lxa},
which implies that $L_{\rm Xa}$ is more severely contaminated by
non-thermal/beamed emission than is $L_{\rm IR}$. The limits on LERGs
tend to lie on, or below, the correlation for the detected objects.
Compton-thick objects (i.e. those with $N_{\rm H} > 10^{24}$
cm$^{-2}$) would have X-ray upper limits that are below/to the
right of the regression line in Fig.\ \ref{li-lxa}, because our
X-ray upper limits in this case would be based on an incorrect, low
$N_{\rm H}$ value. For the FRII LERGs, with $L_{\rm IR} \ga 10^{43}$
erg s$^{-1}$, we see no objects for which this seems likely to be
true, although several NLRG upper limits do lie a long way below the
line. We discuss the implications of this plot for the nature of LERGs
in Section \ref{lergs} and the case of the NLRG outliers in Section
\ref{comptonthick}.

The approximately linear correlation between mid-IR and X-ray power
after correction for absorption is very similar to that found in
lower-power objects (mostly radio-quiet Seyferts) by Gandhi \etal\
(2009) using 12.3 $\mu$m ground-based IR observations: above $L_{\rm
IR} \sim 10^{43}$ erg s$^{-1}$ our objects lie in an identical region
of the IR/X-ray luminosity plot, and our regression slope is
consistent with theirs within the joint errors.

In Fig.\ \ref{li-lxu} we see a large amount of scatter for a given
$L_{\rm IR}$, with quasars and broad-line objects clearly having
significantly higher X-ray luminosity; this is consistent with the
idea that here we are plotting a beamed quantity ($L_{\rm Xu}$, see
above) against one that is basically unaffected by beaming.

\subsection{Emission-line power}
\label{emline}

Data on the luminosity of the [O{\sc iii}] and [O{\sc ii}] emission
lines are available in the compilation of Willott \etal\ (1999) and in
the more recent work of Buttiglione \etal\ (2009). 88
sample sources have [O{\sc iii}] luminosities and 86 [O{\sc ii}]
(Table \ref{lumtab}). Because of the longer wavelength of the [O{\sc
iii}] line, high-redshift, high-luminosity sources are more likely to
have [O{\sc ii}] data than [O{\sc iii}], a fact that must be borne in
mind in interpreting the results in this section.

\begin{figure*}
\epsfxsize 8.5cm
\epsfbox{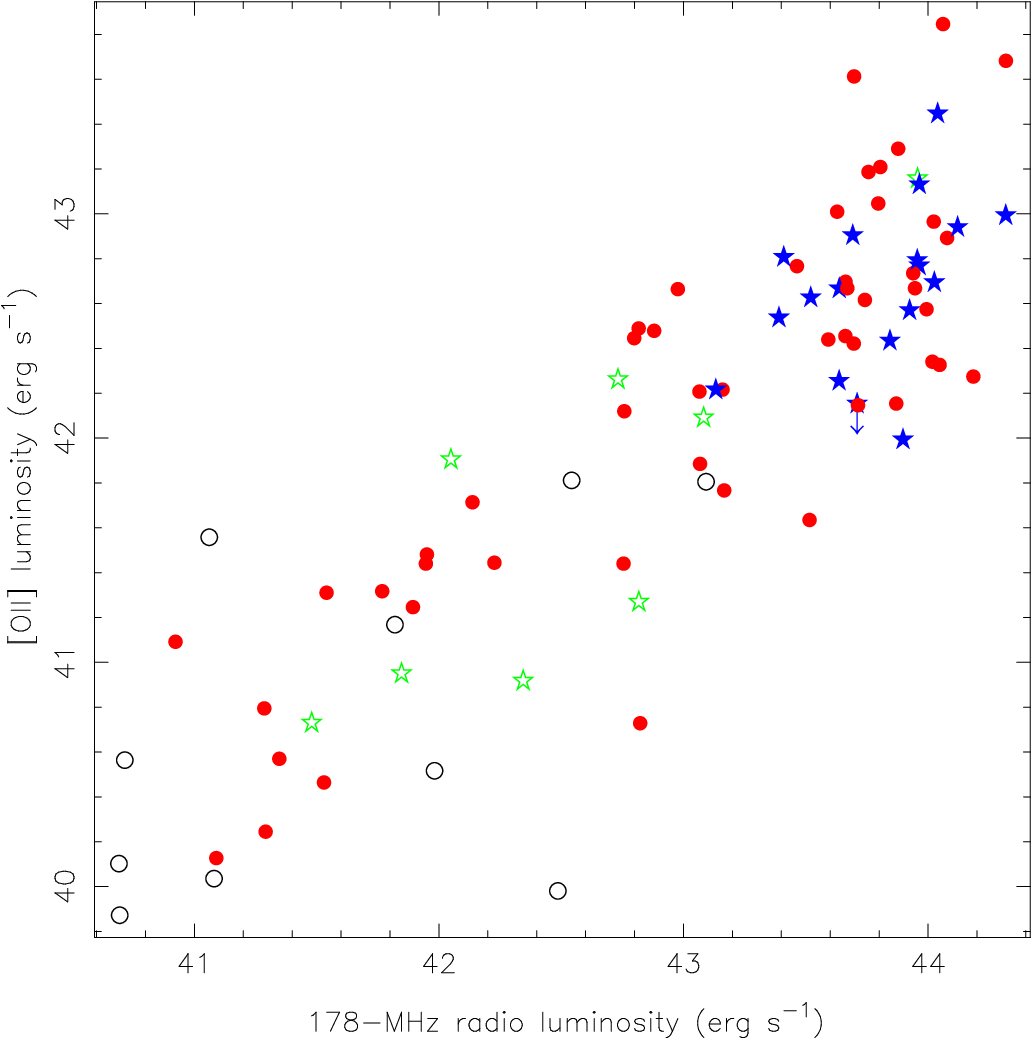}
\epsfxsize 8.5cm
\epsfbox{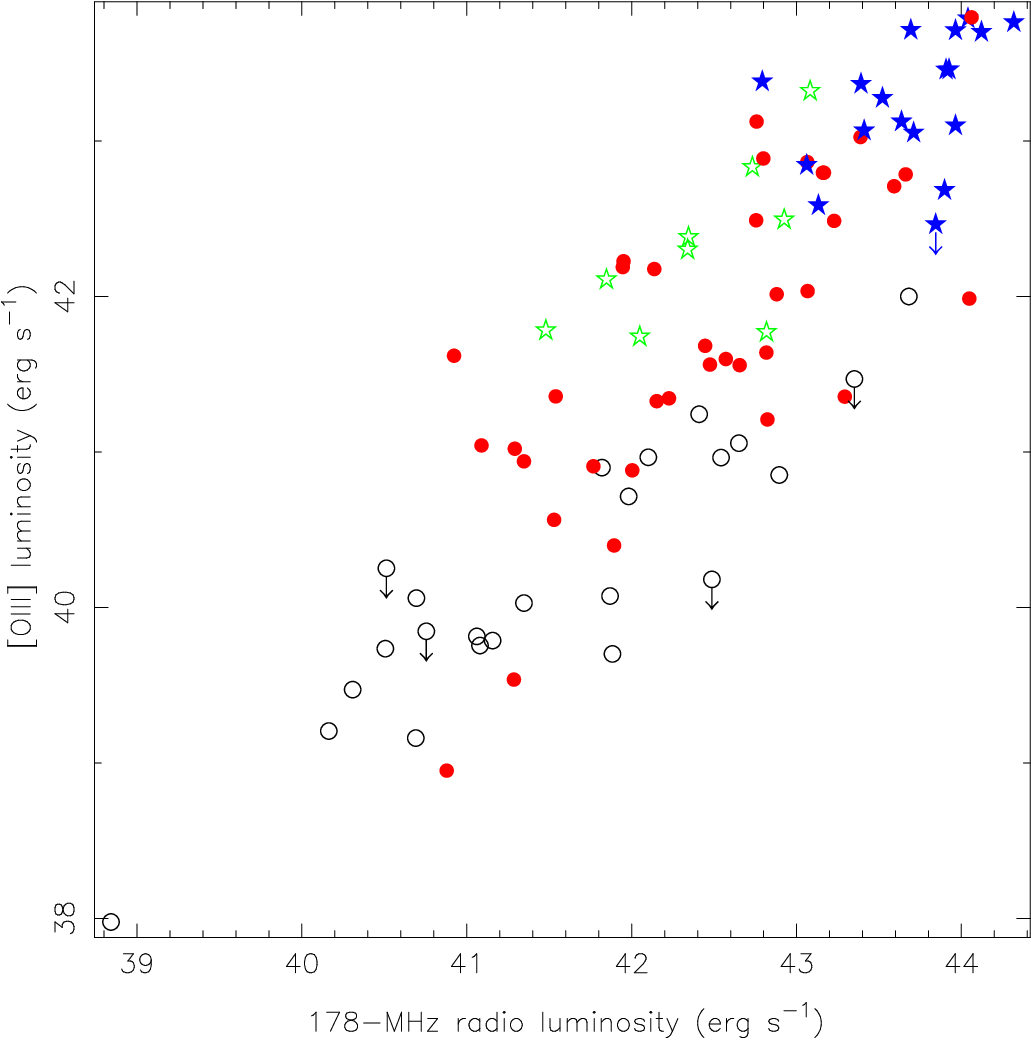}
\caption{Emission-line luminosity as a function of 178-MHz total radio luminosity for the
  $z<1.0$ 3CRR sample. Left panel shows [O{\sc ii}] and right [O{\sc
  iii}]. Symbols as in Fig.\ \ref{lxu-lr}.}
\label{lo-lr}
\end{figure*}

\begin{figure*}
\epsfxsize 8.5cm
\epsfbox{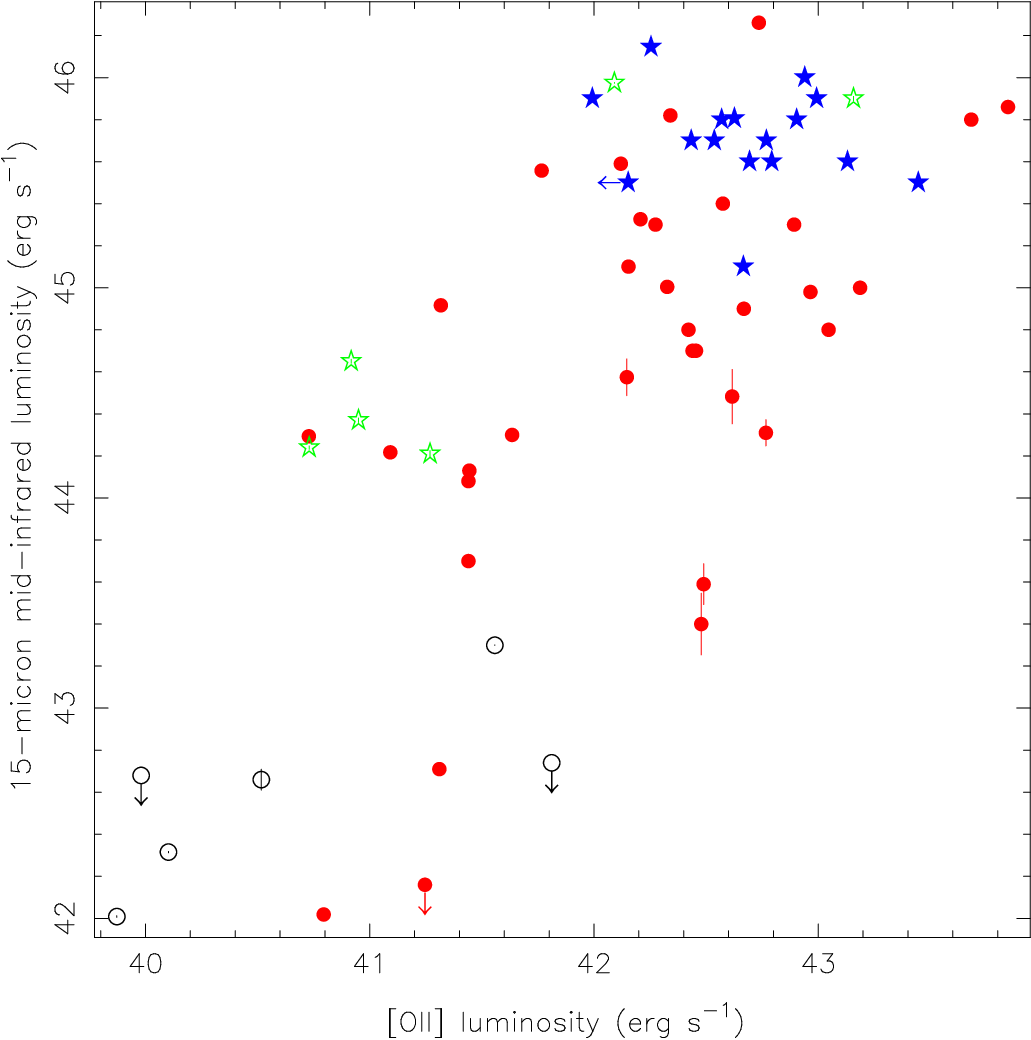}
\epsfxsize 8.5cm
\epsfbox{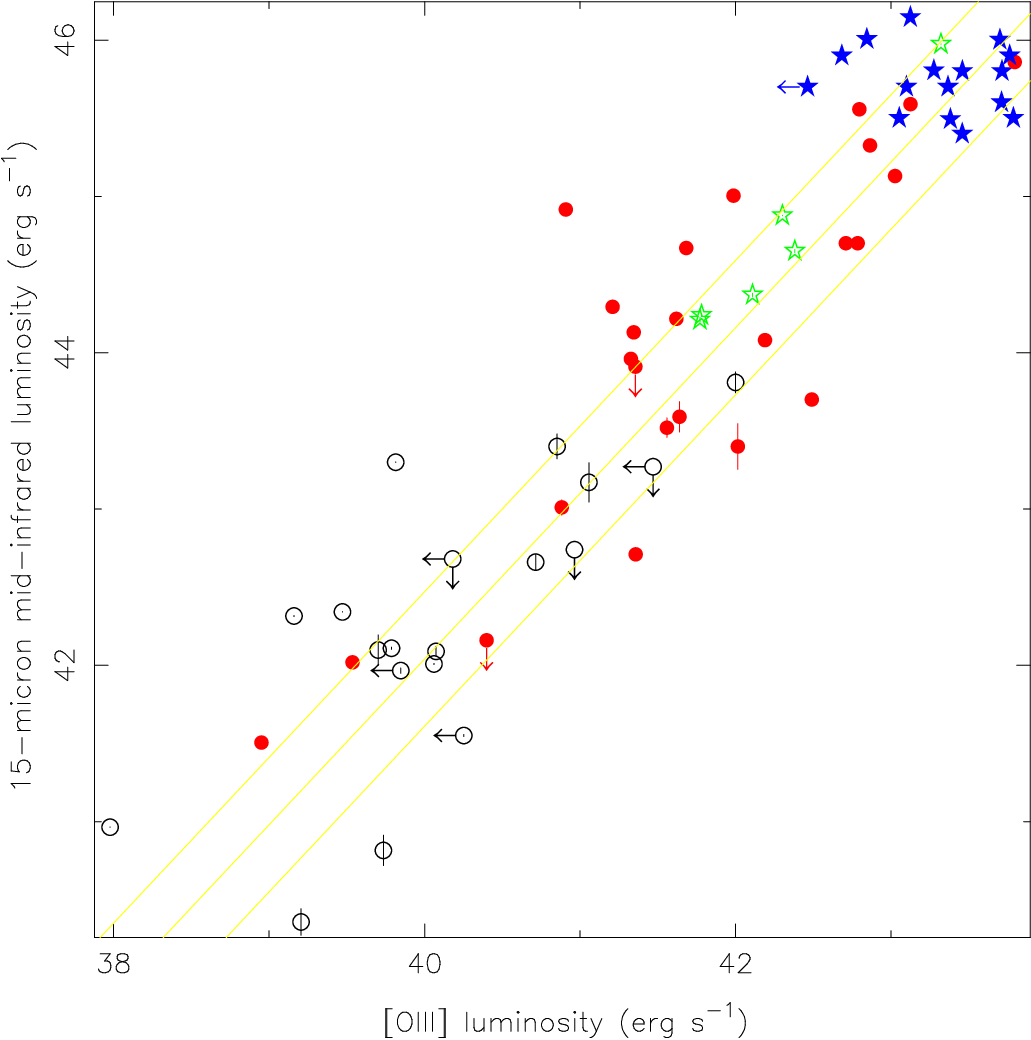}
\caption{15-$\mu$m infrared luminosity plotted
  against emission-line luminosity for the
  $z<1.0$ 3CRR sample. Regression is for all objects. Left panel shows
  [O{\sc ii}] and right [O{\sc iii}]. Symbols as in Fig.\ \ref{lxu-lr}.}
\label{lo-li}
\end{figure*}

\begin{figure*}
\epsfxsize 8.5cm
\epsfbox{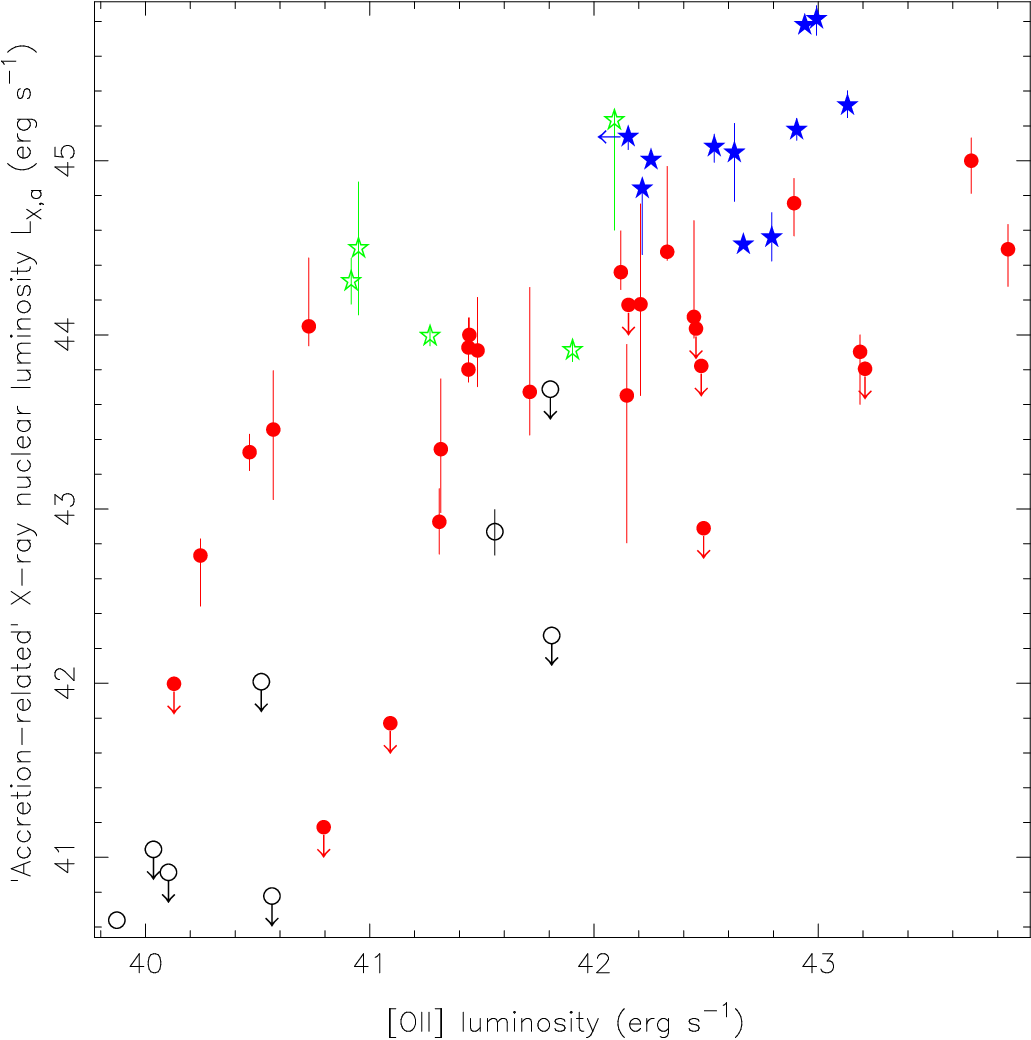}
\epsfxsize 8.5cm
\epsfbox{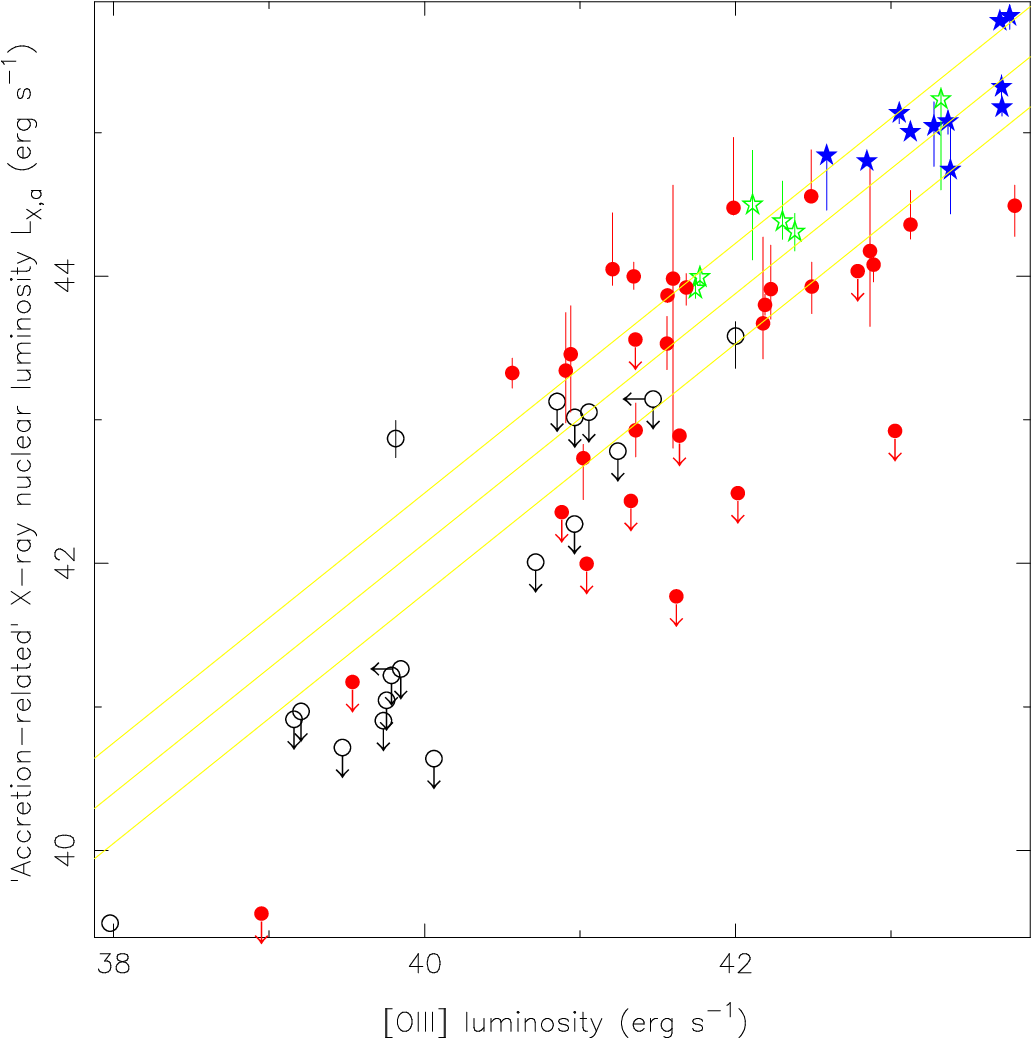}
\caption{X-ray luminosity for the `accretion-related' component, $L_{\rm
  Xa}$ plotted
  against emission-line luminosity for the
  $z<1.0$ 3CRR sample. Regression excludes limits on X-ray
  luminosities. Left panel shows [O{\sc ii}] and right [O{\sc iii}]. Symbols as in Fig.\ \ref{lxu-lr}.}
\label{lo-lxa}
\end{figure*}

Fig.\ \ref{lo-lr} shows the luminosities of the two emission lines
as a function of 178-MHz luminosity for the sample:
this illustrates the well-known fact that broad-line objects have
systematically higher [O{\sc iii}] luminosities than narrow-line
objects of similar luminosity (Jackson \& Browne 1990), while the
[O{\sc ii}] luminosities are similar (Hes, Barthel \& Fosbury 1993).
In addition, we see that the few LERGs with emission-line data lie at
the very bottom of the observed luminosity range. Since line emission
is supposed to be driven by the ionizing luminosity of the accretion
disc, we might expect to see a better correlation with $L_{\rm IR}$
and $L_{\rm Xa}$, and this is in fact what we do see for [O{\sc iii}]
(Figs \ref{lo-li} and \ref{lo-lxa}), with a highly significant
correlation even after accounting for the common correlation with
redshift. Similar results for the [O{\sc iii}]-IR correlation have
recently been obtained for the 2Jy sample of radio galaxies by Dicken
\etal\ (2008). The LERGs in Fig. \ref{lo-lxa} lie on or below
the regression line: that is, the limits on X-ray accretion-related
luminosity are less even than we would predict from their (low)
emission-line luminosities. However, the $L_{\rm [OII]}$ -- $L_{\rm
IR}$ correlation is much weaker (though still significant) and the
$L_{\rm [OII]}$ -- $L_{\rm Xa}$ correlation is not significant after
common correlation with redshift is accounted for. This may be
explained in terms of the relatively weak expected dependence of [OII]
luminosity on ionizing luminosity (e.g. Simpson 1998; Tadhunter \etal\
1998) coupled with the scatter in the $L_{\rm IR}$ -- $L_{\rm Xa}$
relation. The partial correlations between [O{\sc iii}] and
  IR/accretion-related X-ray remain highly significant if the LERGs are
  excluded, while the corresponding correlations involving [O{\sc ii}]
  are not significant.

It is striking that the quasars lie on
what appears to be a reasonable extrapolation of the trend seen for
NLRGs in Figs \ref{lo-li} and \ref{lo-lxa}. We know that the quasars
are systematically brighter than radio galaxies of comparable
luminosity in X-ray (e.g. Belsole \etal\ 2006; Fig.\ \ref{lxa-lr}) IR
(Cleary \etal\ 2007; Fig.\ \ref{li-lr}) and [O{\sc iii}] (Jackson \&
Browne 1990; Fig.\ \ref{lo-lr}) but Figs \ref{lo-li} and \ref{lo-lxa}
show that the typical {\it factors} by which they exceed the expected
luminosities are comparable at the three different wavebands (i.e. a
factor of a few in each case); in simple unified models, this is a
coincidence, since the explanations for the quasar excess are
different in each case. In principle, the receding torus model
(Simpson 1998) provides a more natural explanation for at least the
[O{\sc iii}] and X-ray excesses; we will discuss modified unification
of this kind in more detail in Section \ref{lsummary}.

The relationship between [OIII] and `accretion-related' X-ray emission
plotted here is in excellent agreement with that found for other
classes of AGN. For example, Panessa \etal\ (2006) tabulate 2-10 keV
X-ray and [OIII] luminosities for a sample of low-luminosity Seyferts
and establish a correlation between the two which extends up to
intermediate-luminosity Seyfert 1s and PG quasars. The NLRGs in our
sample lie in an identical region of parameter space to the
high-luminosity objects in their fig. 4 and so on an extension of
their correlation. This strengthens the evidence that the
`accretion-related' X-ray emission in the 3CRR sources really does trace
intrinsic properties of the AGN in a way which is independent of the
properties of the jet.

\begin{figure*}
\epsfxsize 8.5cm
\epsfbox{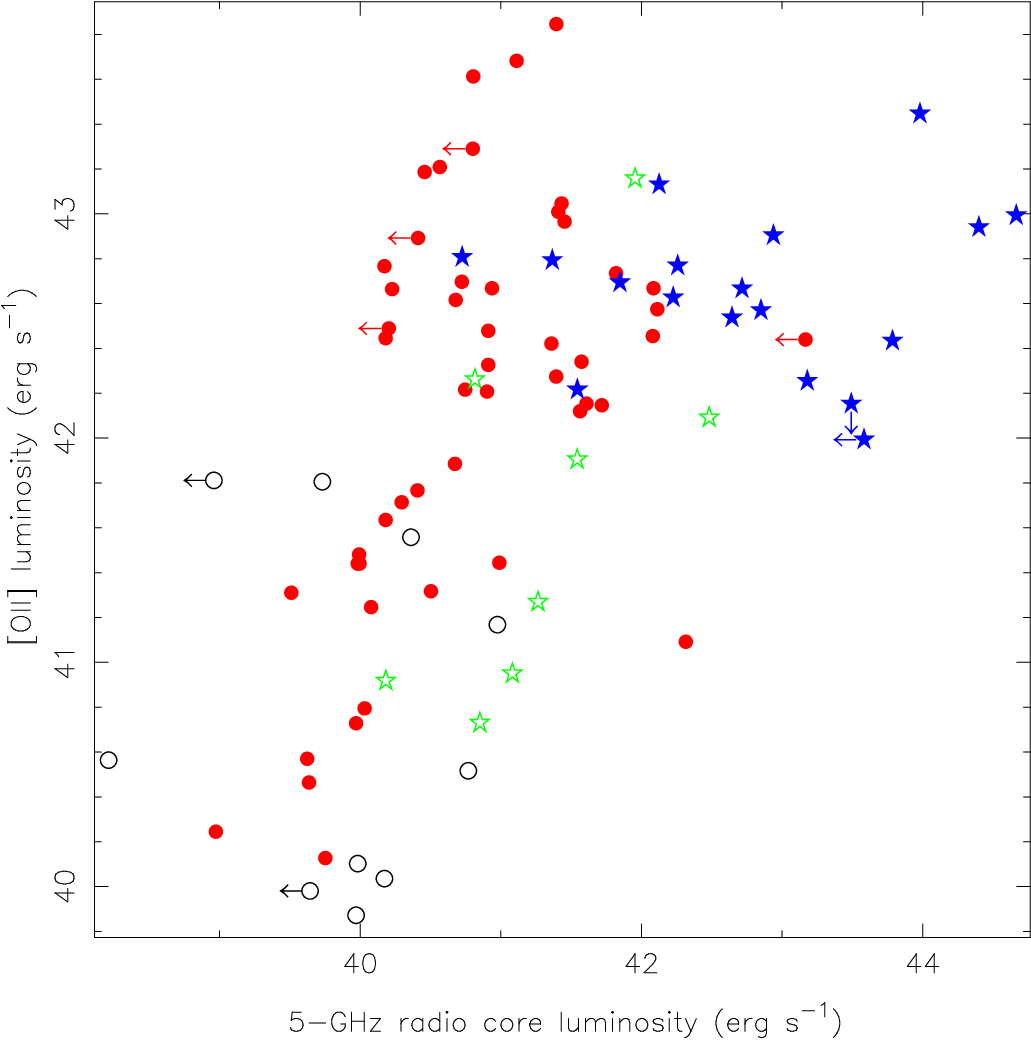}
\epsfxsize 8.5cm
\epsfbox{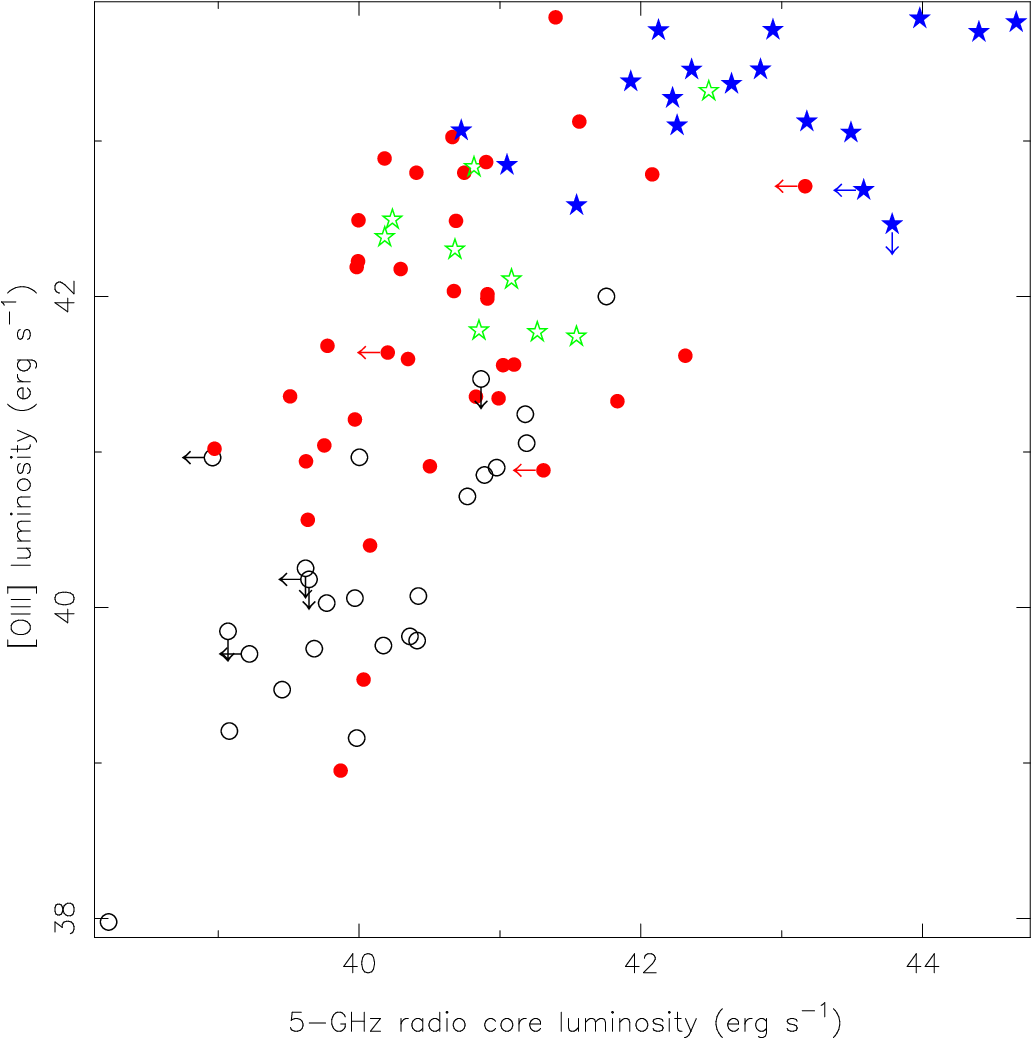}
\caption{Emission-line luminosity as a function of 5-GHz core radio luminosity for the
  $z<1.0$ 3CRR sample. Left panel shows [O{\sc ii}] and right [O{\sc
  iii}]. Symbols as in Fig.\ \ref{lxu-lr}.}
\label{lo-lc}
\end{figure*}

\begin{figure*}
\epsfxsize 8.5cm
\epsfbox{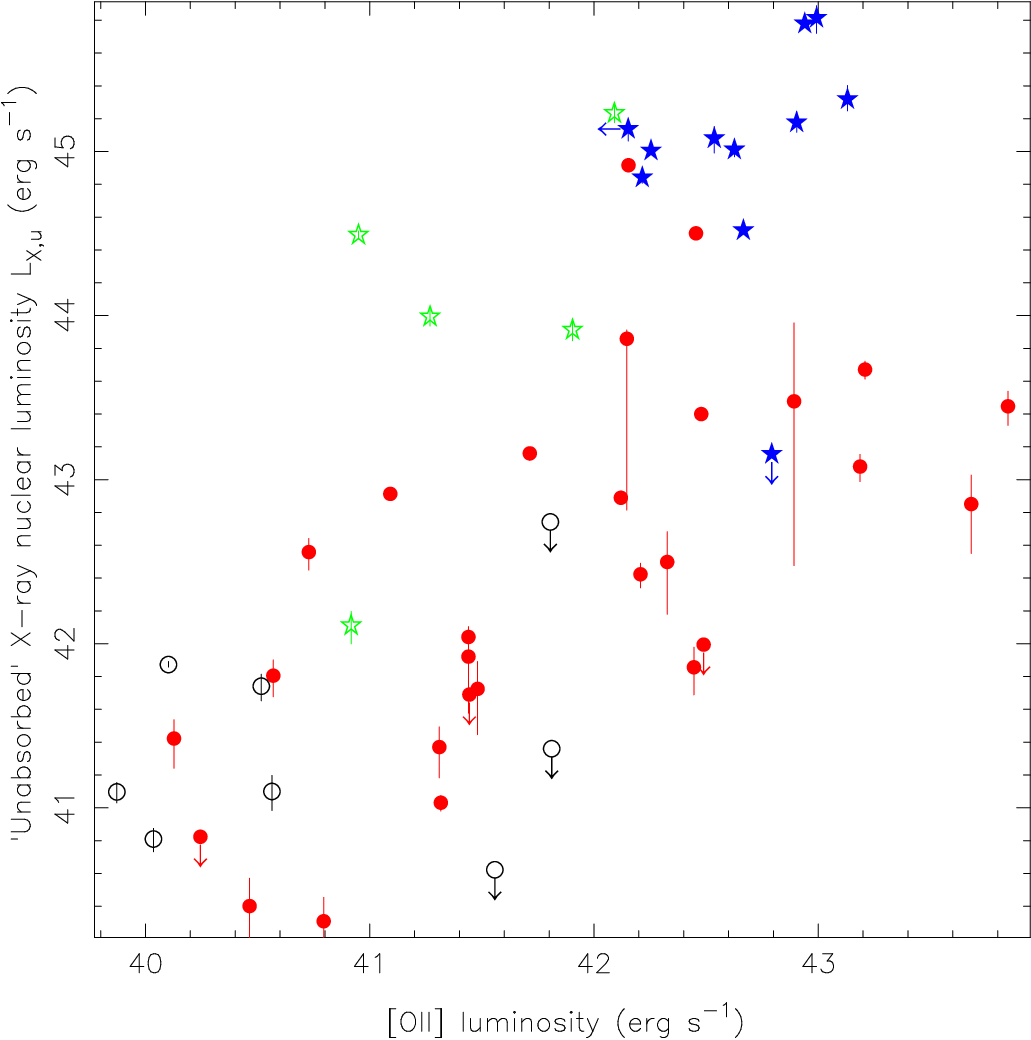}
\epsfxsize 8.5cm
\epsfbox{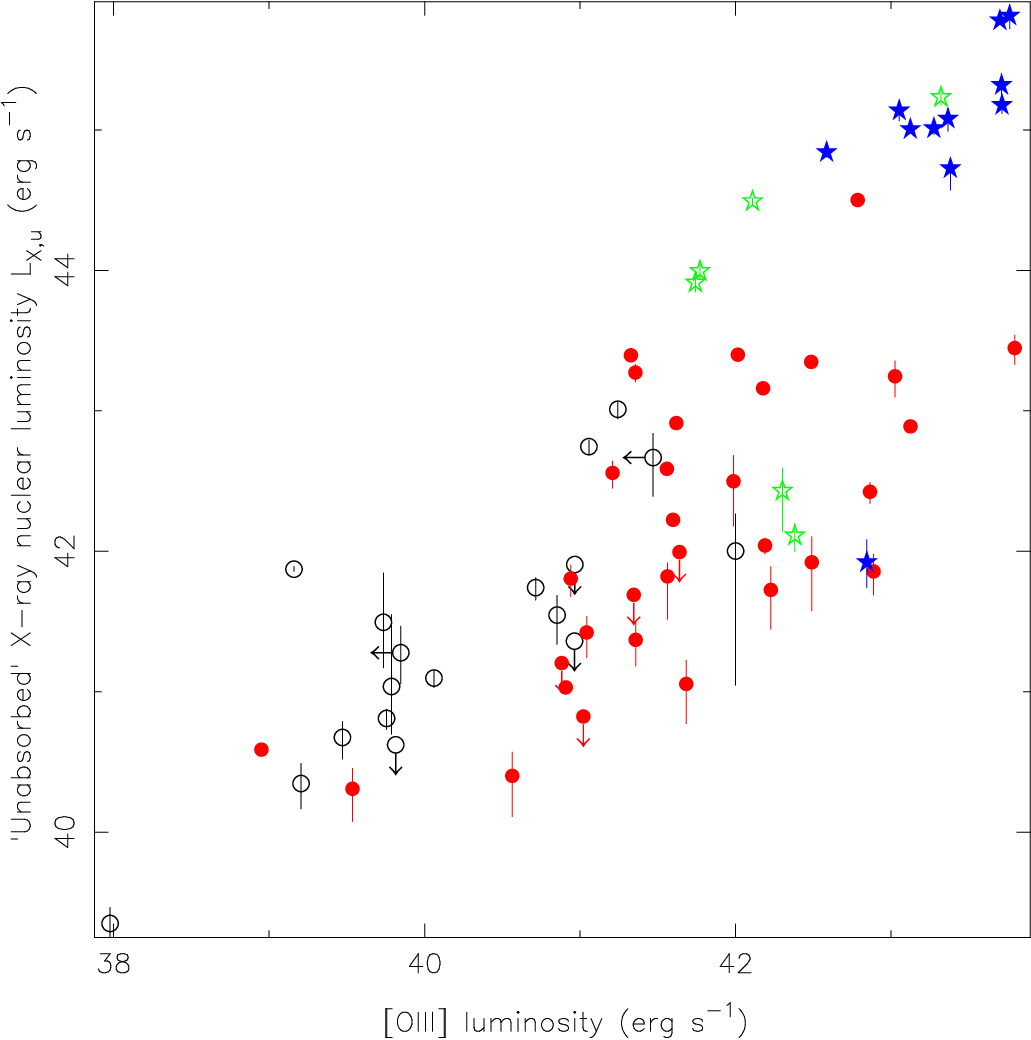}
\caption{X-ray luminosity for the unabsorbed X-ray component, $L_{\rm
  Xu}$ plotted
  against emission-line luminosity for the
  $z<1.0$ 3CRR sample. Left panel shows [O{\sc ii}] and right [O{\sc iii}]. Symbols as in Fig.\ \ref{lxu-lr}.}
\label{lo-lxu}
\end{figure*}

Finally, we examine the relationship between supposed beamed
quantities and optical emission lines. The relationship between radio
core luminosity and emission-line power is shown in Fig. \ref{lo-lc}.
There is a significant correlation overall between the radio core
luminosity and the [O{\sc iii}], but not the [O{\sc ii}] luminosities;
we can interpret this as being driven by the quasars with their higher
emission-line luminosities and core fluxes. Both correlations are
weaker than the corresponding correlations with $L_{178}$, and the
correlation for the NLRGs and LERGs alone is not significant in either
plot. However, if we consider the relationship between emission lines
and $L_{\rm Xu}$, an interesting trend emerges (Fig. \ref{lo-lxu}):
again, the overall [O{\sc iii}] relationship shows a significant
correlation and the [O{\sc ii}] does not, but here there {\it is} a
significant correlation for the NLRGs and LERGs alone between $L_{\rm
  [OIII]}$ and $L_{\rm Xu}$, although it is not as strong as the
correlation for the same objects between $L_5$ and $L_{\rm Xu}$. In
the model in which the scatter in $L_{\rm Xu}$ is mostly due to
beaming, this could perhaps be accounted for by a common correlation
with orientation angle coupled with the orientation-dependent
obscuration of [O{\sc iii}] invoked by Jackson \& Browne (1990) to
explain the position of quasars on these plots. Alternatively it could
indicate some direct connection between $L_{\rm Xu}$ and emission-line
power. We return to this point in Section \ref{jetrelated}.

\subsection{Luminosity correlations and their origins}
\label{lsummary}

Tables \ref{corsum} and \ref{regsum} list the correlations and
regressions discussed in the preceding subsections. To summarize, we
have argued that $L_{\rm 5}$ and $L_{\rm Xu}$ are beamed quantities,
showing a good correlation with each other but generally a poor
correlation, showing separation between radio galaxies and quasars,
with other quantities such as $L_{\rm 178}$ or $L_{\rm IR}$ that are
expected a priori to be unaffected by beaming. The implications of
this result will be discussed in Section \ref{jetrelated}. The total
radio luminosity, $L_{\rm 178}$, is reasonably well correlated with
all quantities ($L_{\rm Xa}$, $L_{\rm IR}$, $L_{\rm [OIII]}$...) that
might be expected to be unbeamed indicators of AGN power, but
generally we find non-linear correlations and relatively large
dispersion in the relationship, which in the standard picture is an
indication of the complex relationship between accretion power, jet
power and radio luminosity (see, e.g., Rawlings \& Saunders 1991 for a
discussion in the context of emission-line power). The best
correlation, with a relatively small dispersion and a slope consistent
with being linear, is that between $L_{\rm Xa}$ and $L_{\rm IR}$ (as
previously noted, using a much smaller sample, by H06). This strongly
supports the idea that in $L_{\rm Xa}$ we have a largely
uncontaminated indicator of AGN power.

In the simplest unified models we might expect all AGN-related
luminosities to correlate linearly with each other. As we have already
noted, however, the situation is more complex in the case of the
emission-line luminosities $L_{\rm [OII]}$ and $L_{\rm [OIII]}$. The
tendency for $L_{\rm [OIII]}$ to be higher in quasars than in radio
galaxies of the same radio luminosity (Fig.\ \ref{lo-lr}) is explained
by Jackson \& Browne (1990) in terms of higher absorption in radio
galaxies, and by Simpson (1998) in terms of the receding torus model,
in which more intrinsically luminous objects have smaller torus
covering fractions and therefore are more likely to be identified as
quasars. Since the covering fraction of the torus determines the
fraction of accretion disc radiation intercepted by it and re-radiated
in the mid-IR, a prediction of the simplest versions of this
model is that the mid-IR luminosity should have a weaker than linear
dependence on the overall AGN luminosity, $L_{\rm IR} \propto L_{\rm
AGN}^{1/2}$ (Dicken \etal\ 2008). If we assume that the
accretion-related X-ray luminosity $L_{\rm Xa}$ scales linearly with
$L_{\rm AGN}$, then our results provide no obvious support for this
model -- the correlation between $L_{\rm IR}$ and $L_{\rm Xa}$ is
consistent with being linear, and a model in which $L_{\rm IR} \propto
L_{\rm Xa}^{1/2}$ is ruled out by our fits. Equivalently, we see no
significant trend in a plot of $L_{\rm IR}/L_{\rm Xa}$ against $L_{\rm
178}$ or $L_{\rm Xa}$: there is a negative trend for the quasars but
this seems mostly likely to be explained by contamination of $L_{\rm Xa}$ by
jet-related emission in the most luminous, core-dominated objects.
Investigating other diagnostic plots of this kind, the only strong
trend we find is that $L_{\rm IR}/L_{\rm OII}$ strongly increases with
$L_{\rm Xa}$, which is in the opposite sense to the prediction of the
simple receding torus model in which $L_{\rm OII}$ should have a
similar dependence on bolometric luminosity to $L_{\rm IR}$ (Tadhunter
\etal\ 1998). We conclude that, while the simplest form of the
receding torus model may provide a better explanation of the
emission-line properties of the radio galaxies, it is not consistent
with our IR and X-ray data unless (1) $L_{\rm Xa}$ does not correlate
linearly with $L_{\rm AGN}$ (which we cannot test directly for our
sample) and/or (2) the IR properties of the torus do not scale in the
simplest possible way. On different grounds Dicken \etal\ (2008)
suggest that (2) is the case for their independent sample of radio
galaxies.

\setcounter{table}{4}
\begin{table*}
\caption{Results of partial correlation analyses described in Section \ref{sec:results}. Note that all
  correlations include all objects in the relevant subsample (column
  4) for which the relevant luminosities are available. The number of
  objects included in the subsample is given in column 5.
  $\tau/\sigma$ gives an indication of the strength of the partial
  correlation in the presence of redshift; we adopt a cutoff of
  $\tau/\sigma > 3$ for a significant correlation.}
\label{corsum}
\begin{tabular}{llrlrlr}
\hline
Abscissa&Ordinate&Figure&Subsample&Number&Correlation?&$\tau/\sigma$\\
\hline
$L_{\rm 178}$&$L_{\rm Xu}$&\ref{lxu-lr}&All&87&N&2.77\\
&&&LERG&26&N&1.13\\
&&&NLRG&40&N&1.56\\
&&&BLRG&6&N&0.09\\
&&&Q&15&N&0.62\\
$L_{\rm 178}$&$L_{\rm Xa}$&\ref{lxa-lr}&All&87&Y&3.75\\
&&&LERG&26&N&0.68\\
&&&NLRG&40&Y&3.37\\
&&&BLRG&6&N&-1.08\\
&&&Q&15&N&0.65\\
$L_{\rm 5}$&$L_{\rm Xu}$&\ref{lxu-lc}&All&87&Y&8.10\\
&&&Q excluded&72&Y&5.84\\
&&&LERG&26&Y&4.08\\
&&&NLRG&40&Y&3.56\\
&&&BLRG&6&N&2.39\\
&&&Q&15&N&2.37\\
&&&NLRG/LERG with [O{\sc iii}]&50&Y&3.97\\
$L_{\rm 5}$&$L_{\rm Xa}$&\ref{lxa-lc}&All&87&Y&4.27\\
&&&Q excluded&72&N&1.45\\
&&&LERG&26&N&1.38\\
&&&NLRG&40&N&-0.08\\
&&&BLRG&6&N&-0.01\\
&&&Q&15&N&2.13\\
$L_{\rm Xu}$&$L_{\rm Xa}$&\ref{lxu-lxa}&NLRG&40&N&-0.13\\
$L_{\rm 178}$&$L_{\rm IR}$&\ref{li-lr}&All&94&Y&4.89\\
$L_{\rm 5}$&$L_{\rm IR}$&\ref{li-lc}&All&94&Y&7.18\\
&&&LERG&24&Y&3.71\\
&&&NLRG&43&N&1.98\\
&&&BLRG&7&N&0.45\\
&&&Q&20&N&0.21\\
$L_{\rm IR}$&$L_{\rm Xu}$&\ref{li-lxu}&All&65&Y&4.92\\
&&&LERG excluded&45&Y&3.19\\
$L_{\rm IR}$&$L_{\rm Xa}$&\ref{li-lxa}&All&65&Y&8.95\\
&&&LERG excluded&45&Y&9.25\\
$L_{\rm 178}$&$L_{\rm [OII]}$&\ref{lo-lr}&All&86&Y&3.86\\
$L_{\rm 178}$&$L_{\rm [OIII]}$&\ref{lo-lr}&All&88&Y&5.31\\
$L_{\rm [OII]}$&$L_{\rm IR}$&\ref{lo-li}&All&62&Y&3.60\\
&&&LERG excluded&56&N&2.66\\
$L_{\rm [OIII]}$&$L_{\rm IR}$&\ref{lo-li}&All&66&Y&7.08\\
&&&LERG excluded&47&Y&6.47\\
$L_{\rm [OII]}$&$L_{\rm Xa}$&\ref{lo-lxa}&All&52&N&1.66\\
&&&LERG excluded&44&N&0.93\\
$L_{\rm [OIII]}$&$L_{\rm Xa}$&\ref{lo-lxa}&All&67&Y&6.15\\
&&&LERG excluded&49&Y&5.22\\
$L_{\rm 5}$&$L_{\rm [OII]}$&\ref{lo-lc}&All&86&N&2.69\\
&&&NLRG and LERG&60&N&1.66\\
$L_{\rm 5}$&$L_{\rm [OIII]}$&\ref{lo-lc}&All&88&Y&4.69\\
&&&NLRG and LERG&61&N&2.31\\
$L_{\rm [OII]}$&$L_{\rm Xu}$&\ref{lo-lxu}&All&52&N&1.78\\
&&&NLRG and LERG&36&N&1.54\\
$L_{\rm [OIII]}$&$L_{\rm Xu}$&\ref{lo-lxu}&All&67&Y&5.11\\
&&&NLRG and LERG&50&Y&3.22\\
\hline
\end{tabular}
\end{table*}

\begin{table*}
\caption{Results of regression analyses described in Section
  \ref{sec:results}. Note that all regressions include all objects in
  the relevant subsample (column 4) for which the relevant
  luminosities are available. The number of objects included in the
  subsample is given in column 5. Errors (formally, credible
  intervals) are the equivalent of $1\sigma$ for one interesting
  parameter only.}
\label{regsum}
\begin{tabular}{llrlrrrr}
\hline
Abscissa&Ordinate&Figure&Subsample&Number&Slope&Intercept&Scatter\\
\hline
$L_{\rm 178}$&$L_{\rm Xa}$&\ref{lxa-lr}&Detected NLRGs&27&
   $0.72^{+0.12}_{-0.38}$&$13.11^{+16.29}_{-5.11}$&$0.32^{+0.04}_{-0.10}$\\
$L_{\rm 5}$&$L_{\rm Xu}$&\ref{lxu-lc}&NLRGs and LERGs&66&
   $1.53^{+0.19}_{-0.26}$&$-19.79^{+10.52}_{-7.67}$&$0.56^{+0.04}_{-0.04}$\\
&&&LERGs&26&
   $1.27^{+0.29}_{-0.39}$&$-9.62^{+15.63}_{-11.44}$&$0.63^{+0.08}_{-0.12}$\\
&&&NLRGs&40&
   $1.72^{+0.30}_{-0.09}$&$-27.55^{+3.55}_{-12.07}$&$0.51^{+0.03}_{-0.07}$\\
$L_{\rm 178}$&$L_{\rm IR}$&\ref{li-lr}&All&94&
   $1.36^{+0.13}_{-0.19}$&$-13.94^{+8.25}_{-5.45}$&$0.59^{+0.03}_{-0.04}$\\
$L_{\rm IR}$&$L_{\rm Xa}$&\ref{li-lxa}&X-ray detected&36&
   $0.97^{+0.23}_{-0.12}$&$0.91^{+5.35}_{-10.13}$&$0.32^{+0.03}_{-0.05}$\\
$L_{\rm 178}$&$L_{\rm [OII]}$&\ref{lo-lr}&All&86&
   $1.02^{+0.10}_{-0.20}$&$-1.73^{+8.46}_{-4.43}$&$0.39^{+0.02}_{-0.04}$\\
$L_{\rm 178}$&$L_{\rm [OIII]}$&\ref{lo-lr}&All&88&
   $1.36^{+0.15}_{-0.13}$&$-15.91^{+5.59}_{-6.47}$&$0.48^{+0.02}_{-0.03}$\\
$L_{\rm [OII]}$&$L_{\rm IR}$&\ref{lo-li}&All&62&
   $1.57^{+0.20}_{-0.27}$&$-21.34^{+11.37}_{-8.28}$&$0.55^{+0.03}_{-0.04}$\\
$L_{\rm [OIII]}$&$L_{\rm IR}$&\ref{lo-li}&All&66&
   $1.06^{+0.05}_{-0.08}$&$-0.36^{+3.40}_{-2.00}$&$0.43^{+0.02}_{-0.03}$\\
$L_{\rm [OIII]}$&$L_{\rm Xa}$&\ref{lo-lxa}&X-ray detected&40&
   $0.87^{+0.08}_{-0.37}$&$7.34^{+15.73}_{-3.43}$&$0.35^{+0.02}_{-0.07}$\\
\hline
\end{tabular}
\end{table*}

\section{Discussion}
\label{discussion}

\subsection{The nature of LERGs and NLRGs revisited}
\label{lergs}

We begin by acknowledging again (see H06) the arbitrary nature of the
optical emission-line classification that we rely on to identify the
LERG population. In the case of the 3CRR galaxies the problem is
particularly acute, as the only information we have in many cases
comes from the qualitative description of poor-quality spectra given
in the original 3CRR catalogue (LRL). Even when this is
not the case, `nuclear' narrow emission lines cannot be unambiguously
associated with the classical narrow-line region, photoionized by an
accretion disc; as discussed by Evans \etal\ (2008), there is likely
to be contamination both with extended emission-line regions affected
by the passage of the jet and with material photoionized by the jet
itself. Thus while the {\it absence} of strong high-excitation emission
lines (neglecting cases where there is strong foreground obscuration)
is a reasonable indicator that there is no radiatively efficient
active nucleus present to photoionize them, the {\it presence} of such
emission lines does not necessarily imply the presence of a
radiatively efficient nucleus.

We therefore initially consider the X-ray properties of the sources
classed as LERGs by LRL. Of the 26 in our sample with X-ray data, only
two show any evidence for a heavily obscured, luminous component in
the X-ray spectrum, and one of these is 3C\,123, whose emission-line
spectrum is known to be severely affected by foreground (Galactic)
reddening (H06). The other is the peculiar object 3C\,293, which is
almost certainly misclassified in LRL given the strong high-excitation
emission lines reported in subsequent work (e.g. van Breugel \etal\
1984). Thus it remains the case (as reported in H06) that a genuine
low-excitation optical spectrum is an excellent predictor of the
absence of a heavily absorbed, `accretion-related' component in the
X-ray. We would argue that both 3C\,293 and 3C\,123 should really be
classed as NLRGs; their IR properties (Fig.\ \ref{li-lxa}) are
entirely consistent with the regression line fitted to the NLRGs.

On the other hand, of the 40 objects classified as NLRGs in our sample
with X-ray data, there are 13 where we have not detected a heavily
absorbed nucleus, i.e., where either the source is not detected at all
or a two-power-law model does not give an improvement in the fit to
the X-ray data. These 13 can be divided into FRIs (3C\,274 = M87,
3C\,84 = Perseus A, 3C\,305, 3C\,338, 3C\,315 and 3C\,346) and FRIIs
(3C\,153, 3C\,300, 3C\,274.1, 3C\,244.1, 3C\,220.1, 3C\,277.2 and
3C\,6.1), in order of 178-MHz radio luminosity. The low-luminosity
cases have already been discussed by Evans \etal\ (2008): as we
pointed out in that paper, we are particularly liable in these systems
to contamination of the optical emission-line spectrum by material not
directly associated with photoionization by the nucleus, and this is
especially true in the cases of objects such as 3C\,274, 3C\,84 and
3C\,338 which lie in the centres of clusters. 3C\,315 is a
non-detection in the 8-ks {\it Chandra} snapshot and it seems possible
that a deeper observation might reveal a hidden nucleus. Of the
high-luminosity cases, 3C\,153 only has poor-quality X-ray data and is
a non-detection in the X-ray. [We note also that the emission-line
classification of 3C\,153 is disputed: Willott \etal\ (1999) classify
it as a LERG on the basis of the high-quality spectroscopy of Lawrence
\etal\ (1996), which would be consistent with the X-ray results, since
no radio nucleus is seen.] More interesting are the six NLRG FRIIs
with comparatively sensitive X-ray data but with no evidence for a
heavily obscured nucleus, 3C\,6.1, 3C\,220.1, 3C\,244.1, 3C\,274.1,
3C\,277.2 and 3C\,300. We return to these objects below.

Mid-IR emission is key to a discussion of obscured nuclei, since
it gives us the ability to distinguish between a very heavily obscured
but still radiatively efficient nucleus and one which is simply not
present, or at least not radiatively efficient (e.g. Whysong \&
Antonucci 2004). An important advance over the situation we described
in H06 is that mid-IR information is available for a large fraction of
the X-ray-observed 3CRR sources, as discussed in Section
\ref{xray-ir}. Fig.\ \ref{li-lxa} is the relevant diagnostic plot.
This shows a very good relationship between mid-IR and X-ray
luminosity for sources where an obscured X-ray component is detected.
Since we assume $N_{\rm H} = 10^{23}$ cm$^{-2}$ in calculating upper
limits on accretion-related luminosity, any object that follows that
relationship but in fact has a higher column density (e.g.
Compton-thick objects with $N_{\rm H} \ga 10^{24}$ cm$^{-2}$) should
appear as an upper limit in X-ray lying below/to the right of the
regression line in Fig.\ \ref{li-lxa}.

Considering the LERGs first, we see that powerful LERGs ($L_{\rm IR} >
10^{43}$ erg s$^{-1}$) tend to lie close to the regression line,
whether they are detections or upper limits in the IR. This implies
that none of these objects is Compton-thick. Conceivably some have
hidden X-ray nuclei that are close to the current upper limits, but
none is a powerful, heavily obscured AGN. This supports similar
conclusions by Ogle \etal\ (2006) and Dicken \etal\ (2008). Low-power
LERGs, including most of the nearby FRI radio galaxies, tend to lie
below the regression line. However, in these cases, contamination of
the mid-IR by galactic and jet-related emission (cf.\ Whysong \&
Antonucci 2004) may be very important. The slit width for the {\it
Spitzer} spectroscopy at zero redshift is 10.5 arcsec, corresponding
to $\sim 4$ kpc at $z \approx 0.02$ (the redshift of a typical FRI in
our sample), so the measured flux densities include a significant
contribution from stars and small-scale dust and are not a reliable
measure of AGN luminosity alone. Some evidence for jet
contamination in the LERGs (only) is provided by the significant
correlation between $L_5$ and $L_{\rm IR}$ for these objects discussed
in Section \ref{radio-ir}. For these objects detailed analysis of the
mid-IR spectroscopic and imaging data (Birkinshaw \etal\ in prep.) is
required to constrain the nature of the IR emission, and accordingly
we do not consider them further.

Several objects classed as NLRGs lie well below the regression line in
Fig.\ \ref{li-lxa} and are labelled in that figure. The
lowest-luminosity of these is M87 (3C\,274), which lies in much the
same region of parameter space as several of the LERG FRIs discussed
above. In this particular case we know unambiguously that the {\it
Spitzer} flux density measurement must be contaminated by non-nuclear
emission, since Whysong \& Antonucci (2004), using ground-based
imaging, measured a point-like flux density of $13 \pm 2$ mJy at 11.7
$\mu$m, explicitly noting that wide-aperture photometry with IRAS gave
a larger result; we also see a clear difference of a factor $\sim 2$
between the fluxes measured in the short-wavelength {\it Spitzer}
spectroscopy (with its 3.7-arcsec slit) and the long-wavelength
spectroscopy at 15 $\mu$m, showing without doubt that the
longer-wavelength data are dominated by extended emission. As with
other low-power objects, jet-related mid-IR emission is also a serious
potential contaminant. The other significant outliers on this plot are
3C\,84 (NGC 1275, Perseus A), 3C\,346, and 3C\,244.1. 3C\,84 is
diversely classified in the literature as a Seyfert 2 (NLRG) or a
Seyfert 1 (BLRG); the detection of broad H$\alpha$ emission
(Fillipenko \& Sargent 1985) would put it in the BLRG class, in which
case it would not be relevant here (for a BLRG we would consider the
much higher unabsorbed X-ray luminosity, $\sim 8 \times 10^{42}$ erg
s$^{-1}$, as our best estimate of the accretion-related X-ray).
Weedman \etal\ (2005) argue on the basis of the silicate feature at 10
$\mu$m that the {\it Spitzer} emission comes from dust rather than
non-thermal (i.e. jet-related) AGN emission, but it is not clear
whether the dust emission is truly nuclear or is related to the gas
and star formation known to be present in the nucleus of NGC 1275. In
any case, {\it INTEGRAL} observations (Bassani \etal\ 2006) suggest
that we are not missing very heavily absorbed emission in 3C\,84.
3C\,346 has an unusually strong unabsorbed X-ray luminosity for its
radio luminosity (lying in the region populated by BLRGs and quasars)
and it seems possible either that it is a previously undentified
broad-line object or that it is a LERG with emission lines related to
the active nucleus: Chiaberge \etal\ (2002) show that it lies in the
same region as broad-line objects and LERGs in their diagnostic plot
relating [OIII] equivalent width to optical/radio core flux density
ratio.

The most interesting of the four outliers is 3C\,244.1, which is a
normal FRII radio galaxy with a very luminous mid-IR detection, lying
more than an order of magnitude in X-ray luminosity below the
regression line. It is also a clear outlier on the plot of X-ray
versus [O{\sc iii}] luminosity (Fig.\ \ref{lo-lxa}). It seems highly
likely that this is a genuine example of a Compton-thick FRII. To make
it lie on the IR/X-ray regression line we determine (using {\sc
xspec}) that we would require $N_{\rm H} > 2.5 \times 10^{24}$
cm$^{-2}$. As we will discuss in the next section, this is not an
unreasonably large column density for a NLRG. Other NLRG FRIIs
without detected heavily absorbed components but with IR observations
(3C\,6.1, 3C\,220.1, 3C\,274.1 and 3C\,300) lie close to the regression line in
Fig.\ \ref{li-lxa} for the assumed limiting column density of
$10^{23}$ cm$^{-2}$, and so could be brought back to the line without
requiring Compton-thick $N_{\rm H}$ values. 3C\,277.2, the remaining
powerful FRII without a detection of heavily absorbed X-rays, has no
{\it Spitzer} data.

\subsection{Absorbing columns in NLRGs and type-II quasars}
\label{comptonthick}

We now have enough examples of objects with significant intrinsic
absorbing columns to begin to ask questions about the {\it
distribution} of $N_{\rm H}$ values. In total we have 40 objects with
measured values of $N_{\rm H}$ (Table \ref{nhvals}). This includes four FRI LERGs where an additional absorbing component, presumably
related to large-scale dust rings in the host galaxy, is required to
allow a good single power-law fit (e.g. Hardcastle \etal\ 2002;
Hardcastle, Sakelliou \& Worrall 2005): as this is a rather different
situation to genuine nuclear absorption we do not include these in the
analysis that follows, though we do include the two powerful LERGs
(3C\,123 and 3C\,293) which we have argued are misclassified in LRL
(Section \ref{lergs}). It also includes three objects classed as BLRGs
(3C\,33.1, 3C\,109, 3C\,381) and four quasars (3C\,47, 3C\,249.1,
3C\,325, 3C\,351) where a two-power-law model is required for a good
fit to the X-ray spectrum. Of the broad-line objects, only 4 have a
best-fitting column density $>10^{22}$ cm$^{-2}$; one of these
(3C\,381) is almost certainly not a genuine BLRG (see Section
\ref{381}). On the other hand, no NLRG with detected obscured
emission has a best-fitting $N_{\rm H}$ value below $10^{22}$
cm$^{-2}$.

\begin{figure}
\epsfxsize 8.5cm
\epsfbox{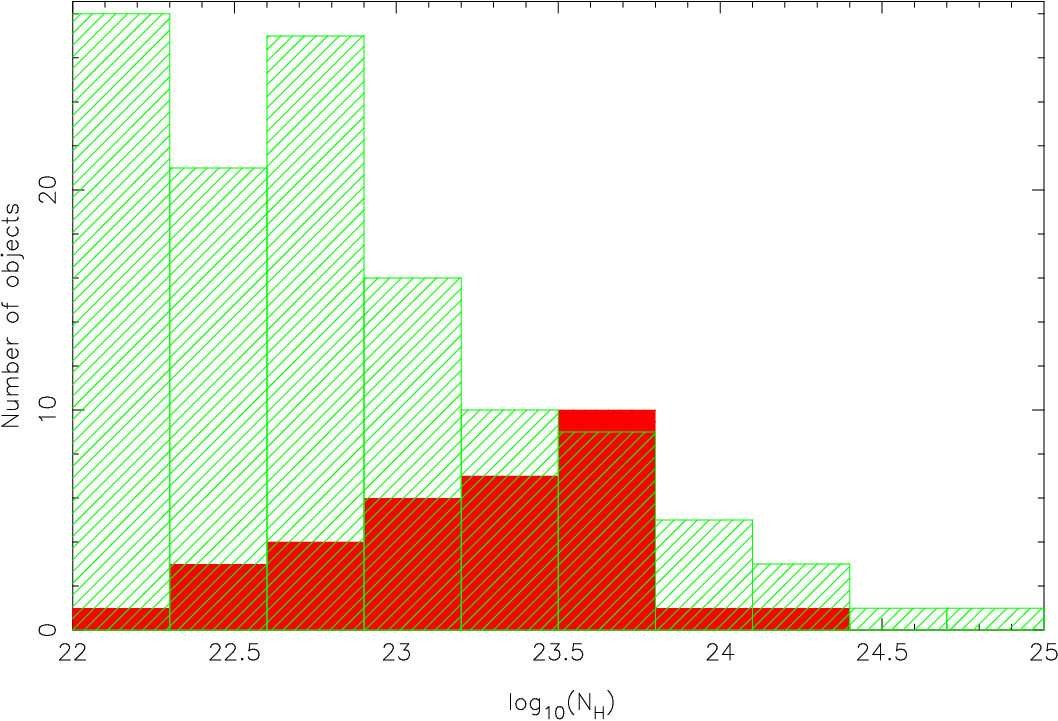}
\vskip 6pt
\epsfxsize 8.5cm
\epsfbox{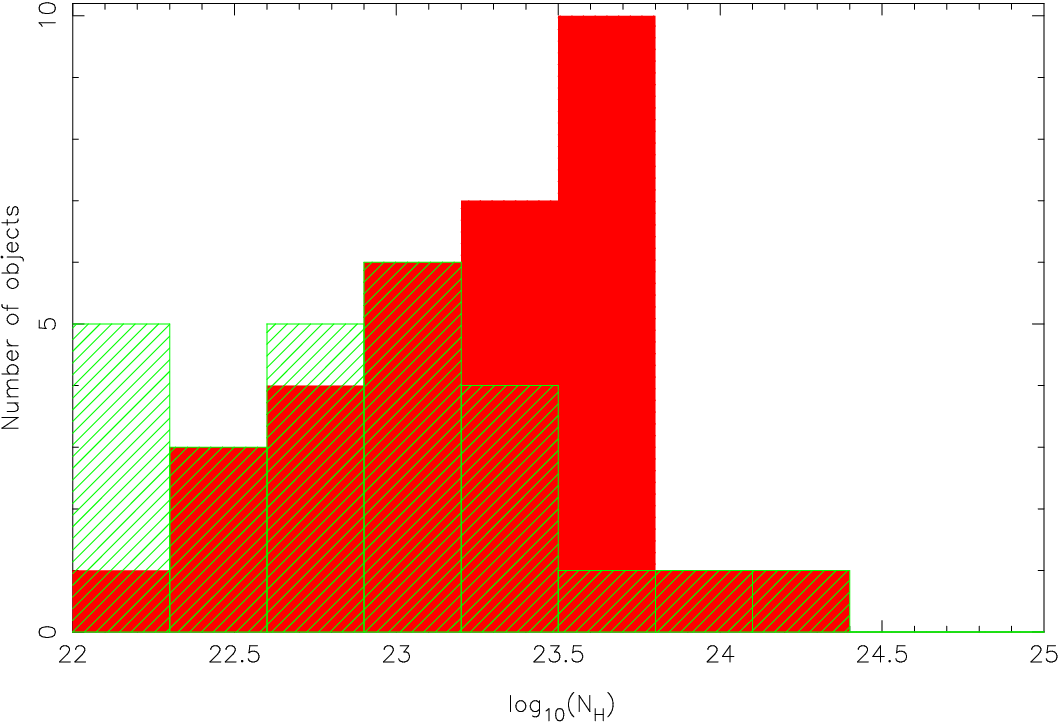}
\caption{Top panel: distribution of column densities in 3CRR NLRGs with $N_{\rm
  H} > 10^{22}$ cm$^{-2}$ (red) and in candidate type 2 quasars from
  the COSMOS sample with the same $N_{\rm H}$ selection criterion (green).
  Bottom panel: the same, but showing only the $z<1.0$ COSMOS objects.}
\label{nhhist}
\end{figure}

It is interesting to compare the distribution of $N_{\rm H}$ values
for these radio-selected obscured AGN with the distribution obtained
for X-ray-selected objects (X-ray selection, coupled with
spectroscopic or photometric redshifts, is necessary if we are to
obtain a large database of $N_{\rm H}$ values). Fig.\ \ref{nhhist}
shows a histogram of our $N_{\rm H}$ data for NLRGs (plus 3C\,123 and
3C\,293) with $N_{\rm H} > 10^{22}$ cm$^{-2}$ superposed on the column
densities of candidate type 2 quasars (selected on the basis that
$L_{\rm X, unabs, 0.5-10} > 10^{44}$ erg s$^{-1}$ and $N_{\rm
H}>10^{22}$ cm$^{-2}$) from the COSMOS survey (Mainieri \etal\ in
prep.)\footnote{We use these unpublished data because of the large
number of obscured objects that have been detected; we note
that the $N_{\rm H}$ distributions of the COSMOS objects are in good
qualitative agreement with the published results of other surveys,
e.g. Tajer \etal\ (2007)}. The most striking result here is the almost
complete absence of absorbed radio-loud sources with low column
densities (less than a few $\times 10^{22}$ cm$^{-2}$), a difference
that is clearly significant on a Kolmogorov-Smirnov test. This is very
hard to explain as a selection effect, since the plot includes all the
NLRGs with measured $N_{\rm H}$ values. The handful of NLRGs with no
measured $N_{\rm H}$ values are clearly not enough to make up for the
discrepancy. Although some (about 1/4) of the NLRGs in our sample would
not have passed the luminosity selection applied to the COSMOS
objects, it is hard to see how this could significantly affect the
results. Similarly, although the X-ray selection of the COSMOS objects
does imply some bias away from heavily obscured objects, very large
numbers of objects would have had to be missed in order to make the
true distribution of type 2 quasars match that of the radio galaxies.
Because the COSMOS sample extends to high redshifts, they include some
high-column density objects whose $N_{\rm H}$ we would not have been
able to measure (see discussion of 3C\,244.1, above) but this cannot
account for the difference in distributions either, as we can see by
restricting the comparison to the $z<1.0$ COSMOS objects (Fig.\
\ref{nhhist}, bottom panel). The only difference would seem to be
radio luminosity; while some of the COSMOS objects have radio
detections, the maximum radio luminosity in the sample is 2-3 orders
of magnitude below that found in 3CRR objects, and most are
radio-quiet. Accordingly we propose that this is a real physical
difference between the populations of radio- and X-ray selected type 2
quasars; at least the most luminous radio-loud objects (as represented
by the 3CRR sample) have significantly higher obscuring column
densities than the population as a whole. This could have implications
for the nature of the cold accreting material in the powerful 3CRR
objects. More detailed analysis must await the availability of full
data on the COSMOS objects or other large samples of X-ray selected
type 2 quasars.

Our column density distribution for the 3CRR objects may also be
compared to the work on nearby Seyferts by e.g. Risaliti, Maiolino \&
Salvati (1999) or Cappi \etal\ (2006). Although the sample sizes of
these papers are generally small, the selection criteria are more
similar to those we employ, since they involve X-ray spectroscopic
observations of samples selected at other wavebands. Statistical
comparisons here are difficult, since there are many Compton-thick
low-luminosity Seyferts (1/3 or more of Seyfert 2s; Cappi \etal\
2006), implying lower limits on the X-ray-measured $N_{\rm H}$ values,
whereas (as we argue above) few radio galaxies for which data are
available are Compton-thick. Qualitatively it seems that the
distribution of {\it measured} $N_{\rm H}$ values for 3CRR objects,
peaking at a few $\times 10^{23}$ cm$^{-2}$ is similar to the {\it
measured} $N_{\rm H}$ distribution determined by Risaliti \etal\
(1999) for objects denoted `strict Seyfert 2s'. However, the Seyfert
population as a whole, including Seyfert 1s and intermediate Seyfert
types, shows a continuous distribution of $N_{\rm H}$ values down to
values of $\sim 10^{20}$ cm$^{-2}$ (Cappi \etal\ 2006), rather like
the COSMOS quasar-2 candidates, and we clearly do not observe this in
the 3CRR objects.

There is no evidence for variation in the distribution of $N_{\rm H}$
values for the 3CRR objects with redshift. Dividing the sample with
measured $N_{\rm H} > 10^{22}$ cm$^{-2}$ at the median redshift of
0.2, there are no significant differences in the $N_{\rm H}$
distribution on a K-S test. This is unsurprising since our NLRG sample
is dominated by low-redshift objects; we do not have enough objects at
high redshift and luminosity to search for cosmological or
(equivalently, in this sample) luminosity evolution effects. This
situation would change if the $z>1.0$ 3CRR radio galaxies with {\it
Chandra} or {\it XMM-Newton} observations were included in our
analysis, and we expect to return to this point in a future paper.

\begin{table}
\caption{Measured intrinsic column densities for 3CRR sources. Errors are
  directly derived from X-ray fits, except for the sources where
  $N_{\rm H}$ was measured by Belsole \etal\ (2006), where we have
  scaled the $1\sigma$ errors quoted in their paper.}
\label{nhvals}
\begin{tabular}{lrllrrr}
\hline
Source&$z$&FR&Type&\multicolumn{3}{c}{Absorbing column
  (cm$^{-2}$)}\\
&&class&&Value&\multicolumn{2}{c}{90\% conf. range}\\
\hline
3C\,20&0.174&II&NLRG&18.17&14.25&23.55\\
3C\,33&0.0595&II&NLRG&38.80&33.00&45.60\\
3C\,33.1&0.181&II&BLRG&4.15&2.79&5.71\\
3C\,47&0.425&II&Q&10.70&5.13&28.34\\
3C\,61.1&0.186&II&NLRG&56.03&35.85&82.77\\
3C\,79&0.2559&II&NLRG&24.89&16.62&35.91\\
3C\,83.1B&0.0255&I&LERG&3.19&2.53&4.03\\
3C\,98&0.0306&II&NLRG&11.80&9.50&14.70\\
3C\,109&0.3056&II&BLRG&0.46&0.36&0.58\\
3C\,123&0.2177&II&LERG&3.07&2.12&4.60\\
3C\,132&0.214&II&NLRG&4.71&2.89&7.46\\
3C\,171&0.2384&II&NLRG&8.53&7.27&9.98\\
3C\,184&0.994&II&NLRG&48.70&28.80&84.89\\
3C\,184.1&0.1187&II&NLRG&3.67&2.39&5.24\\
3C\,192&0.0598&II&NLRG&51.63&35.06&84.47\\
3C\,223&0.1368&II&NLRG&5.67&2.03&13.45\\
4C\,73.08&0.0581&II&NLRG&53.56&25.16&109.04\\
3C\,228&0.5524&II&NLRG&5.92&0.00&56.50\\
3C\,234&0.1848&II&NLRG&28.09&26.04&30.67\\
3C\,249.1&0.311&II&Q&0.37&0.11&0.81\\
3C\,265&0.8108&II&NLRG&16.80&4.96&34.07\\
3C\,272.1&0.0029&I&LERG&0.18&0.11&0.27\\
3C\,280&0.996&II&NLRG&9.70&1.97&19.90\\
3C\,284&0.2394&II&NLRG&161.69&114.45&540.78\\
3C\,285&0.0794&II&NLRG&32.10&27.46&37.62\\
3C\,292&0.71&II&NLRG&26.40&16.69&47.46\\
3C\,293&0.0452&I&LERG&13.12&9.28&18.18\\
3C\,295&0.4614&II&NLRG&40.96&32.54&48.71\\
3C\,296&0.0237&I&LERG&1.52&0.59&2.40\\
3C\,321&0.096&II&NLRG&88.18&57.20&134.50\\
3C\,325&0.86&II&Q&2.93&2.16&3.87\\
3C\,330&0.5490&II&NLRG&23.60&1.56&50.41\\
3C\,349&0.205&II&NLRG&1.16&0.93&1.37\\
3C\,351&0.371&II&Q&0.85&0.74&0.99\\
3C\,381&0.1605&II&BLRG&30.51&23.58&39.19\\
3C\,433&0.1016&II&NLRG&9.30&8.12&10.47\\
3C\,436&0.2145&II&NLRG&36.18&22.97&56.06\\
3C\,452&0.0811&II&NLRG&57.40&49.80&66.50\\
3C\,457&0.428&II&NLRG&34.23&29.91&38.98\\
3C\,465&0.0293&I&LERG&0.45&0.06&1.05\\

\hline
\end{tabular}
\end{table}

\subsection{The unabsorbed component of the X-rays}
\label{jetrelated}

NLRG (and some BLRG/quasar) X-ray spectra almost universally show a
`soft excess' over a simple absorbed power-law model, represented by
the unabsorbed X-ray component described in the previous sections.
This excess can usually be fitted by power-law models, although
generally the statistics are quite poor and the power-law index is
often unconstrained.

Three classes of model for this emission are
encountered in the literature:
\begin{enumerate}
\item It is thermal or line emission either from the IGM of the host system or
  from photoionized material close to the nucleus.
\item It is non-thermal, power-law emission arising from the central AGN and
  visible to us either via scattering or in partial-covering models.
\item It is non-thermal, power-law emission related not directly to
  the central AGN but to the unresolved nuclear jet (either from
  synchrotron or from inverse-Compton emission).
\end{enumerate}

In model (i), we need to distinguish between thermal/line emission
arising from the IGM and from material close to the nucleus. The
former is expected in all observations where a spatially large
aperture has to be used (i.e. essentially all {\it XMM}, {\it ROSAT}
or {\it ASCA} observations of moderate-redshift radio galaxies in our
sample) but of course is expected to be removed by small-aperture
spectroscopy with local background subtraction, as is possible with
{\it Chandra}. More interesting is the case where line emission arises
on scales comparable to that of the NLR, as is known in some Seyferts
even in the presence of radio jets (e.g. Evans \etal\ 2006b); grating
spectroscopy shows that the soft X-rays can be completely dominated by
emission lines from material photoionized by the AGN (e.g. Guainazzi
\& Bianchi 2007). In radio galaxies the clearest evidence that this
can be important is seen in {\it XMM} spectroscopy of the non-3CRR
BLRG 3C\,445 (Sambruna, Reeves \& Braito 2007) where many strong
emission-line features are seen below 2 keV. Even in this model,
though, a soft power-law component (represented by Sambruna \etal\
using a partial covering model) is required and this power-law
emission dominates the soft part of the spectrum. The only argument
that the soft X-ray emission is {\it dominated} by lines in a radio
galaxy comes from the analysis of 3C\,234 by Piconcelli \etal\ (2008).
However, as we show in Section \ref{3C234}, it is perfectly possible
to obtain good fits to the {\it XMM} data on this object with models
that are dominated by a power law at soft energies, while still
containing line emission. The general picture that we derive from
  detailed individual observations of this emission-line component, i.e., that
  it is present in some sources but not dominant, is consistent with
  the observed correlation
  between $L_{\rm Xu}$ and the [O{\sc iii}] emission-line luminosity
  in LERGs and NLRGs (Section \ref{emline}); the fact that this
  correlation is present could be interpreted as
  indicating that an emission-line component of the soft X-ray
  spectrum is important in the population as a whole, but the fact
  that it is weaker than the correlation with $L_5$
  (even in a matched sample: Table \ref{corsum}),
  while $L_5$ and $L_{\rm OIII}$ are not correlated at all, suggests
  that it does not dominate the X-ray emission. Thus we can
conclude that, while thermal/line emission at soft energies can be
energetically important and must be considered as a contaminant of any
non-thermal soft component, it remains most likely at present that all
radio galaxies require a power-law component that must be understood
in terms of models (ii) or (iii).

Distinguishing between models (ii) and (iii) with observations of
individual sources is more difficult. The principal argument in favour
of model (ii) is that partial covering/reflection models have been
successfully invoked in radio-quiet systems where a jet is not
present; therefore, it can be argued, invoking a jet-related X-ray
emission component is disfavoured by Occam's razor. The principal
arguments in favour of model (iii) are as follows:

\begin{itemize}
\item As shown above, in our previous work on the 3CRR {\it
  Chandra}/{\it XMM} dataset, and in many other papers (e.g. Worrall \&
  Birkinshaw 1994; Edge \& R\"ottgering 1995; Worrall 1997; Canosa
  \etal\ 1999; Hardcastle \& Worrall 1999; Balmaverde \etal\ 2006)
  there is a strong correlation between the nuclear radio emission and
  the unabsorbed X-ray component (which extends to the optical in the
  case of FRI LERGs: Hardcastle \& Worrall 2000). We know that the
  radio emission is beamed and originates in the jet, so it is very
  hard to escape from a model in which the X-ray does likewise. The
  unabsorbed component is not significantly correlated with {\it
  total} radio power (Fig.\ \ref{lxu-lr}) and in general very poorly correlated
  with other indicators of total AGN power where available (e.g. Fig.\
  \ref{li-lxu}) which is consistent with the idea that it is strongly
  related to beaming.
\item The unabsorbed component is not significantly correlated in
  NLRGs with the absorbed component (Section \ref{rxc} and
  Fig.\ \ref{lxu-lxa}): we might expect a significant correlation (albeit with
  scatter) in any model in which the two had the same origin, arguing
  against model (ii).
\item At least in low-power/low-excitation radio galaxies, a beamed
  component of X-ray emission is required for unification with BL Lac
  objects to operate, and the level of the observed component in radio galaxies is in agreement with
  widely accepted effective beaming speeds in unified models
  (Hardcastle \etal\ 2003). 
  Similarly, a jet-related component of the X-ray emission in
  core-dominated quasars is required to explain their X-ray
  properties, and we might expect this to be present in the X-ray
  spectra of the non-aligned counterparts of core-dominated quasars,
  the NLRGs and BLRGs/lobe-dominated quasars. FRII NLRGs lie on a radio/X-ray
  correlation indistinguishable from LERGs and FRIs (Fig.
  \ref{lxu-lc}) and so it seems very natural to suggest that we are
  observing the same jet-related component, with the same role in
  unified models, in both cases.
\end{itemize}

It seems to us very hard to construct a version of models (i) or (ii)
that naturally predicts the strong correlation of $L_{\rm Xu}$ with
beamed quantities, the generally poor correlation with unbeamed quantities,
  and the agreement between the properties of LERG and NLRG, that
we see in large samples such as the present one; nor is it obvious how
such models would be compatible with unification of either low-power
or high-power radio sources. Model (iii), in our view, remains clearly
the strongest. Any challenge to model (iii) {\it must} address these
sample-wide results rather than showing that individual objects may be
modelled in some other way.

\subsection{Interpreting the `fundamental plane' of black-hole activity}
\label{sec:fp}

Recently two independent groups (Merloni, Heinz \& di Matteo 2003;
Falcke, K\"ording \& Markoff 2004) have presented evidence for a link
between the accretion and jet properties of stellar-mass
(XRB/microquasar) and supermassive (AGN) black hole systems, in the
form of a non-linear relationship between black hole mass, X-ray and
radio core luminosity, defining what has become known as the
`fundamental plane' of black-hole activity. The ability to extrapolate
from the properties of Galactic black-hole systems, where exquisitely
detailed timing studies are possible, to those of powerful AGN would
be of great value in our understanding of both classes of object.

However, in deciding where radio-loud AGN should lie on the
fundamental plane plot, great care must be taken to compare like with
like. We know from the work described above that the nuclear X-ray
luminosity of radio-loud AGN has two components, one related to the
jet, the other (present only in some systems) related to a luminous
accretion disc. Depending on the type of object being considered, one
or the other may dominate, or both may contribute more or less equally
to, the 2-10 keV X-ray luminosity used by Merloni \etal\ and Falcke
\etal\ to establish the fundamental plane relationship (a point also
made by K\"ording, Falcke \& Corbel 2006). In fact, the two groups
took rather different approaches to their selection of AGN: Falcke
\etal\ used AGN where they believed the X-ray emission to be
jet-dominated (including FRI radio galaxies and BL Lac objects) while
Merloni \etal\ explicitly excluded jet-dominated systems like BL Lacs, and instead
considered a rather mixed set of AGN, including Seyfert 1 and 2s and
some powerful FRIIs like Cygnus A where we might expect the
accretion-related X-ray component to dominate at 2-10 keV, as well as
some FRIs where the present work shows that the jet-related component
is the only one present. This choice is the key to the interpretation
of the `fundamental plane' relationship: is it telling us about the
nature of jets (as Falcke \etal\ 2004 would suggest) or about the
relationship between accretion power and jet production (as considered
by Merloni \etal\ 2003)?

Our data on powerful radio galaxies can help to resolve these
questions. NLRGs in particular offer us a very useful tool to
distinguish between the two possible interpretations, since they
appear to have {\it both} powerful jets and radiatively efficient
accretion (see H07), and the X-ray data allow us to separate the two
contributions to $L_{\rm X}$ (see Section \ref{jetrelated}).
Unfortunately, what we lack for these objects is good black-hole mass
determinations. For want of anything better, we adopt the same
approach in this paper as we did in H07, and use the Marconi \& Hunt
(2003) relationship between K-band absolute magnitude and $M_{\rm
BH}$, taking the K-band luminosities of 3CRR objects either from
Willott \etal\ (2003) or from 2MASS. This allows us to plot LERGs and
NLRGs on the `fundamental plane' relation, albeit with large
uncertainties given the scatter in the $L_K$ -- $M_{\rm BH}$
relationship. (BLRGs and quasars are excluded since their black-hole
masses cannot be estimated in this way.)

\begin{figure}
\epsfxsize 8.5cm
\epsfbox{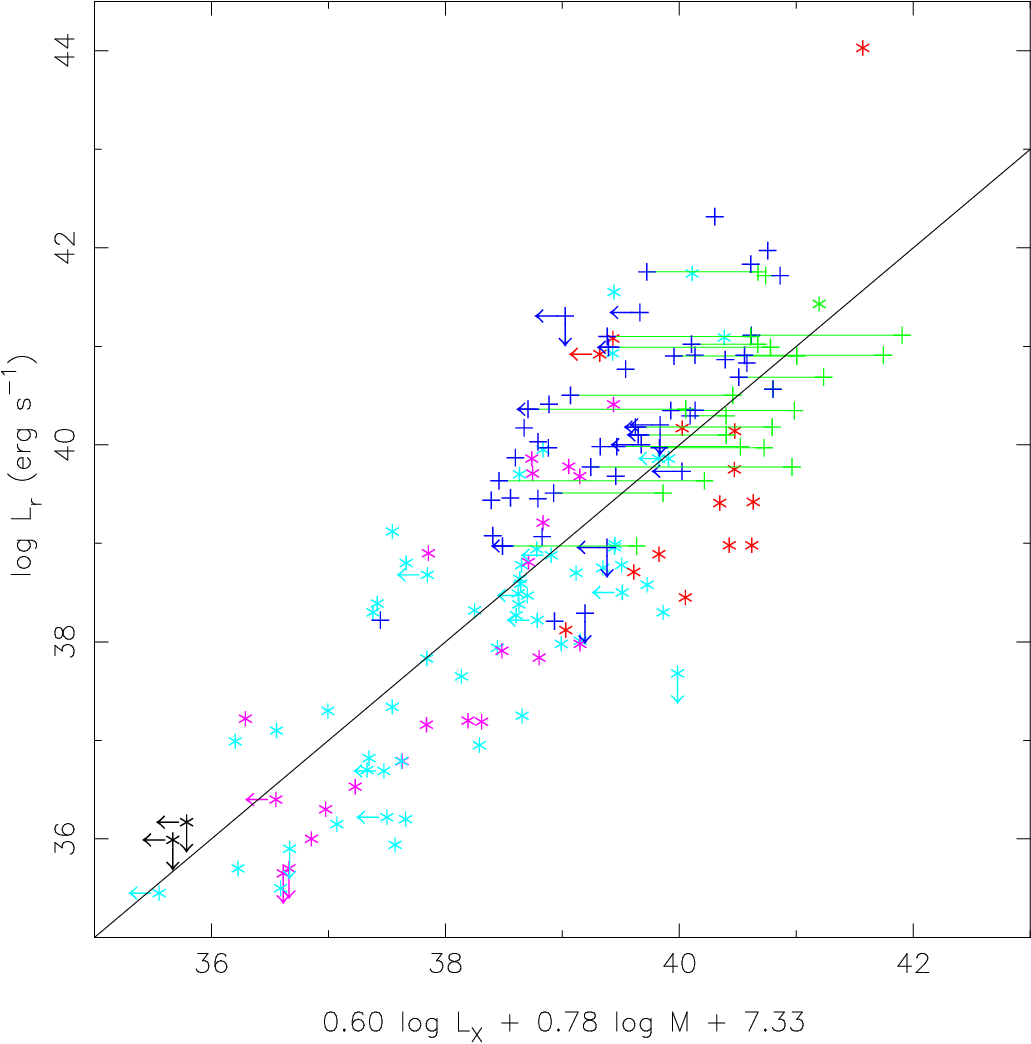}
\caption{A projection of the `fundamental plane' of black-hole
  activity for AGN including 3CRR radio galaxies, according to the
  relation of Merloni \etal\ (2003). Crosses mark data points from the
  present work, stars points from Merloni \etal . Coloured stars
  indicate different source types in the sample of Merloni \etal\ ;
  red points are quasars, cyan Seyferts, magenta LLAGN -- including
  some radio galaxies --, black normal galaxies, and the one green
  star the FRII NLRG Cygnus A. Green crosses indicate
  accretion-related X-ray luminosities and blue crosses jet-related
  luminosities: for sources where both are measured, the two are
  linked by a green line. The radio luminosity plotted in all cases is
  the nuclear 5-GHz radio luminosity: $M$ is the black-hole mass in
  solar masses and $L_{\rm X}$ is as discussed in the text.}
\label{fp}
\end{figure}

Fig.\ \ref{fp} shows the 3CRR sources with appropriate data plotted on
a plane equivalent to that of Merloni \etal\ (2003)'s fig.\ 5. For the
29 objects (mostly LERGs) where we have only a single unabsorbed
luminosity measurement, we take this to be $L_{\rm X}$ and plot a
single point. For the 22 objects (mostly NLRGs) with an estimated
$M_{\rm BH}$ and a measured `accretion-related' X-ray luminosity,
$L_{\rm Xa}$, we compute $L_{\rm X}$ using both quantities and plot
the two results joined by a horizontal line. (For clarity, and for
consistency with other treatments of the fundamental plane, no error
bars are plotted.) It is clear that the
choice of $L_{\rm X}$ value makes a significant difference to the
answers obtained. The best-fitting relation of Merloni \etal\ lies
significantly closer to the points denoting the accretion-related
luminosities than it does to the jet-related luminosities; the latter,
almost without exception, lie above and to the left of the regression
line. Given that the AGN in the analysis of Merloni \etal\ are
dominated by Seyferts and quasars, which we would expect to show
mostly accretion-related emission, this is not at all surprising. The
Merloni \etal\ (2003) version of the fundamental plane involves an
accretion and not a jet origin for the X-rays.

\begin{figure}
\epsfxsize 8.5cm
\epsfbox{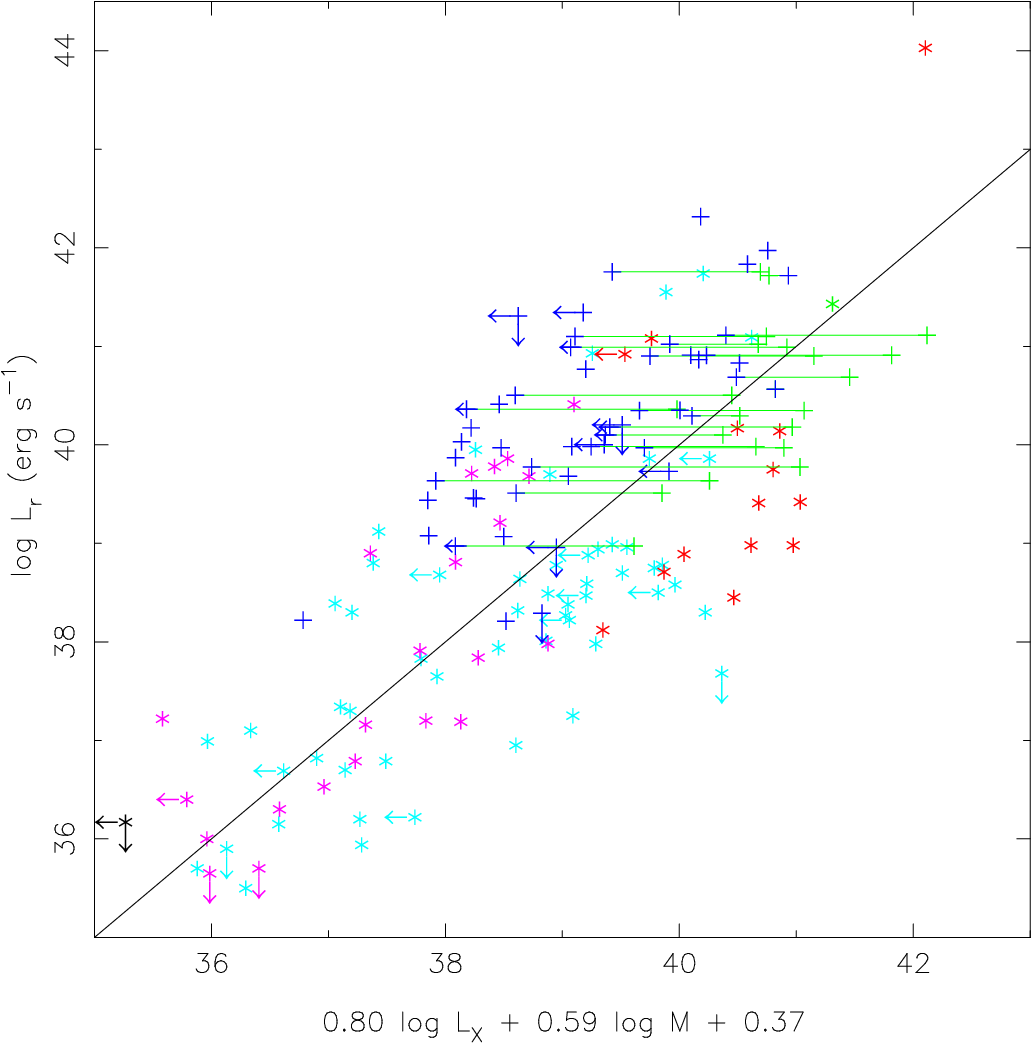}
\caption{A projection of the `fundamental plane' of black-hole activity for AGN
  including 3CRR radio galaxies, according to the relation of K\"ording
  \etal\ (2006). Symbols as in Fig.\ \ref{fp}.}
\label{fp2}
\end{figure}

What is slightly more surprising is that the version of the
fundamental plane relationship fitted to low-hard state XRB, LLAGN and
FRI radio galaxies by K\"ording \etal\ (2006), which was explicitly
intended to provide information about the nature of jets, fares no
better at predicting the level of the jet-related X-ray emission of
the LERGs (including most 3CRR FRIs) and NLRGs in our current sample
(Fig. \ref{fp2}). K\"ording \etal\ used X-ray luminosities
extrapolated from the optical rather than directly measured X-ray
luminosities for the FRIs to which they fit their fundamental plane
relations, on the basis that the X-ray emission from FRI nuclei may
partly or wholly be inverse-Compton rather than synchrotron in origin.
However, it is clear that this extrapolation from the optical must
give X-ray flux densities which substantially exceed the true ones.

We conclude that systems where the jet and accretion-related emission
can be separated give us a clear indication that the fundamental plane
relationship as currently derived is a consequence of a relationship
between accretion power and jet power, rather than arising purely from
the nature of relativistic jets. This has the consequence that
jet-dominated objects, such as FRI radio galaxies and BL Lac objects,
should {\it not} be included in fits that attempt to derive parameters
of the fundamental plane, and care should also be taken in using
objects, such as NLRGs, that may contain a substantial contribution
from the jet in the X-ray spectrum. The `fundamental plane conspiracy'
identified by K\"ording \etal\ (2006) -- in which objects with clearly
different X-ray emission processes tend to lie close to the
fundamental plane -- is more easily understood if the fundamental
plane relation originates in accretion. Even for jet-dominated
objects, the X-ray emission represents a fraction of the total
(invisible) accretion power which, we infer, is not vastly different
from the fraction represented by accretion-related X-ray emission
(Fig. \ref{lxu-lxa}).

\section{Summary and outlook}

We have analysed new and archival X-ray and infra-red data to produce
the most complete database yet of nuclear luminosities for the 3CRR
(LRL) sample of radio-loud active galaxies. Our principal results may
be summarized as follows:

\begin{enumerate}
\item As in all our (and others') previous work on the subject, we
  have found a good correlation between the unabsorbed component of
  X-ray luminosity, present in essentially every radio source, and the
  5-GHz core radio luminosity (Section \ref{rxc}). We argue in Section
  \ref{jetrelated} that the evidence that at least some, and in many
  cases all, of this soft component of radio-source X-ray spectra
  originates in the jet is now very hard to evade, although we do
    not rule out an additional significant role for unresolved X-ray
    emission-line material (either photoionized by the AGN or
    shock-ionized by the jet) as seen in detailed observations of some
    powerful FRIIs.
\item Using the new infra-red data (Section \ref{xray-ir}), we have
  shown that it is very unlikely that low-excitation FRII radio
  galaxies can be heavily obscured (e.g. Compton-thick) normal AGN
  (Section \ref{lergs}). This closes a loophole in the argument of H06
  regarding these objects: they must really have active nuclei that
  are either radiatively efficient, but intrinsically much fainter
  than their radio luminosity would imply, or radiatively inefficient. For
  low-power FRIs we cannot make such a definite statement because of
  contamination of the {\it Spitzer} spectra by extended emission,
  although individual sources have been discussed by others (e.g.
  Whysong \& Antonucci 2004) with the same conclusions.
\item On the other hand, we have shown (Section \ref{xray-ir}) that
  the heavily absorbed nuclear X-ray component, present in the vast
  majority of NLRG FRIIs in our sample, is very well correlated with
  the 15-$\mu$m mid-IR luminosity from {\it Spitzer}. One case where
  there is no absorbed nuclear X-ray emission but strong IR emission
  is to our knowledge the best candidate so far for a Compton-thick
  NLRG (Section \ref{comptonthick}). The fraction of Compton-thick
  NLRGs is clearly significantly lower than that of Compton-thick local
  Seyferts, while the radio galaxies' column density distribution is
  inconsistent with that of X-ray-selected type 2 quasars in the sense
  of having many objects with relatively {\it high} column densities.
  The reason for the differences in these column density distributions
  remains unclear.
\item We argue that the X-ray and IR data taken together are not
  consistent with the simplest possible models of the IR emission from a
  receding torus (Section \ref{lsummary}).
\item Our discussion of the positioning of radio galaxies on the
  fundamental plane relationships of Merloni \etal\ (2003) and Falcke
  \etal\ (2004), presented in Section \ref{sec:fp}, suggests that this
  relationship is primarily related to accretion rather than to
  properties of the jets. Accordingly, care should be taken to use
  only accretion-related quantities when constructing such plots.
\end{enumerate}

Although the 3CRR sample continues to suffer from incomplete
observations and bias at all non-radio wavebands, we have shown in
this paper that there has been a huge advance in the availability of
X-ray data of sufficient quality to allow nuclear spectroscopy over
the past few years, allowing significant advances in our ability to
draw scientific conclusions. Considering the whole 3CRR sample
(including $z>1.0$ objects not discussed in the present paper) the
fraction of objects observed already significantly exceeds the
fraction observed by {\it ROSAT} at the end of its lifetime
(Hardcastle \& Worrall 1999). There is a realistic prospect of being
able to carry out unbiased, complete studies of the nuclei of these
objects by the end of the {\it Chandra} and {\it XMM} missions; we and
others have recently been awarded observations which will go a long
way towards providing the necessary X-ray data, particularly at
$z<0.5$ and $z>1.0$. Unfortunately, the short lifetime of {\it
Spitzer} prevents the same statement being made for the crucial mid-IR
waveband, but far-IR observations with {\it Herschel} will provide
important further constraints on models.

\section*{Acknowledgements}

MJH thanks the Royal Society for a research fellowship. DAE
acknowledges partial financial support from the NASA {\it XMM-Newton}
GO program, through grants NNX07AQ52G and NNX08AX30G.
We are very grateful to Vincenzo Mainieri for providing data on the
COSMOS sample of X-ray-selected type 2 quasars in advance of
publication, to Elmar K\"ording for discussion of the fundamental
plane relation, to Matt Jarvis for discussion of the expectations from
receding-torus models, and to Chiranjib Konar for allowing us to use
{\it XMM} data on 3C\,457 obtained for another purpose in this paper.
We thank an anonymous referee for helpful comments on the paper.
This work is partly based on observations obtained with {\it
XMM-Newton}, an ESA science mission with instruments and contributions
directly funded by ESA Member States and the USA (NASA).

\appendix
\section{Notes on individual sources}

Here we discuss any interesting features of previously unreported
X-ray spectra, and give references to previous analyses.

\subsection{3C6.1}

Tabulated {\it Chandra} observing time here combines two observations
taken on 2002 Aug 26 and 2002 Oct 15. There is no evidence for
variation in the nuclear X-ray flux over this time period.

\subsection{3C\,48}

A detailed analysis of the {\it Chandra} data for this quasar is given
by Worrall \etal\ (2004). For simplicity here we fit a single
power-law model which gives an adequate estimate of the total X-ray
luminosity of the source.

\subsection{3C\,76.1}

Details of the {\it XMM} analysis for this FRI radio galaxy are given by Croston
\etal\ (2008).

\subsection{3C\,132}

The short {\it Chandra} exposure of this source required us to bin the
spectrum by a factor of 10, rather than our usual 20, in order to
obtain enough bins for spectral fitting.

\subsection{3C\,171}

Full details of the long {\it Chandra} observation of this object will
be presented in a future paper (Hardcastle, Harris \& Massaro in
prep.).

\subsection{3C\,220.1}

The {\it Chandra} data for this object were previously analysed by Worrall
\etal\ (2001).

\subsection{3C\,228}

Tabulated {\it Chandra} observing time here combines two observations
taken on 2001 Apr 23 and 2001 Jun 03. There is no evidence for
variation in the nuclear X-ray flux over this time period.

The evidence for an absorbed nuclear X-ray component in this source is
marginal. Adding a second component to the fit clearly improves it,
and the normalization of the second component is not consistent with
zero at the 90 per cent confidence level: however, the column density
of the intrinsic absorber is not constrained (Table \ref{nhvals}).
Accordingly there is a large uncertainty on the luminosity of the
absorbed component.

\subsection{3C\,234}
\label{3C234}

\begin{figure}
\epsfxsize 8.5cm
\epsfbox{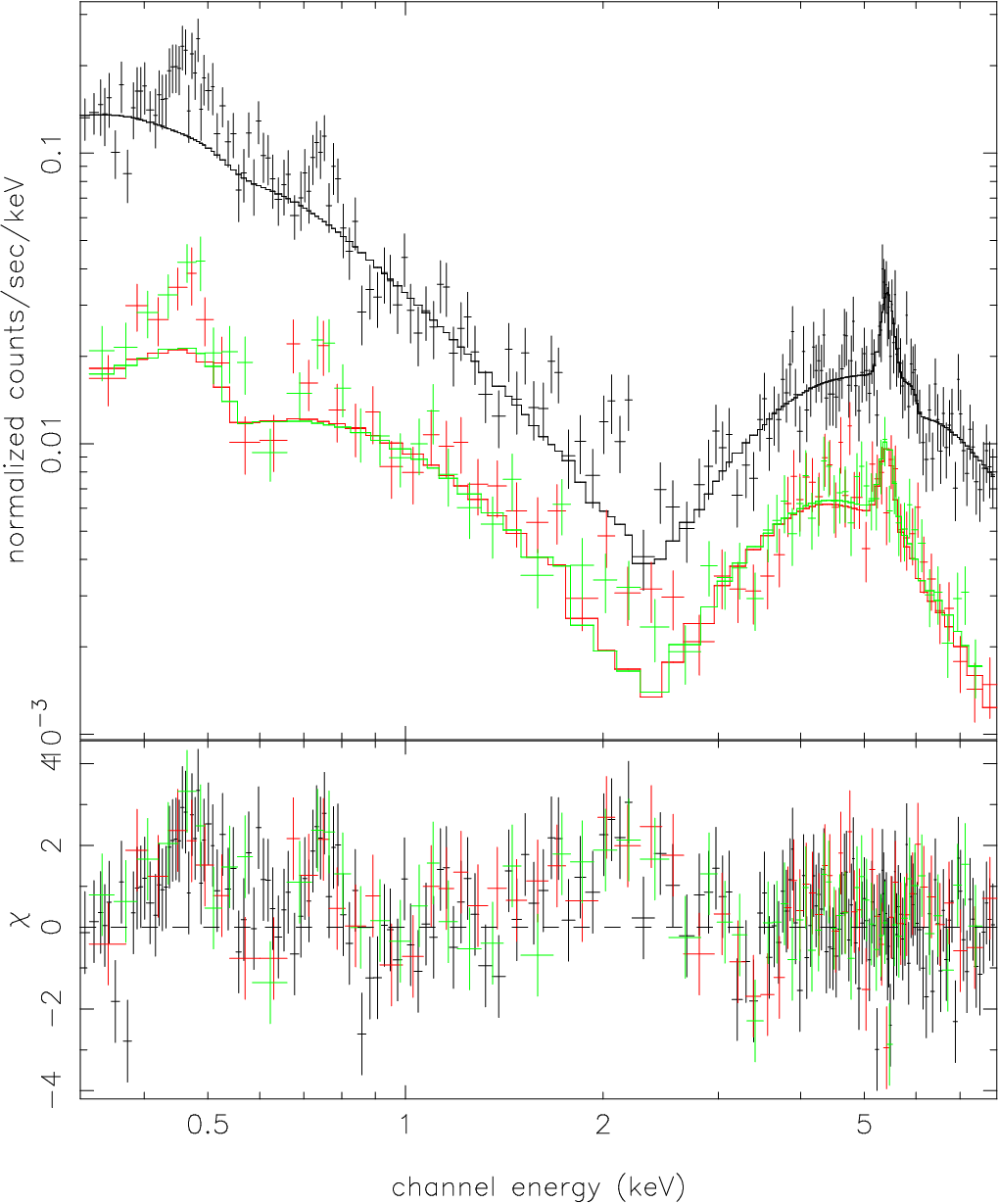}
\epsfxsize 8.5cm
\epsfbox{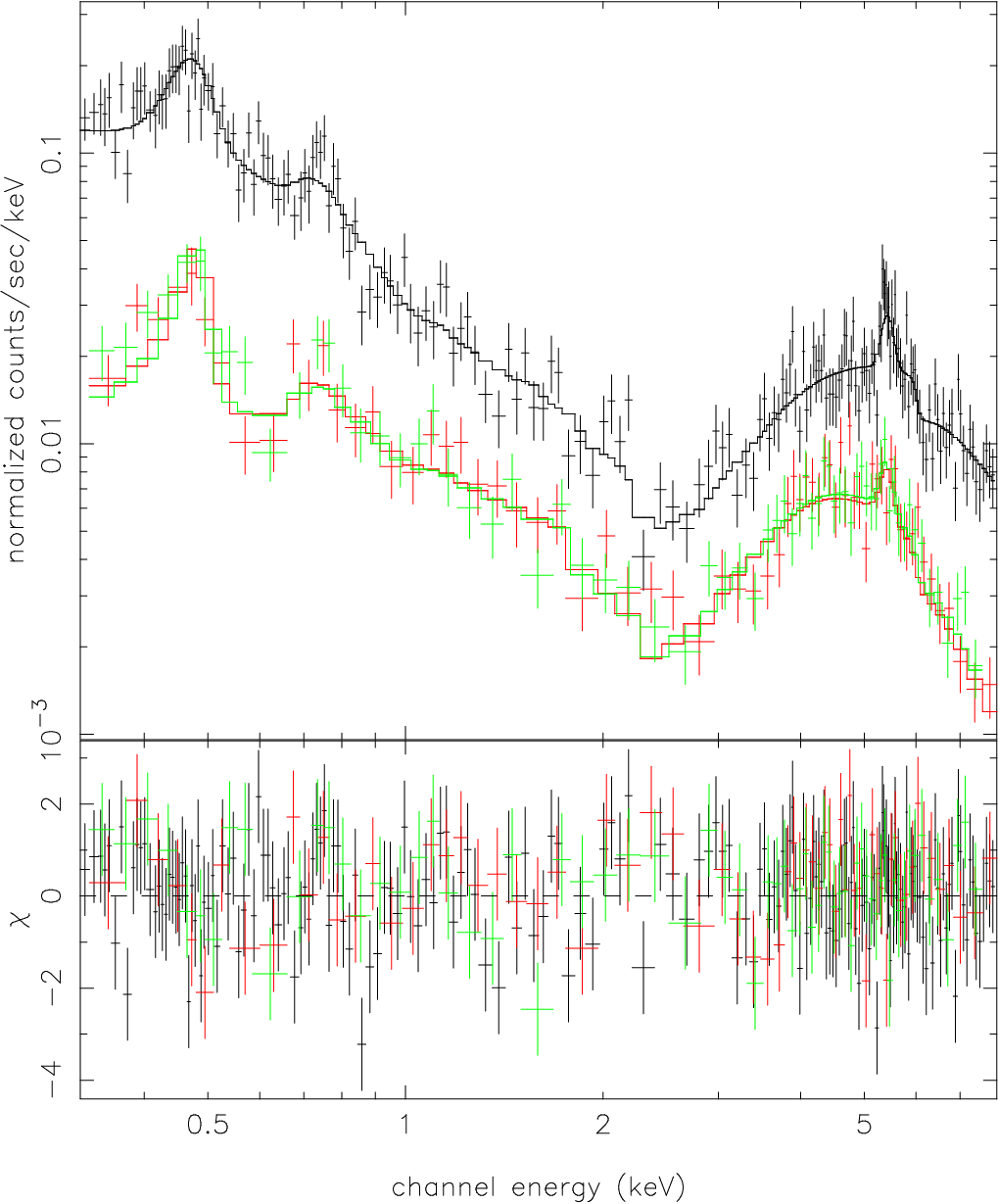}
\caption{The {\it XMM} spectrum of 3C\,234 with two fitted models:
  top, our standard double power-law model (with a Gaussian to model
  the prominent iron line); bottom, the same model with the addition
  of two MEKAL components.}
\label{234-spectrum}
\end{figure}

3C\,234 is a NLRG that famously shows polarized broad emission lines
in the optical (Antonucci 1982; Antonucci \& Barvainis 1990); thus it
is one of only a few objects to show unambiguous {\it optical}
evidence for the presence of the `hidden quasar' predicted by unified
models. The long {\it XMM-Newton} observation of 3C\,234 has been
analysed by Piconcelli \etal\ (2008), who argue that there are strong
soft emission lines in the spectrum and therefore that the soft excess
in this object cannot be jet-related (see Section \ref{jetrelated}).
We extracted spectra from the data in the standard way, fitting as
usual in the well-calibrated 0.3-8.0 keV band, and confirm that our
standard model (one power law with Galactic absorption only, one power
law with Galactic and intrinsic absorption, and a Gaussian
representing the Fe K$\alpha$ feature) does not give an acceptable fit
to the data: there are significant residuals at soft energies (Fig.\
\ref{234-spectrum}). However, we find an acceptable fit ($\chi^2 =
362$ for 320 degrees of freedom) if we add a single MEKAL model to the
data, and a good fit ($\chi^2 = 332$ for 318 degrees of freedom) with
two MEKAL components, having $kT = 0.61 \pm 0.07$ and $0.15 \pm 0.01$
keV (with fixed abundances of 0.35 solar). The addition of the MEKAL
models not only removes the soft residuals but also reduces those at
the crossover between the two power laws at $\sim 2$ keV. Crucially,
in these models, the power-law component dominates the soft X-ray
emission from 3C\,234, providing 60 per cent of the X-ray flux between
0.5 and 2.0 keV in the two-MEKAL model, while there is no evidence for
soft residuals in the fit (Fig.\ \ref{234-spectrum}). In other words,
we find that the power-law component does dominate at such energies, and
so the majority of the soft X-ray emission in 3C\,234 can be
associated with the jet. Indeed, we note that the temperature of one
of the MEKAL models fitted by Piconcelli \etal\ is 8 keV, which
effectively mimics a power law. We therefore cannot agree with their
rejection of jet-related models for the soft excess in 3C\,234 or in
radio-loud AGN in general, as discussed in more detail in Section
\ref{jetrelated}.

It is still of interest to ask whether these MEKAL models are
physically realistic and, if so, what their physical origin is. As our
25-arcsec extraction circle has a radius of 70 kpc at the redshift of
3C\,234, one obvious possibility is thermal emission from the
large-scale environment. The bolometric X-ray luminosities of the two
MEKAL models are $6 \times 10^{43}$ erg s$^{-1}$ (soft) and $4 \times
10^{42}$ erg s$^{-1}$ (hard) though of course most of the luminosity
of the softer MEKAL ($kT = 0.15$ keV) is emitted outside the {\it XMM}
band and so is not directly observed. The hotter of the two MEKAL
models is therefore entirely consistent in temperature and luminosity
terms with being thermal emission from the IGM of a poor group
environment (e.g. Helsdon \& Ponman 2000) as has been observed for
other NLRGs (e.g. Hardcastle \etal\ 2007b). However, the very luminous, soft
MEKAL component certainly cannot have this origin: it may well arise from
AGN-photoionized gas as suggested by Piconcelli et al. A deep {\it
Chandra} observation of this system would have the capability to
resolve some or all of the components seen here and so show which of
them are truly associated with the active nucleus. Alternatively, it
may be that the spectrum of 3C\,234 is similar to that of 3C\,445
(Sambruna \etal\ 2007) and would be better modelled with a combination
of numerous emission lines arising from scattering from warm gas.

For consistency with our earlier fitting the results presented in the
main body of the paper are those derived from the model with only one
additional MEKAL component, which gives an acceptable fit to the data.
The derived parameters for the two power-law components are very
little affected by the addition of the second component.

\subsection{3C\,244.1}

The {\it XMM-Newton} data for this source were unfortunately very badly
affected by flaring, leaving us with only a short exposure time even
after increasing our usual flare filtering threshold. The
background-subtracted data were binned by a factor 10 to allow model
fitting. However, we obtain very similar fits (including a very
similar upper limit on the heavily absorbed component) if we fit to
data which has not been flare-filtered at all, so we are confident
that the unusually strong limit on this component is not an artefact
of the low count rate in the filtered data. Neither a heavily absorbed
power-law component nor an Fe K$\alpha$ line are seen in the X-ray data.

\subsection{3C\,300}

The {\it XMM} observations were taken in three roughly equal parts, on
28 Dec 2007, 30 Dec 2007, and 01 Jan 2008. There is no evidence for
nuclear X-ray variability in this period. The data were affected by
flaring but as they were originally quite deep the constraints on any
absorbed nuclear component are good.

\subsection{3C\,325}

Tabulated {\it Chandra} observing time here combines two observations
taken on 2005 Apr 14 and 2002 Apr 17. There is no evidence for
variation in the nuclear X-ray flux over this time period.

\subsection{3C\,349}

Two {\it XMM} observations were carried out for this source, on 2007
Aug 07 and 2007 Oct 03. There was no evidence for variability in the
data, and so the results tabulated come from a joint fit to the two
datasets. The best-fitting column density for the obscured component
is unusually low for a NLRG.

\subsection{3C\,381}
\label{381}

3C\,381 is classed as a BLRG in LRL, but as pointed out by Hardcastle
\etal\ (1997) the evidence for this classification is very tenuous.
Jackson \& Rawlings (1997) classify it as a NLRG, based on the
spectroscopy of Saunders \etal\ (1989), whose published spectrum shows
no evidence for broad H$\alpha$. Our X-ray spectrum gives a
well-constrained, high absorbing column. We infer that the
classification of this source as a BLRG by LRL is likely to have been
incorrect. For consistency, we retain its BLRG classification in the
plots and tables in the present paper.

\subsection{3C\,433}

The {\it Chandra} data for this object have been presented by Miller
\& Brandt (2009).

\subsection{3C\,442A}

X-ray data for 3C\,442A are as described by Hardcastle \etal\ (2007).
This source is known to be variable in the X-ray: we use the X-ray
data from the 2005 Jul 27 observation. As discussed by Hardcastle
\etal\ (2007), intrinsic absorption may explain the flat photon index
we obtain, but for consistency we use the Galactic value here.

\subsection{3C\,457}

Details of the deep {\it XMM} observation of this object will be
presented by Konar \etal\ (2009). The very flat photon index for the
soft component probably indicates some excess absorption over the
Galactic value used here; this issue will be discussed in more detail
by Konar \etal

\setcounter{table}{3}
\renewcommand\thetable{\arabic{table}}
\onecolumn
\begin{longtable}{llrrrrrrrrrrrrr}
\caption{Luminosities for sources in the $z<1.0$ 3CRR sample}\\
\hline
Source&Type&$z$&$L_{178}$&$L_5$&$L_{\rm Xu}$&&&$L_{\rm Xa}$&&&$L_{\rm
  IR}$&&$L_{\rm [OIII]}$&$L_{\rm [OII]}$\\
\hline\endhead
4C\,12.03 &E&0.156&42.10&40.00&$<$41.91&--&--&$<$43.02&--&--&--&--&40.97&--\\
3C\,6.1   &N&0.8404&43.87&41.61&44.92&44.89&44.94&$<$44.17&--&--&45.100&0.010&--&42.15\\
3C\,16    &E&0.405&43.09&39.73&$<$42.74&--&--&$<$43.69&--&--&--&--&--&41.81\\
3C\,19    &N&0.482&43.25&40.14&--&--&--&--&--&--&--&--&--&--\\
3C\,20    &N&0.174&42.82&39.97&42.56&42.45&42.64&44.05&43.94&44.44&44.293&0.004&41.21&40.73\\
3C\,22    &B&0.938&43.96&41.95&--&--&--&--&--&--&45.900&0.010&--&43.16\\
3C\,28    &E&0.1952&42.54&$<$38.96&$<$41.36&--&--&$<$42.27&--&--&$<$42.740&--&40.96&41.81\\
3C\,31    &E&0.0167&40.31&39.45&40.67&40.52&40.79&$<$40.72&--&--&42.341&0.002&39.47&--\\
3C\,33    &N&0.0595&41.95&39.98&42.04&41.98&42.09&43.80&43.73&43.87&44.080&0.012&42.19&41.44\\
3C\,33.1  &B&0.181&42.34&40.68&42.43&42.14&42.59&44.38&44.26&44.66&44.878&0.002&42.30&--\\
3C\,34    &N&0.689&43.70&40.80&--&--&--&--&--&--&--&--&--&43.61\\
3C\,35    &E&0.0677&41.35&39.77&--&--&--&--&--&--&--&--&40.03&--\\
3C\,41    &N&0.795&43.66&40.72&--&--&--&--&--&--&--&--&--&42.70\\
3C\,42    &N&0.395&43.07&40.67&--&--&--&--&--&--&--&--&42.04&41.89\\
3C\,46    &N&0.4373&43.16&40.75&--&--&--&--&--&--&--&--&42.80&42.22\\
3C\,47    &Q&0.425&43.52&42.23&45.01&44.97&45.04&45.05&44.77&45.21&45.805&0.004&43.28&42.63\\
3C\,48    &Q&0.367&43.64&43.18&45.00&45.00&45.01&45.00&45.00&45.01&46.145&0.002&43.12&42.25\\
3C\,49    &N&0.6207&43.44&41.54&--&--&--&--&--&--&--&--&--&--\\
3C\,55    &N&0.735&44.02&41.57&--&--&--&--&--&--&45.820&0.013&--&42.34\\
3C\,61.1  &N&0.186&42.76&40.00&41.92&41.58&42.10&43.93&43.74&44.10&43.700&0.030&42.49&41.44\\
3C\,66B   &E&0.0215&40.69&39.97&41.10&41.03&41.15&$<$40.64&--&--&42.008&0.004&40.06&39.87\\
3C\,67    &B&0.3102&42.73&40.82&--&--&--&--&--&--&--&--&42.83&42.26\\
3C\,76.1  &E&0.0324&40.75&39.07&41.28&41.05&41.47&$<$41.26&--&--&41.966&0.017&$<$39.85&--\\
3C\,79    &N&0.2559&43.07&40.90&42.42&42.34&42.49&44.18&43.65&44.75&45.326&0.004&42.86&42.21\\
3C\,83.1B &E&0.0255&40.88&39.46&41.13&41.03&41.21&$<$40.42&--&--&42.205&0.004&--&--\\
3C\,84    &N&0.0177&40.92&42.32&42.91&42.88&42.94&$<$41.77&--&--&44.217&--&41.62&41.09\\
3C\,98    &N&0.0306&41.29&38.97&$<$40.82&--&--&42.73&42.44&42.83&--&--&41.02&40.24\\
3C\,109   &B&0.3056&43.08&42.48&45.23&45.18&45.29&45.23&44.60&45.29&45.975&0.001&43.32&42.09\\
4C\,14.11 &E&0.206&42.41&41.18&43.01&42.94&43.07&$<$42.78&--&--&--&--&41.24&--\\
3C\,123   &E&0.2177&43.68&41.76&42.00&41.05&42.27&43.58&43.36&43.68&43.810&0.067&42.00&--\\
3C\,132   &N&0.214&42.52&40.10&$<$41.99&--&--&43.25&43.04&43.40&--&--&--&--\\
3C\,138   &Q&0.759&43.92&42.85&--&--&--&--&--&--&45.800&0.010&43.46&42.57\\
3C\,147   &Q&0.5450&44.04&43.98&--&--&--&--&--&--&45.500&0.010&43.79&43.45\\
3C\,153   &N&0.2769&42.82&$<$40.20&$<$41.99&--&--&$<$42.89&--&--&43.590&0.097&41.64&42.49\\
3C\,171   &N&0.2384&42.80&40.18&41.86&41.69&41.98&44.08&43.96&44.18&--&--&42.89&42.45\\
3C\,172   &N&0.5191&43.46&40.17&--&--&--&--&--&--&44.310&0.062&--&42.77\\
3C\,173.1 &E&0.292&42.90&40.89&41.55&41.34&41.69&$<$43.13&--&--&43.400&0.079&40.85&--\\
3C\,175   &Q&0.768&43.96&42.26&--&--&--&--&--&--&45.700&0.010&43.10&42.77\\
3C\,175.1 &N&0.920&43.95&42.09&--&--&--&--&--&--&--&--&--&42.67\\
3C\,184   &N&0.994&44.08&$<$40.41&43.48&42.48&43.95&44.76&44.57&44.90&45.300&0.010&--&42.89\\
3C\,184.1 &N&0.1187&41.95&39.99&41.73&41.45&41.89&43.91&43.70&44.22&--&--&42.23&41.48\\
DA240   &E&0.0356&41.08&40.17&40.81&40.73&40.87&$<$41.05&--&--&--&--&39.76&40.04\\
3C\,192   &N&0.0598&41.54&39.51&41.37&41.18&41.49&42.93&42.74&43.12&42.710&0.028&41.36&41.31\\
3C\,196   &Q&0.871&44.63&41.84&--&--&--&--&--&--&46.000&0.010&--&--\\
3C\,200   &E&0.458&43.21&41.97&43.58&43.52&43.64&$<$43.78&--&--&44.100&0.010&--&--\\
4C\,14.27 &N&0.3920&43.05&$<$39.68&--&--&--&--&--&--&--&--&--&--\\
3C\,207   &Q&0.684&43.71&43.49&45.14&45.06&45.19&45.14&45.06&45.19&45.500&0.010&43.05&$<$42.15\\
3C\,215   &Q&0.411&43.13&41.54&44.84&44.81&44.87&44.84&44.46&44.87&--&--&42.59&42.22\\
3C\,217   &N&0.8975&43.88&$<$40.80&--&--&--&--&--&--&--&--&--&43.29\\
3C\,216   &Q&0.668&43.84&43.78&--&--&--&--&--&--&45.700&0.010&$<$42.46&42.43\\
3C\,219   &B&0.1744&42.82&41.26&43.99&43.94&44.04&43.99&43.94&44.04&44.210&0.016&41.77&41.27\\
3C\,220.1 &N&0.61&43.66&42.08&44.50&44.48&44.52&$<$44.04&--&--&44.700&0.010&42.79&42.46\\
3C\,220.3 &N&0.685&43.74&$<$40.02&--&--&--&--&--&--&45.100&0.010&--&--\\
3C\,223   &N&0.1368&42.14&40.29&43.16&43.12&43.19&43.67&43.43&44.27&--&--&42.18&41.71\\
3C\,225B  &N&0.58&43.74&40.68&--&--&--&--&--&--&44.483&0.129&--&42.62\\
3C\,226   &N&0.82&43.94&41.82&--&--&--&--&--&--&46.261&0.006&--&42.74\\
4C\,73.08 &N&0.0581&41.35&39.62&41.80&41.68&41.90&43.46&43.05&43.79&--&--&40.94&40.57\\
3C\,228   &N&0.5524&43.71&41.72&43.86&42.81&43.91&43.65&42.81&43.94&44.574&0.087&--&42.15\\
3C\,234   &N&0.1848&42.76&41.56&42.89&42.87&42.91&44.36&44.26&44.60&45.590&0.005&43.13&42.12\\
3C\,236   &E&0.0989&41.82&40.98&--&--&--&--&--&--&--&--&40.90&41.17\\
4C\,74.16 &?&0.81&43.82&41.11&--&--&--&--&--&--&--&--&--&--\\
3C\,244.1 &N&0.428&43.39&40.66&43.25&43.10&43.35&$<$42.92&--&--&45.130&0.009&43.03&--\\
3C\,247   &N&0.7489&43.63&41.41&--&--&--&--&--&--&--&--&--&43.01\\
3C\,249.1 &Q&0.311&42.79&41.93&44.72&44.57&44.77&44.74&44.43&45.04&45.493&0.001&43.38&--\\
3C\,254   &Q&0.734&43.96&42.13&45.32&45.25&45.40&45.32&45.25&45.40&45.600&0.010&43.71&43.13\\
3C\,263   &Q&0.652&43.69&42.94&45.18&45.12&45.24&45.18&45.12&45.24&45.800&0.010&43.71&42.90\\
3C\,263.1 &N&0.824&44.02&41.45&--&--&--&--&--&--&44.980&0.016&--&42.97\\
3C\,264   &E&0.0208&40.69&39.98&41.87&41.86&41.89&$<$40.91&--&--&42.315&0.003&39.16&40.10\\
3C\,265   &N&0.8108&44.06&41.40&43.45&43.33&43.54&44.49&44.28&44.63&45.860&0.010&43.80&43.85\\
3C\,268.1 &N&0.9731&44.19&41.39&--&--&--&--&--&--&45.300&0.010&--&42.27\\
3C\,268.3 &B&0.371&42.92&40.24&--&--&--&--&--&--&--&--&42.49&--\\
3C\,272.1 &E&0.0029&38.84&38.22&39.35&39.24&39.46&$<$39.49&--&--&40.964&0.001&37.98&--\\
A1552   &E&0.0837&41.59&40.34&--&--&--&--&--&--&--&--&--&--\\
3C\,274   &N&0.0041&40.88&39.87&40.59&40.56&40.61&$<$39.56&--&--&41.506&0.003&38.95&--\\
3C\,274.1 &N&0.422&43.29&40.83&43.27&43.21&43.33&$<$43.56&--&--&$<$43.910&--&41.36&--\\
3C\,275.1 &Q&0.557&43.64&42.72&44.52&44.51&44.54&44.52&44.51&44.54&45.100&0.010&--&42.67\\
3C\,277.2 &N&0.766&43.81&40.57&43.67&43.61&43.72&$<$43.81&--&--&--&--&--&43.21\\
3C\,280   &N&0.996&44.32&41.11&42.85&42.55&43.03&45.00&44.81&45.13&45.800&0.010&--&43.68\\
3C\,284   &N&0.2394&42.57&40.35&42.22&42.19&42.26&43.98&42.80&44.63&--&--&41.60&--\\
3C\,285   &N&0.0794&41.53&39.64&40.40&40.11&40.57&43.33&43.22&43.43&--&--&40.56&40.46\\
3C\,286   &Q&0.849&44.03&41.85&--&--&--&--&--&--&45.600&0.010&--&42.69\\
3C\,288   &E&0.246&42.81&41.34&$<$41.41&--&--&$<$42.48&--&--&$<$43.250&--&--&--\\
3C\,289   &N&0.9674&43.99&42.11&--&--&--&--&--&--&45.400&0.010&--&42.57\\
3C\,292   &N&0.71&43.60&40.82&43.62&43.32&43.80&44.40&44.26&44.51&44.800&0.010&--&--\\
3C\,293   &E&0.0452&41.06&40.36&$<$40.62&--&--&42.87&42.74&43.00&43.299&0.001&39.81&41.56\\
3C\,295   &N&0.4614&44.05&40.91&42.50&42.18&42.68&44.48&44.43&44.97&45.004&0.005&41.99&42.33\\
3C\,296   &E&0.0237&40.51&39.68&41.49&41.17&41.85&$<$40.90&--&--&40.816&0.097&39.74&--\\
3C\,299   &N&0.367&42.98&40.23&--&--&--&--&--&--&--&--&--&42.66\\
3C\,300   &N&0.272&42.88&40.91&43.40&43.38&43.42&$<$42.49&--&--&43.400&0.146&42.02&42.48\\
3C\,303   &B&0.141&42.05&41.54&43.91&43.85&43.97&43.91&43.85&43.97&--&--&41.74&41.90\\
3C\,305   &N&0.0417&41.09&39.75&41.42&41.24&41.54&$<$42.00&--&--&--&--&41.04&40.13\\
3C\,309.1 &Q&0.904&44.12&44.40&45.78&45.76&45.79&45.78&45.76&45.79&46.000&0.010&43.70&42.94\\
3C\,310   &E&0.0540&41.87&40.42&--&--&--&--&--&--&42.089&0.032&40.07&--\\
3C\,314.1 &E&0.1197&41.88&$<$39.22&--&--&--&--&--&--&42.098&0.097&39.70&--\\
3C\,315   &N&0.1083&42.00&$<$41.31&$<$41.20&--&--&$<$42.36&--&--&43.010&0.048&40.88&--\\
3C\,319   &E&0.192&42.49&$<$39.64&--&--&--&--&--&--&$<$42.680&--&$<$40.18&39.98\\
3C\,321   &N&0.096&41.77&40.50&41.03&40.98&41.07&43.34&42.98&43.75&44.916&0.001&40.91&41.32\\
3C\,326   &N&0.0895&41.89&40.08&--&--&--&--&--&--&$<$42.160&--&40.40&41.25\\
3C\,325   &Q&0.86&43.96&41.37&$<$43.16&--&--&44.56&44.43&44.70&45.600&0.010&--&42.79\\
3C\,330   &N&0.5490&43.76&40.46&43.08&42.99&43.15&43.90&43.60&44.00&45.000&0.010&--&43.19\\
NGC\,6109 &E&0.0296&40.62&39.44&40.35&39.81&40.57&$<$40.82&--&--&--&--&--&--\\
3C\,334   &Q&0.555&43.39&42.64&45.08&44.99&45.15&45.08&44.99&45.15&45.700&0.010&43.37&42.54\\
3C\,336   &Q&0.927&43.91&42.36&--&--&--&--&--&--&45.400&0.010&43.46&--\\
3C\,341   &N&0.448&43.17&40.41&--&--&--&--&--&--&45.558&0.002&42.80&41.77\\
3C\,338   &N&0.0303&41.29&40.03&40.31&40.08&40.45&$<$41.17&--&--&42.018&0.007&39.54&40.79\\
3C\,340   &N&0.7754&43.67&40.94&--&--&--&--&--&--&44.900&0.010&--&42.67\\
3C\,337   &N&0.635&43.52&40.18&--&--&--&--&--&--&44.300&0.010&--&41.63\\
3C\,343   &Q&0.988&43.90&$<$43.58&--&--&--&--&--&--&45.900&0.010&42.68&41.99\\
3C\,343.1 &N&0.750&43.59&$<$43.17&--&--&--&--&--&--&44.700&0.010&42.71&42.44\\
NGC\,6251 &E&0.024&40.43&40.35&42.77&42.75&42.79&$<$41.58&--&--&42.873&--&--&--\\
3C\,346   &N&0.162&42.15&41.83&43.40&43.38&43.41&$<$42.44&--&--&43.960&0.004&41.33&--\\
3C\,345   &Q&0.594&43.34&44.59&45.64&45.58&45.71&45.64&45.58&45.71&--&--&--&--\\
3C\,349   &N&0.205&42.48&41.10&41.82&41.52&41.92&43.87&43.82&43.91&--&--&41.56&--\\
3C\,351   &Q&0.371&43.06&41.05&41.92&41.74&42.08&44.80&44.77&44.82&46.005&0.001&42.84&--\\
3C\,352   &N&0.806&43.80&41.43&--&--&--&--&--&--&44.800&0.010&--&43.05\\
3C\,380   &Q&0.691&44.32&44.67&45.81&45.72&45.89&45.81&45.72&45.89&45.900&0.010&43.76&42.99\\
3C\,381   &B&0.1605&42.34&40.18&42.11&42.00&42.20&44.31&44.18&44.44&44.650&0.010&42.38&40.92\\
3C\,382   &B&0.0578&41.48&40.85&--&--&--&--&--&--&44.240&0.008&41.78&40.73\\
3C\,386   &E&0.0177&40.51&39.62&--&--&--&--&--&--&41.550&0.007&$<$40.25&--\\
3C\,388   &E&0.0908&41.98&40.77&41.74&41.65&41.81&$<$42.01&--&--&42.660&0.049&40.71&40.52\\
3C\,390.3 &B&0.0569&41.85&41.08&44.49&44.47&44.52&44.50&44.12&44.88&44.370&0.011&42.11&40.95\\
3C\,401   &E&0.201&42.65&41.19&42.74&42.69&42.79&$<$43.05&--&--&43.170&0.125&41.06&--\\
3C\,427.1 &E&0.572&43.83&40.53&$<$42.45&--&--&$<$43.24&--&--&$<$43.800&--&--&--\\
3C\,433   &N&0.1016&42.45&39.77&41.06&40.77&41.22&43.92&43.80&44.02&44.670&0.004&41.68&--\\
3C\,436   &N&0.2145&42.65&41.02&42.59&42.55&42.62&43.53&43.35&43.72&43.520&0.062&41.56&--\\
3C\,438   &E&0.290&43.35&40.87&42.67&42.39&42.84&$<$43.14&--&--&$<$43.270&--&$<$41.47&--\\
3C\,441   &N&0.708&43.70&41.36&--&--&--&--&--&--&44.800&0.010&--&42.42\\
3C\,442A  &E&0.027&40.71&38.21&41.10&40.98&41.20&$<$40.78&--&--&--&--&--&40.56\\
3C\,449   &E&0.0171&40.16&39.08&40.35&40.17&40.49&$<$40.97&--&--&40.358&0.084&39.21&--\\
3C\,452   &N&0.0811&42.23&40.99&$<$41.69&--&--&44.00&43.91&44.10&44.130&0.010&41.35&41.44\\
NGC\,7385 &E&0.0243&40.44&39.90&--&--&--&--&--&--&--&--&--&--\\
3C\,454.3 &Q&0.859&43.70&45.07&46.37&46.24&46.47&46.37&46.24&46.47&--&--&--&--\\
3C\,455   &Q&0.5427&43.41&40.72&--&--&--&--&--&--&--&--&43.07&42.81\\
3C\,457   &N&0.428&43.23&40.69&43.35&43.30&43.40&44.56&44.52&44.88&--&--&42.49&--\\
3C\,465   &E&0.0293&41.16&40.41&41.04&40.70&41.55&$<$41.22&--&--&42.109&0.007&39.79&--\\

\hline
\label{lumtab}
\end{longtable}

\clearpage

\end{document}